\documentclass{article}

\usepackage{PRIMEarxiv}

\usepackage[utf8]{inputenc} 
\DeclareUnicodeCharacter{202F}{\,}
\usepackage[T1]{fontenc}    
\usepackage{url}            
\usepackage{booktabs}       
\usepackage{amsfonts}       
\usepackage{nicefrac}       
\usepackage{microtype}      
\usepackage{lipsum}
\usepackage{fancyhdr}       
\usepackage{graphicx}       
\usepackage{xcolor}
\graphicspath{{media/}}     

\usepackage{array}
\usepackage{mathtools}
\usepackage{color}
\usepackage{amsmath}
\usepackage[indent,bf]{caption}
\usepackage{rotating}
\usepackage{booktabs}
\usepackage{setspace}
\usepackage{ragged2e}
\usepackage{longtable}
\usepackage[colorlinks=true,pdftex,bookmarks=true,bookmarksnumbered=true,breaklinks=true,citecolor=blue]{hyperref}
\usepackage{pdfsync}
\usepackage{enumerate}
\usepackage{titlesec}
\usepackage{etoolbox}
\usepackage{longtable}
\usepackage{scrextend}
\usepackage[T1]{fontenc}
\usepackage{amssymb}
\usepackage{bm}
\usepackage{lineno}
\usepackage{subcaption}
\usepackage{caption}
\usepackage{float}
\usepackage{epstopdf}
\usepackage{pdfpages}
\allowdisplaybreaks[1]
\usepackage{fancyhdr}
\usepackage{setspace}
\usepackage{xcolor}
\usepackage[capitalise]{cleveref}

\usepackage{multirow}

\DeclareMathAlphabet\mathbfcal{OMS}{cmsy}{b}{n}

\usepackage{algorithmicx}
\usepackage[ruled]{algorithm}
\usepackage{algpseudocode}
\usepackage{algpascal}

\title{A Domain Decomposition-based Solver for Acoustic Wave propagation in Two-Dimensional Random Media
}
 \author{
  Sudhi Sharma Padillath Vasudevan \thanks{This research is part of the doctoral thesis submitted to Department of Civil and Environmental Engineering at Carleton University in 2023.} \\
  Department of Civil and Environmental Engineering, \\
  Carleton University,  \\
  Ottawa, Canada.\\
  \texttt{sudhipv@cmail.carleton.ca} \\
}

\definecolor{ss}{rgb}{0.0, 0.0, 0.0}

\begin{document}
\maketitle

\begin{abstract}
An acoustic wave propagation problem with a log normal random field approximation for wave speed is solved using a sampling-free intrusive stochastic Galerkin approach. The stochastic partial differential equation with the inputs and outputs expanded using polynomial chaos expansion (PCE) is transformed into a set of deterministic PDEs and further to a system of linear equations.  Domain decomposition (DD)-based solvers are utilized to handle the overwhelming computational cost for the resulting system with increasing mesh size, time step and number of random parameters. A conjugate gradient iterative solver with a two-level Neumann-Neumann preconditioner is applied here showing their efficient scalabilities.
\end{abstract}



\section{Introduction}

\textcolor{ss}{Uncertainty quantification (UQ) can be regarded as the process of identifying, propagating and quantifying the uncertainty in computational models \cite{book_ghanem,book_raulphsmith}. Uncertainty in these models can be associated with the assumptions used in modelling, inherent variabilities in parameters, initial and/or boundary conditions, the associated data for calibration etc. Inclusion of these uncertainties into the computational modelling framework provides a response with quantified uncertainty which is essential for high consequential systems such as aerospace, military applications, medicine, epidemiology, economics etc.}

\textcolor{ss}{This article focuses on the parametric uncertainty (uncertainty in model parameters including initial or boundary conditions) and their quantification. A functional dependence of uncertainty in input to output is not explicitly available for most physical models. Moreover, this dependence is nonlinear even for a linear system with Gaussian randomness. For example, consider the case of a linear spring with Gaussian variability in stiffness. The uncertainty in output can be found to be non Gaussian because of the nonlinear dependence of model parameter to output.}

\textcolor{ss}{Monte Carlo Simulations provides an estimate of the uncertainty in output by propagating identically distributed independent samples of input through the model. The mean estimate converges to the truth with increasing number of samples. However, the slow rate of convergence necessitates a large number of samples which are prohibitive for high resolution PDE models. PCE based surrogates are widely used to model parametric uncertainties in PDEs replacing MCS \cite{PC_challenges,askey_xiu,xiu_review,compare_eldred_NISP,hazra_TDgPC}. Random quantities are expanded in this method using orthogonal polynomials of random variables forming a subspace. Projections on to this subspace minimizing the error provides the deterministic coefficients in the expansion which are used for computing the response statistics. Non-intrusive spectral projection (NISP) or collocation methods approximates the computation of these deterministic coefficients using sampling or quadrature utilizing the deterministic code as a black box
 \cite{compare_eldred_NISP,knio_book,compare_phipps}. Even though this allows the modeller to reuse existing deterministic model, increasing number of stochastic parameters leads to exponential increase in the number of samples or quadrature points necessary for accuracy. This becomes prohibitive for high resolution PDE models.}

\textcolor{ss}{Uncertainty for high resolution PDE models can also be tackled using Intrusive/sampling-free approaches \cite{book_ghanem}. Different from the sampling approach requiring large number of realizations for convergence, intrusive method directly propagates the input uncertainty through the model. This is achieved by altering the original PDE by a stochastic PDE. This extra effort in one-time conversion can be compensated by an accurate solution compared to NISP and a reduction in execution time for the same set of expansion parameters \cite{compare_phipps}. Spectral stochastic finite element method (SSFEM)/ stochastic Galerkin method developed by Ghanem and Spanos \cite{book_ghanem} is a popular intrusive approach for dealing with stochastic PDEs having random parameters. SSFEM essentially transforms the original stochastic PDE into a set of coupled deterministic PDEs by Galerkin projection. Finite element discretizations of these deterministic PDEs lead to a large set of linear or nonlinear equations. This stochastic system contains a coupled set of $(P+1) \times N$ equations, where $P$ is the number of terms in the spectral expansion of output and $N$ the number of deterministic degrees of freedom in the finite element discretization. Solving this ensuing large system of equations becomes a formidable task considering the coupled structure and computational requirements for inversion. We utilize a DD-based algorithm to tackle this problem by distributing the memory and computations to many processes. A non-overlapping DD-based Neumann-Neumann preconditioner is used along with the conjugate gradient iterative solver which accelerates its convergence. Increasing the number of random variables or order of expansion decreases the sparsity and increases the condition number of the stochastic system which can be handled using these two-level preconditioners. Thus, application of DD-based methods provides a natural way to decompose the overwhelmingly large system of equations from the intrusive method and allows to construct preconditioners to iterative solvers. Wave propagation problems are solved using collocation and stochastic Galerkin methods having random coefficients in \cite{elasticw_sto2d,xiu_wave,SG_app_SWE,SG_application_porousmedia,wave_UQ,wave_MultiLevelMC}.
However, they do not utilize domain decomposition methods or large number of random variables which this article focuses on.}

\textcolor{ss}{On the other hand, "embarrassingly parallel" algorithms can be easily applied for sampling-based approaches by solving a single or many samples in individual cores. However, for high resolution PDE models a single core may not be able to handle the memory requirement even for the deterministic system. Application of DD-based methods to reduce the memory requirement can also be applied here by decomposing the deterministic problem in to subdomains before sampling. However, for each sample evaluation, the communication cost between processes can significantly impact the overall time to solution and reduce the efficiency of the method.}

\textcolor{ss}{Even though the deterministic time dependent problem has an improved condition number compared to the static case with addition of mass and damping matrices, significant challenges still exist in terms of stochastic problems with respect to parallel scaling.}

\begin{itemize}
\item[1.] \textcolor{ss}{The random input parameters of the model are nonlinearly transformed to the output and thus small changes to the inputs can have drastic effects in the output for a stochastic time-dependent problem.}

\item[2.] \textcolor{ss}{Wave propagation problem deals with values in the range sufficiently high as the amplitude of forcing or initial condition and significantly small in the range of zero. These varying data ranges and the CFL conditions necessitate a sufficiently small time step and spatial discretization. It is also important to note that the quantification of uncertainty can become erroneous without a sufficiently discretized deterministic model.}

\item[3.] \textcolor{ss}{The uncertainty for a time dependent problem varies drastically with time for the wave propagation problem. The probability density functions (pdfs) at different time steps of the time dependent problem can be non-Gaussian and with multiple peaks (see \ref{sec:pdfs}).
To capture these uncertainties we need higher order expansions which increases the dimensionality of the problem.}
\end{itemize}

From the above considerations, the main contributions of the article are listed as,

\begin{itemize}
\item[1.] \textcolor{ss}{Development of the mathematical framework for a DD-based two-level parallel scalable solver for the time dependent problem in a deterministic and stochastic setting.}
\item[2.] \textcolor{ss}{Application of sampling-free intrusive stochastic Galerkin method for a time dependent acoustic wave propagation problem with a random field representation of wave speed.}
\item[3.] \textcolor{ss}{Demonstrating the scalabilities of the solver with respect to random parameters.}
\end{itemize}

\textcolor{ss}{This article is organized as follows. The deterministic acoustic wave propagation and the deterministic two-level Neumann-Neumann preconditioner formulation are explained first. The stochastic wave propagation with a random process representation of wave speed is formulated using the intrusive stochastic Galerkin method along with the probabilistic version of the preconditioner. The numerical section illustrates the strong and weak scalabilities of the solver for the stochastic problem. Many related numerical investigations are added in the appendices as follows. A stochastic one-dimensional axial bar vibration problem is considered to study the different uncertainty quantification methods for time dependent problem. The intrusive sampling-free stochastic Galerkin and non-intrusive methods are compared against Monte Carlo solutions. The verifications of solutions for deterministic and stochastic problems in two dimensions are also presented.}

\section{Acoustic Wave Propagation in Deterministic Media}\label{sec:det_acousticwave}
Acoustic wave equation models the propagation of pressure waves through a fluid medium \cite{Reynolds}. The initial disturbance causes the movement of particles to transfer momentum to adjacent particles which continues between particles to set the propagation of waves. The speed of particles in motion is a characteristic of the medium called characteristic speed of propagation or wave speed. This characteristic speed is a function of the temperature, pressure and other characteristics of the fluid medium. In the deterministic setting, this wave speed is assumed a constant whereas in some cases it might be necessary to include the spatio-temporal variation of the wave speed into consideration for accurately predicting the wave mechanics. The amplitude of oscillation for each particle is measured as the sound pressure whose strength depends on the initial disturbance. 
The initial-boundary value problem for a wave propagating through a two-dimensional domain ${D}$ with a constant wave speed $c$ can be written as \cite{zampieri2006implicit,Reynolds}:
\begin{align}\label{Eq.acoustic_deter}
\frac{\partial^2 u (\mathbf{x},t) }{\partial t^2} + \eta \frac{\partial u(\mathbf{x},t)}{\partial t} - c^2 \; \nabla^2 u(\mathbf{x},t) &= f(\mathbf{x},t) \quad  \textrm{in} \quad {D} \times (0,T) \\
u(\mathbf{x},t) &= \Phi(\mathbf{x},t) \quad \textrm{on} \quad {\it{\Gamma}}_D \times (0,T) \\
\nabla u(\mathbf{x},t) \cdot \mathbf{\hat{n}} & = \Psi(\mathbf{x},t ) \quad \textrm{on} \quad {\it{\Gamma}}_N \times (0,T) \\
u(\mathbf{x},t) &= u_0(\mathbf{x}) \quad \textrm{in} \quad D \\
\frac{\partial u(\mathbf{x},t)}{\partial t} &= v_0(\mathbf{x}) \quad \textrm{in} \quad D 
\end{align}
where $u$ is the acoustic pressure $\Phi(\mathbf{x},t)$, $\Psi(\mathbf{x},t )$ are the functional forms of Dirichlet and Neumann boundary conditions and $u_0(\mathbf{x})$ and $v_0(\mathbf{x})$ are the initial pressure and initial velocity of the particle. $\it{\Gamma}_D \cup \it{\Gamma}_N $ is the complete boundary formed by the union of Dirichlet and Neumann parts and $\mathbf{\hat{n}}$ represents the outward normal to the boundary at $\mathbf{x}$. The assumption of constant wave speed is not valid for heterogeneous media. It is possible to represent this wave speed using random fields to obtain a probabilistic representation of the wave. The probabilistic formulation is presented later. The weak form for the deterministic wave equation is as follows \cite{zampieri2006implicit}:

\begin{equation}
(\frac{\partial^2 u}{\partial t^2} , v) + \eta (\frac{\partial u}{\partial t} , v) + a(u, v) = (f, v) \quad \forall v \in V
\end{equation}
where,
\begin{align*}
    a(u, v) &= c^2 \int_{D} {\nabla u \cdot \nabla v} \, d \mathbf{x},\\
(f, v) &= \int_{D} fv \, d \mathbf{x},
\end{align*}
and the test function $v(\mathbf{x})$ lies on $V = \{ v \in H^1({\Omega}) : v = 0 \, \mathrm{on} \, \it{\Gamma}_D \}$.
The above weak form is semi-discretized using the finite element method leading to \cite{zampieri2006implicit,bathe_1}
\begin{equation} \label{eq:acoustic_ode}
  \bf{M} \ddot{u} + \bf{C} \dot{u} + \bf{K} u = f  
\end{equation}
where $\bf{M},\bf{C},\bf{K}$ are the mass, damping and stiffness matrices respectively and $\bf{f}$ is the force vector and $\bf{u}$ is the acoustic pressure. The damping matrix $\bf{C}$ is assumed to be the Rayleigh damping generated by combining mass and stiffness matrices with appropriate coefficients \cite{humar} (see subsection \ref{subsec:rayleigh}). Depending on the wavelength or frequency, spatial mesh resolution $\Delta h$, and the time integration step $\Delta t$ should be selected carefully (i.e. high-frequency waves having short wavelengths require small $\Delta h$ and $\Delta t$.
The system of ODEs in Eq.~(\ref{eq:acoustic_ode}) can be solved numerically using an appropriate time integration scheme. Explicit time integration schemes provide the solution of the current time step in terms of the previous time step. However, these schemes are conditionally stable and result in smaller time step requirements following the Courant-Friedrichs-Lewy condition for a one-dimensional problem as \cite{bathe_1,farhat_nasa,book_nonlinear}:
\begin{equation}\label{Eq.CFL}
\frac{c \,\, \Delta t}{\Delta h} = C_{CFL} \, \leq C_{max}
\end{equation}
where $c$ is the maximum wave speed in the domain and $C_{CFL}$ is the Courant number and $C_{max}$ is the maximum allowable value. This condition becomes more restrictive for two-dimensional and three-dimensional problems.
Implicit time integration schemes provide the solution for the current time step by solving a system of equations involving both the current and previous time steps. The implicit scheme is computationally more expensive than the explicit method but has an unconditional stability which allows larger time steps and thus reduces total time consumption \cite{bathe_1, bathe_2}. The Newmark-beta implicit scheme which has a second-order convergence and an unconditional stability property is selected in this study for the time integration \cite{book_hughes,bathe_book}. The acoustic pressure and its higher order time-derivatives (such as velocity $v$ and acceleration $a$) at a given time step for this scheme can be written as\cite{book_hughes}:
\begin{align}
\mathbf{u}_{n+1} = \mathbf{u}_n + \Delta t \mathbf{v}_n + (1-2\zeta)  \Delta t^2 \frac{\mathbf{a}_n}{2} + \zeta \Delta t^2 \mathbf{a}_{n+1} \\
\mathbf{v}_{n+1} = \mathbf{v}_n + (1-\gamma)  \Delta t \mathbf{a}_n + \gamma \Delta t \mathbf{a}_{n+1}
\end{align}
where $\gamma = \frac{1}{2}$ and $\zeta = \frac{1}{4}$ are constants. Consequently the time discretized ODEs Eq.~(\ref{eq:acoustic_ode}) leads to the following linear systems as:
\begin{equation}\label{Eq:Kuf}
\mathbf{\tilde{K}} \mathbf{u}_{n+1} = \mathbf{\tilde{f}}_{n+1},
\end{equation} 
where $\mathbf{\tilde{K}}$ is called the transient operator or dynamic stiffness matrix expressed as \cite{farhat_nasa}:
\begin{equation}\label{Eq:K_T}
\mathbf{\tilde{K}} = \bf{M} \frac{\mathrm{1}}{\zeta \mathrm{\Delta t^2}} + \bf{C} \frac{\gamma}{\zeta \mathrm{\Delta t} } + \bf{K}.
\end{equation}
with the forcing $\mathbf{\tilde{f}_{n+1}}$ computed as:
\begin{equation}
\mathbf{\tilde{f}_{n+1}} = \bf{f}_{n+1} + \bf{f_m} + \bf{f_c}
\end{equation}
where,
\begin{align}\label{Eq:fparts}
    \bf{f_m} &= \bf{M}  \bf{\tilde{u}_m},  \\ \label{Eq:fparts2}
    \bf{f_c} &= \bf{C}   \bf{\tilde{u}_c}, 
\end{align}
and
\begin{align}\label{Eq:UmCm}
    \bf{\tilde{u}_m} &=  \frac{\bf{u_n} + \mathrm{\Delta t}  \bf{v_n} } {\zeta \Delta t^2}+\frac{(1-2\zeta)\bf{a_n}}{2\zeta}, \\ \label{Eq:UmCm2}
    \bf{\tilde{u}_c} &=  \gamma \Delta t  \bf{\tilde{u}_m}  - \bf{v_n} -  (\mathrm{1}-\gamma) \bf{a_n}  \mathrm{\Delta t}.
\end{align}

Note that the Eq.~(\ref{Eq:Kuf}) at each time step involves the solution of only a static system with varying forcing terms. The transient stiffness operator in Eq.~(\ref{Eq:K_T}) does not depend on time and can be pre-assembled and reused reducing computational time and storage requirements. The coefficient matrix in Eq.~(\ref{Eq:Kuf}) is symmetric and positive-definite which can be solved using a conjugate-gradient-based iterative solver with a non-overlapping DD-based preconditioner as described in the next section.

\section{Deterministic Two-level Neumann-Neumann Preconditioner}

\begin{figure}[htbp]
\centering   
\includegraphics[width=2.6in,height=2.6in]
{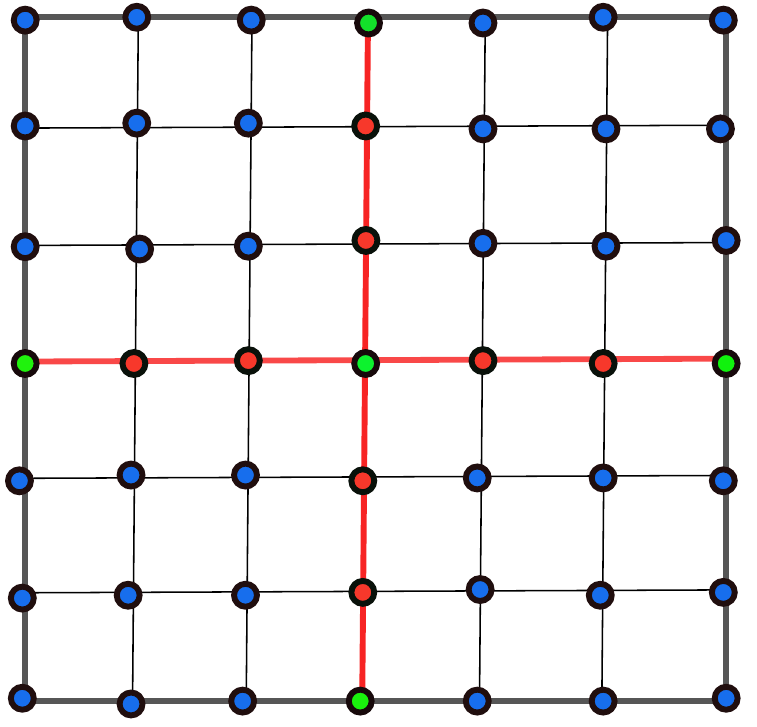}
\caption{A unit square domain with four subdomains showing interior (blue), interface (red + green ), corner (green) and remaining (red) nodes.}\label{fig.interface_corner}
\end{figure}

Domain decomposition methods rely on the idea of dividing the original domain into overlapping or non-overlapping subdomains with smaller sub-problems which can be solved in parallel using high-performance computing systems. All the neighbouring subdomains share a common boundary with a set of nodes called interface nodes as shown in Fig.~\ref{fig.interface_corner}. The interface nodes can also be decomposed into corner and remaining nodes as shown which will be used for constructing the coarse grid as explained later. The remaining nodes are interior to each subdomain and are called interior nodes. 

Eliminating the interior unknowns from each of these subdomains generates a dense linear system generally referred to as the interface Schur complement system \cite{book_DD_tarek,dd_TFchan,book_DD_widlund}. The Schur complement system has a better condition number than that of the original system and thus takes less computational cost to solve \cite{dd_TFchan,subber_thesis}. The direct solution to the Schur complement system and back substitution to find the interior unknowns is generally known as the direct substructuring method \cite{dd_TFchan}. For large Schur complement systems, an iterative solver with an efficient preconditioner is used. Several variants of non-overlapping DD-based preconditioners are developed which scale with respect to mesh size and number of subdomains. This article utilizes a two-level Neumann-Neumann preconditioner which is analogous to the Balancing Domain Decomposition by constraints (BDDC) method \cite{cros_2LNN,bddc_mandel}.

At each time step, a linear system as in Eq.~(\ref{Eq:Kuf}) is solved using the DD-based preconditioner as:
\begin{equation}
\mathbf{\tilde{K}} \mathbf{u} = \mathbf{\tilde{f}}.
\end{equation} \label{Eq:Kuf-2}
The nodes inside each subdomain are decomposed as interior $\bf{u}_{I}^s$, and interface $\bf{u}_{\it{\Gamma}}^s$ nodes as \cite{subber_thesis}:

\begin{equation}
\mathbf{u}^s =
\begin{Bmatrix}
     {\bf{u}}_{I}^s
\\[0.3em]
     {\bf{u}}_{{\it{\Gamma}}}^s
\\[0.3em]
\end{Bmatrix}.
\end{equation}
where the subscript $I$ and $\Gamma$ denote the interior and interface. A restriction operator $\bf{R}_{s}$, consisting of zeros and ones mapping the global interface nodes to the subdomain interface nodes is constructed as \cite{subber_thesis}:
\begin{equation}
\bf{u}_{\it{\Gamma}}^s = \bf{R}_{s} {u}_{\it{\Gamma}}.
\end{equation}
The equilibrium equations for each of the subdomains can be written in terms of interior and interface nodes as \cite{subber_thesis}:
\begin{equation}
{\begin{bmatrix}
    {\tilde{\mathbf{K}}}_{II}^s & {\tilde{\mathbf{K}}}_{I \it{\Gamma}}^s
\\[0.3em]
     {\tilde{\mathbf{K}}}_{\it{\Gamma} I}^s & {\tilde{\mathbf{K}}}_{\it{\Gamma} \it{\Gamma}}^s
\end{bmatrix}}
\begin{Bmatrix}
     {\mathbf{u}}_{I}^s
\\[0.3em]
     \mathbf{u}_{\it{\Gamma}}^s
\\[0.3em]
\end{Bmatrix}  =
\begin{Bmatrix}
     \tilde{\mathbf{f}}_{I}^s
\\[0.3em]
     \tilde{\mathbf{f}}_{\it{\Gamma}}^s
\\[0.3em]
\end{Bmatrix}.
\end{equation}
where $ \tilde{\mathbf{f}}_{I}^s$ and $\tilde{\mathbf{f}}_{\it{\Gamma}}^s$ correspond to the interior and interface forces respectively which contain contributions from mass and damping terms of the previous time step as shown in Eqs. (\ref{Eq:fparts}) and (\ref{Eq:fparts2}). The decomposed force vector with these contributions is written as:

\begin{equation}\label{Eq.fmis}
{\begin{Bmatrix}
   \tilde{\bf{f}}_{mI}^s
\\[0.3em]
    \tilde{\bf{f}}_{m\it{\Gamma}}^s
\\[0.3em]
\end{Bmatrix} }=
\begin{bmatrix}
    \mathbf{M}_{II}^s & \mathbf{M}_{I \it{\Gamma}}^s
\\[0.3em]
     \mathbf{M}_{\it{\Gamma} I}^s & \mathbf{M}_{\it{\Gamma} \it{\Gamma}}^s
\end{bmatrix}
\begin{Bmatrix}
     \tilde{\bf{u}}_{mI}^s
\\[0.3em]
     \tilde{\bf{u}}_{m\it{\Gamma}} ^s
\\[0.3em]
\end{Bmatrix},
\end{equation}
and
\begin{equation}\label{Eq.fcis}
{\begin{Bmatrix}
   \tilde{\bf{f}}_{cI}^s
\\[0.3em]
    \tilde{\bf{f}}_{c\it{\Gamma}}^s
\\[0.3em]
\end{Bmatrix} }=
\begin{bmatrix}
    \mathbf{C}_{II}^s & \mathbf{C}_{I \it{\Gamma}}^s
\\[0.3em]
     \mathbf{C}_{\it{\Gamma} I}^s & \mathbf{C}_{\it{\Gamma} \it{\Gamma}}^s
\end{bmatrix}
\begin{Bmatrix}
     \tilde{\bf{u}}_{cI}^s
\\[0.3em]
     \tilde{\bf{u}}_{c\it{\Gamma}} ^s
\\[0.3em]
\end{Bmatrix},
\end{equation}
where $\tilde{\bf{u}}_{mI}^s$, $\tilde{\bf{u}}_{\it{m \Gamma}}^s$,$\tilde{\bf{u}}_{cI}^s$ and $\tilde{\bf{u}}_{\it{c\Gamma}}^s$ for corresponding interior and interface nodes can be calculated as in Eqs. (\ref{Eq:UmCm}) and (\ref{Eq:UmCm2}). The global system combining all subdomains is written as:

\begin{equation}\label{Eq:GAssembly}
{\begin{bmatrix}
     \tilde{\mathbf{K}}_{II}^{\it{1}} & \dots  &   0    \ \  &  \tilde{\mathbf{K}}_{I {\it{\Gamma}}}^{\it{1}} \mathbf{R}_{\it{1}}
\\[0.3em]
     \vdots   & \ddots & \vdots \ \  &  \vdots
\\[0.3em]
     0        & \dots  &  \tilde{\mathbf{K}}_{II}^{\it{n_s}} \ \ &  \tilde{\mathbf{K}}_{I {\it{\Gamma}}}^{\it{n_s}} \mathbf{R}_{\it{n_s}}
\\[0.3em]
    \bf{R}^{\it{T}}_{\it{1}} \tilde{\mathbf{K}}_{\it{\Gamma I}}^{\it{1}} & \dots  &  \mathbf{R}^T_{\it{n_s}} \tilde{\mathbf{K}}_{{\it{\Gamma}} I}^{\it{n_s}} \ \ & \displaystyle\sum_{s=1}^{\it{n_s}} \mathbf{R}^{\it{T}}_s \tilde{\mathbf{K}}_{{\it{\Gamma}} {\it{\Gamma}}}^s \mathbf{R}_{\it{s}}
\end{bmatrix}}
\begin{Bmatrix}
     \bf{u}_{\it{I}}^{\it{1}}
\\[0.3em]
     \vdots
\\[0.3em]
      \bf{u}_{\it{I}}^{\it{n_s}}
\\[0.3em]
      \bf{u}_{\it{\Gamma}}
\\[0.3em]
\end{Bmatrix}  =
\begin{Bmatrix}
     \tilde{\mathbf{f}}_I^{\it{1}}
\\[0.3em]
     \vdots
\\[0.3em]
     \tilde{\mathbf{f}}_{I}^{\it{n_s}}
\\[0.3em]
     \displaystyle\sum_{s=1}^{\it{n_s}} \bf{R}^{\it{T}}_s  \tilde{f}_{\it{\Gamma}}^{\it{s}}
\end{Bmatrix}.
\end{equation}

After applying Gaussian block elimination the above equation can be recast as \cite{SG_DD_sarkar}:

\begin{equation}\label{Eq:SchurAssembly}
{\begin{bmatrix}
     \tilde{\mathbf{K}}_{II}^{\it{1}} & \dots  &   0    \ \  &  \tilde{\mathbf{K}}_{I {\it{\Gamma}}}^{\it{1}} \mathbf{R}_{\it{1}}
\\[0.3em]
     \vdots   & \ddots & \vdots \ \  &  \vdots
\\[0.3em]
     0        & \dots  &  \tilde{\mathbf{K}}_{II}^{\it{n_s}} \ \ &  \tilde{\mathbf{K}}_{I {\it{\Gamma}}}^{\it{n_s}} \mathbf{R}_{\it{n_s}}
\\[0.3em]
   0 & \dots  &  0 \ \ & \displaystyle\sum_{s=1}^{\it{n_s}} \mathbf{R}^{\it{T}}_s \mathbf{S}_s \mathbf{R}_{\it{s}}
\end{bmatrix}}
\begin{Bmatrix}
     \bf{u}_{\it{I}}^{\it{1}}
\\[0.3em]
     \vdots
\\[0.3em]
      \bf{u}_{\it{I}}^{\it{n_s}}
\\[0.3em]
      \bf{u}_{\it{\Gamma}}
\\[0.3em]
\end{Bmatrix}  =
\begin{Bmatrix}
     \tilde{\mathbf{f}}_I^{\it{1}}
\\[0.3em]
     \vdots
\\[0.3em]
     \tilde{\mathbf{f}}_{I}^{\it{n_s}}
\\[0.3em]
     \displaystyle\sum_{s=1}^{\it{n_s}} \bf{R}^{\it{T}}_s  \mathbf{g}_{\it{\Gamma}}^{\it{s}}
\end{Bmatrix}.
\end{equation}
where $\bf{S}_{\it{s}}$ and $\bf{g}_{\it{\Gamma}}^{\it{s}}$ are the subdomain level Schur complement and corresponding force vector defined as:

\begin{align}
\mathbf{S}_s  &=  \tilde{\mathbf{K}}_{{\it\Gamma} {\it\Gamma}}^s - \tilde{\mathbf{K}}_{\it{\Gamma} I}^s [\tilde{\mathbf{K}}_{II}^{s}]^{-1} \tilde{\mathbf{K}}_{I \it{\Gamma}}^s \\
\mathbf{g}_{\it{\Gamma}}^s &= \tilde{\mathbf{f}}_{\it{\Gamma}}^s - \tilde{\mathbf{K}}_{\it{\Gamma} I}^s [\tilde{\mathbf{K}}_{II}^{s}]^{-1} \tilde{\mathbf{f}}_I^s 
\end{align}

It is possible to combine the subdomain level interface problems into a global system of equations as \cite{subber_thesis}:

\begin{equation}
\mathbf{S} \mathbf{u}_{\it{\Gamma}} = \mathbf{g}_{\it{\Gamma}}
\end{equation}
where
\begin{align}
\mathbf{S} &= \sum_{s=1}^{n_s}  \mathbf{R}^T_s \mathbf{S}_s \mathbf{R}_s \\
\mathbf{g}_{\it{\Gamma}} &= \sum_{s=1}^{n_s} \mathbf{R}^T_s  \mathbf{g}_{\it{\Gamma}}^s.
\end{align}
Note that by eliminating the interior unknowns from the global system of equations, a smaller dense interface problem is generated (to be solved before interior unknowns can be computed). A two-level Neumann-Neumann preconditioner can be constructed for the interface problem by decomposing the interface into corner and remaining nodes as \cite{subber_thesis}:

\begin{equation}\label{eq:uruc}
\begin{Bmatrix}
     {{\mathbf{u}^s_r}}
\\[0.3em]
     {{\mathbf{u}^s_c}}
\\[0.3em]
\end{Bmatrix}  =
\begin{Bmatrix}
     {{\mathbf{R}_s^r}}
\\[0.3em]
     {{\mathbf{R}_s^c}}
\\[0.3em]
\end{Bmatrix}
\mathbf{u}^s_{\it\Gamma},
\end{equation}
where $\mathbf{u}^s_c$ are the corner nodes (also known as cross points shown in Fig.~\ref{fig.interface_corner}) and $\mathbf{u}^s_r$ are the remaining nodes. The corner nodes are identified as the nodes on the interface which are shared among more than two subdomains and the end nodes of interface edges. The remaining interface nodes are on the interface shared between only two subdomains. A local Dirichlet problem on these subdomains can be solved by assuming the values of the interface unknowns. It is thus possible to calculate the residual force developed on the subdomain boundaries owing to the erroneous initial guess of interface values. This can be written in terms of the Schur complement of each subdomain as \cite{subber_thesis}:
\begin{equation}
\mathbf{r}_{\it{\Gamma}} ^s = \mathbf{S}_s \mathbf{u}_{\it{\Gamma}} ^s - \mathbf{g}_{\it{\Gamma}}^s.
\end{equation}
The contributions from all the subdomains are aggregated to obtain the global residual force vector from which the incremental solution for the next iteration ($j$) is computed as \cite{subber_thesis}:
\begin{equation}
\mathbf{r}_{\it{\Gamma}_j} = \sum_{s=1}^{n_s} \mathbf{R}^T_s \mathbf{r}_{\it{\Gamma}} ^s.
\end{equation}
The residual force vector for each subdomain can then be computed as \cite{subber_thesis}:
\begin{equation}
\bf{r}_{\it{\Gamma}} ^s = {\bf D}_{s} {\bf{R}}_{s} \bf{r}_{\it{\Gamma}_j}.
\end{equation}
where the partition of unity matrix, $\mathbf{D}_s$ is computed to distribute the subdomain contributions without redundancy as \cite{book_DD_tarek}:
\begin{equation}\label{Eq.det_scaling}
\sum_{s=1}^{n_s} {\bf{R}}^T_{s} \ {\bf D}_{s} \ {\bf{R}}_{s} = {\bf{I}}.
\end{equation}
The increment in solution at the boundary due to the residual vector can be computed from the subdomain Schur complement problem as defined earlier. Generally, it is not necessary to assemble the subdomain level Schur complement explicitly and compute its inverse. Only an effect of this matrix is calculated by solving a Neumann problem as \cite{subber_thesis}:

\begin{equation}\label{eq:N-NNC2}
{\begin{bmatrix}
   \bf{\tilde{K}}^{\it{s}}_{\it{ii}} & \bf{\tilde{K}}^{\it{s}}_{\it{ir}} & \bf{\tilde{K}}^{\it{s}}_{\it{ic}}
\\[0.3em]
     \bf{\tilde{K}}^{\it{s}}_{\it{ri}}   & \bf{\tilde{K}}^{\it{s}}_{\it{rr}} & \bf{\tilde{K}}^{\it{s}}_{\it{rc}}
\\[0.3em]
     \bf{\tilde{K}}^{\it{s}}_{\it{ci}}  & \bf{\tilde{K}}^{\it{s}}_{\it{cr}} & \bf{\tilde{K}}^{\it{s}}_{\it{cc}}
\end{bmatrix}}
\begin{Bmatrix}
     \bf{x}^{\it{s}}_{\it{i}}
\\[0.3em]
     \bf{u}^{\it{s}}_{\it{r}}
\\[0.3em]
      \bf{u}^{\it{s}}_{\it{c}}
\\[0.3em]
\end{Bmatrix}  =
\begin{Bmatrix}
     \bf{0}
\\[0.3em]
    \bf{\tilde{f}}^{\it{s}}_{\it{r}}
\\[0.3em]
    \bf{\tilde{f}}^{\it{s}}_{\it{c}}
\\[0.3em]
\end{Bmatrix},
\end{equation}
where,
\begin{equation}
\begin{Bmatrix}
      \bf{\tilde{f}}^{\it{s}}_{\it{r}}
\\[0.3em]
      \bf{\tilde{f}}^{\it{s}}_{\it{c}}
\\[0.3em]
\end{Bmatrix}  =
\begin{Bmatrix}
     \bf{R}_{\it{s}}^{\it{r}}
\\[0.3em]
     \bf{R}_{\it{s}}^{\it{c}}
\\[0.3em]
\end{Bmatrix}
\bf{r^{\it{s}}_{\it\Gamma}}.
\end{equation}
Eliminating the variable $\bf{x}^{\it{s}}_{\it{i}}$ by solving a Dirichlet problem one obtains \cite{subber_thesis}:
\begin{equation}\label{eq:N-NNC3}
{
\begin{bmatrix}
   {\mathbf{S}_{rr} ^s}  & {\mathbf{S}_{rc} ^s \mathbf{B}_c ^s }
\\[0.3em]
     {\displaystyle \sum_{s=1}^{n_s} \mathbf{B}^{sT}_c \mathbf{S}_{cr} ^s } &  { \displaystyle \sum_{s=1}^{n_s} \mathbf{B}^{sT}_c \mathbf{S}_{cc} ^s \mathbf{B}^{sT}_c}
\\[0.3em]
\end{bmatrix}
}
{
\begin{Bmatrix}
     {{\mathbf{u}^s_r}}
\\[0.3em]
      {{\mathbf{u}^s_c}}
\\[0.3em]
\end{Bmatrix} 
}  =
\begin{Bmatrix}
    \tilde{\mathbf{f}}^s_r
\\[0.3em]
    \displaystyle \sum_{s=1}^{n_s} \mathbf{B}^{sT}_c \tilde{\mathbf{f}}^s_c
\\[0.3em]
\end{Bmatrix},
\end{equation}
where each Schur complement matrix can be defined for the corresponding corner and remaining nodes as:
\begin{equation}\label{eq:Schur_alphabeta}
\mathbf{S}^s_{\alpha \beta} = \mathbf{A}^s_{\alpha \beta} - \mathbf{A}^s_{\alpha i} [\mathbf{A}^s_{ii}]^{-1} \mathbf{A}^s_{i\beta},
\end{equation}
where $\alpha$ and $\beta$ are dummy variables with values representing respectively $r$ and $c$ and $\mathbf{B}_c^s$ is the boolean restriction operator mapping global corner nodes to local corner nodes as \cite{subber_thesis}:
\begin{equation}
{\bf{u}}_c ^{s}= {\bf{B}}_c ^{s} {\bf{u}}_c.
\end{equation}
Elimination of the remaining nodes from the Eq.~(\ref{eq:N-NNC3}) leads to the coarse problem of our preconditioner as \cite{subber_thesis}:
\begin{equation}\label{eq:coarsegrid}
\mathbf{F}_{cc} \mathbf{u}_c = \mathbf{d}_c,
\end{equation}
where the operator $\mathbf{F}_{cc}$ and corresponding force vector $\mathbf{d}_{c}$ for corner nodes is defined as:
\begin{align}\label{eq:Fcc}
\mathbf{F}_{cc} &= \sum_{s=1}^{n_s} {\mathbf{B}_c^s}^T (\mathbf{S}_{cc}^s - \mathbf{S}_{cr}^s [\mathbf{S}_{rr}^s]^{-1} \mathbf{S}_{rc}^s)\mathbf{B}_c^s, \\
\mathbf{d}_{c} &= \sum_{s=1}^{n_s} {\mathbf{B}_c^s}^T (\tilde{\mathbf{f}}^s_c - \mathbf{S}_{cr}^s [\mathbf{S}_{rr}^s]^{-1} \tilde{\mathbf{f}}^s_r).
\end{align}
The remaining interface unknown $\mathbf{u}_r^s$ can be computed in parallel as follows:
\begin{equation}\label{eq:urs}
\mathbf{S}^s_{rr} \mathbf{u}_r^s = \tilde{\mathbf{f}}^s_r  - \mathbf{S}^s_{rc} \mathbf{B}_c^s \mathbf{u}_c.
\end{equation}
Further, the global interface solution can be computed as:
\begin{equation}\label{eq:ugb}
\mathbf{u}_{\it\Gamma} =  \sum_{s=1}^{n_s}  \mathbf{R}_{s}^{T} \mathbf{D}_{s} ({\mathbf{R}^{r}_{s}}^{T} \mathbf{u}_{r}^{s} + {\mathbf{R}_{s}^c}^{T} \mathbf{u}^s_c).
\end{equation}
Finally, the two-level Neumann-Neumann preconditioner can be expressed as \cite{subber_thesis}:
\begin{equation}\label{eq:NNCP}
\mathbf{M}_{NNC}^{-1} = \sum_{s=1}^{n_s} \mathbf{R}_s^T \mathbf{D}_s({\mathbf{R}_s^r}^T [\mathbf{S}^s_{rr}]^{-1} \mathbf{R}_s^r) \mathbf{D}_s \mathbf{R}_s  + \mathbf{R}_0^T [\mathbf{F}_{cc}]^{-1} \mathbf{R}_0,
\end{equation}
where $\mathbf{R}_0$ is defined as:
\begin{equation}
\mathbf{R}_0 = \sum_{s=1}^{n_s} {\mathbf{B}_c^s}^T ({\mathbf{R}_s^c} - \mathbf{S}_{cr}^s [\mathbf{S}^s_{rr}]^{-1} \mathbf{R}_s^r) \mathbf{D}_s \mathbf{R}_s.
\end{equation}
A conjugate gradient iterative solver equipped with the above two-level Neumann-Neumann preconditioner is used to solve the system in Eq.~(\ref{Eq:Kuf}).
A pseudo-code for acoustic wave propagation using a two-level Neumann-Neumann preconditioner is outlined in Algorithm \ref{algorithm:1}. More details on the implementation and algorithms for preconditioned conjugate gradient iterative solver with the above preconditioner can be found in \cite{subber_thesis, ajit_thesis, ajit_CMAM}.

\alglanguage{pseudocode}
\begin{algorithm}
\caption{: \textcolor{blue}{Pseudo Code for Acoustic Wave Propagation in a Deterministic Medium}}
\label{algorithm:1}
\begin{algorithmic}[1]
\State {\bf{Initial Condition} :}  $n =0$, $u_0, v_0, a_0$
\For{each process, $ s = 0, 1, 2, ... n_s $}
\State Assemble $K^s$, Stiffness Matrix  
\State Assemble $M^s$, Mass Matrix 
\State Assemble $C^s$, Damping Matrix 
\State Assemble $f^s$, Force Vector
\State Decompose $u_0, v_0, a_0 $  as $u_{\Gamma,0}^{s},v_{\Gamma,0}^{s},a_{\Gamma,0}^{s},u_{I,0}^{s},v_{I,0}^{s},a_{I,0}^{s}$
\State \textcolor{blue}{\bf{Compute Transient Stiffness Matrix:}}   $\tilde{K}^s = M ^s\frac{1}{\zeta \Delta t^2} + C^s \frac{\gamma}{\zeta \Delta t} + K^s $ (see Eq.~(\ref{Eq:K_T}) )
\EndFor \vspace{2 mm}
 \While {n $<T_N$} 
 
 \State $t = (n+1) \Delta t$
 
 \For {$s=1:n_s$}
  
  \State \textcolor{blue}{{\bf{Compute Mass Component of Force:}}}  $f_{mI,n}^s$, $f_{m\it{\Gamma},n}^s$ (see Eq.~(\ref{Eq.fmis}) )
  \State \textcolor{blue}{{\bf{Compute Damping Component of Force:}}}  $f_{cI,n}^s$, $f_{c\it{\Gamma},n}^s$ (see Eq.~( \ref{Eq.fcis}) )
  
  \State $\tilde{f}_{\Gamma,n+1} ^{s} = f^s _{\Gamma,n+1} + f_{m\Gamma,n} ^{s}+ f_{c\Gamma,n} ^{s} $\;
  \State $\tilde{f}_{I,n+1} ^{s} = f^{s} _{I,n+1} + f_{mI,n} ^{s}+ f_{cI,n} ^{s} $\;
 
 \State $ g_\Gamma = g_\Gamma + R_s ^T [\tilde{f} ^s _{\Gamma} - \tilde{K} ^s _{\Gamma I}[\tilde{K} ^s _{II}]^{-1} \tilde{f} ^s _{I}] R_s $
 
 \EndFor

 \State \textcolor{blue}{\bf{Apply Conjuagte Gradient Iterative solver with Two level Neumann-Neumann Preconditioner}} : $\tilde{K} u_{n+1} = \tilde{F}_{n+1}$ (see Eq.~(\ref{eq:NNCP}) )
 \State Compute $u_{\it{\Gamma},n+1}, u_{I,n+1} ^s$
\State Update initial solution with new solution as $u_{\it{\Gamma},0}^{s} = R_s u_{\it{\Gamma},n+1}, \quad u_{I,0}^{s} = u_{I,n+1}^{s}$
 \State  $n = n + 1$

 \EndWhile

\State {\bf{Output} :}  $U(ndof,ndof,T_N)$
\end{algorithmic}
\end{algorithm}

\section{Acoustic Wave propagation in Random Media}\label{sec:acwave_random_c2}

The speed of propagation of acoustic waves through a medium is dependent on the specific characteristics of the medium such as pressure, temperature, density etc., which can vary significantly. Thus, considering the wave speed as a constant parameter as described in earlier sections is not appropriate for highly consequential applications such as medical science, military, aerospace etc., where quantified variability in predictions are important. In this section, the uncertain nature of wave speed is considered as a random field. The stochastic PDE for acoustic waves considering the wave speed as a random field can be written as \cite{motamed2013stochastic}:
\begin{align}\label{Eq.acous_sto}
\frac{\partial^2 u(\mathbf{x},t,\theta)}{\partial t^2} + \eta \frac{\partial u(\mathbf{x},t,\theta)}{\partial t} - \nabla . ( c_s(\mathbf{x},\theta)  \nabla u (\mathbf{x}, t, \theta)) &= f (\mathbf{x}, t) \quad  \textrm{in} \quad D \times (0,T) \times \Omega \\
u(\mathbf{x},t, \theta) &= \Phi(\mathbf{x},t,\theta) \quad  \textrm{on} \quad  {\it{\Gamma}}_D \times (0,T) \times \Omega \\
\nabla u(\mathbf{x},t,\theta) \cdot \mathbf{\hat{n}}  &= \Psi(\mathbf{x},t,\theta) \quad  \textrm{on} \quad {\it{\Gamma}}_N \times (0,T) \times \Omega \\
u(\mathbf{x},t, \theta) &= u_0(\mathbf{x}, \theta)\quad   \textrm{in} \quad  D \times \Omega\\
\frac{\partial u(\mathbf{x},t, \theta)}{\partial t} &= v_0(\mathbf{x},\theta) \quad  \textrm{in} \quad  D \times \Omega 
\end{align}
where $u$ is the random acoustic pressure field and $c(\mathbf{x}, \theta)$ is the square of the wave speed represented as a log-normal random field. $\Omega$ represents the set of all possible outcomes associated with the probability space $(\Omega, \sigma, P)$ and $\theta$ is the random aspect of the problem \cite{xiu_wave,wave_UQ}. $\Phi(\mathbf{x},t,\theta)$, $\Psi(\mathbf{x},t,\theta)$ are the random Dirichlet and Neumann boundary conditions and $u_0(\mathbf{x},\theta)$ and $v_0(\mathbf{x},\theta)$ are the random initial pressure and velocity of the particle. $\it{\Gamma}_D \cup \it{\Gamma}_N $ is the complete boundary formed by the union of Dirichlet and Neumann parts and $\mathbf{\hat{n}}$ represents the outward unit normal to the boundary at $\mathbf{x}$. For the current model, initial and boundary conditions are considered deterministic (same as in Section \ref{sec:det_acousticwave}) even though they can also be considered random.

Spatial discretization of the weak form leads to a system of stochastic ODEs as:
\begin{equation}
\label{Eq.randODEchapt2}
\bf{M} (\theta) \ \ddot{\bf{u}}(\theta) +  {\bf{C}} (\theta) \dot{\bf{u}}(\theta) + {\bf{K}} (\theta) \bf{u}(\theta) \approx \bf{f}
\end{equation}
where $\bf{M}, \bf{C}, \bf{K}$ are the random mass, damping and stiffness matrices (although in this special case, $M$ is considered deterministic). The square of wave speed, $c_s$ in the above problem is expanded as a log-normal random field (enforcing positivity) computed as the exponential of a Gaussian field with a known covariance function. The KLE provides a means to expand any input random process having a known covariance function in terms of uncorrelated orthogonal random variables. An exponential covariance kernel is assumed for the spectral expansion of the underlying Gaussian process in two dimensions as \cite{book_ghanem}:
\begin{equation}\label{Eq.CovarKer2D_chapt2}
\mathcal{H}(x_1,x_2;y_1,y_2) = \sigma^2\ e^{-\left({\frac{|x_1-x_2|}{b_x}  } + {\frac{|y_1-y_2|}{b_y}} \right) }
\end{equation}
where $\sigma^2$ is the variance of the process and $b_x$ and $b_y$ are the correlation lengths in $x$ and $y$ directions. The decomposition of this covariance kernel in terms of eigenvalues and eigenfunctions can be written as \cite{book_ghanem}:
\begin{equation}\label{Eq.KLEchapt2}
g(\mathbf{x},\theta) \approx {g}_0(\mathbf{x})+\sum_{n=1}^{L}{\xi_n(\theta)\sqrt{\lambda_{n}}\phi(\mathbf{x})}
\end{equation}
where $g_0(\mathbf{x})$ is the mean field, $\lambda$ and $\phi$ are the eigenvalues and eigenfunctions of the covariance kernel and $\xi_{1},\xi_{2},...,\xi_{L}$ are the uncorrelated Gaussian random variables having zero mean and unit variance. The log-normal field for the square of wave speed $c_s(\mathbf{x},\theta)$ is computed by taking the exponential of the expansion in Eq.~(\ref{Eq.KLEchapt2}) and can be approximated using PCE as\cite{lognormal_roger}:
\begin{equation}\label{Eq.c0pce}
c_s(\mathbf{x},\theta) \approx  \sum_{i=0}^{M} c_i(\mathbf{x})  \frac{\langle\Psi_i(\bm{\eta})\rangle}{\langle\Psi_i^2(\bm{\eta})\rangle}\Psi_i(\bm{\xi})
\end{equation}
where 
\begin{equation}\label{Eq.LNL0}
c_0(\mathbf{x}) \approx \mathrm{exp}\left[ g_0(\mathbf{x}) + \frac{1}{2}  \sum_{i=1}^{L} g_i^2(\mathbf{x}) \right]
\end{equation}
is the mean of the log normal expansion and each $ \langle\Psi_i(\bm{\eta})\rangle$ are the expectation of functionals centered around $ g_i(\mathbf{x}) $ \cite{ajit_thesis,subber_thesis,mohammed_thesis}. Each $ \langle\Psi_i(\bm{\xi})\rangle$ is the orthogonal polynomial chaos basis satisfying \cite{book_ghanem},
\begin{equation}\label{Eq.PCEOrtho}
\langle\Psi_i(\bm{\xi}),\Psi_j(\bm{\xi})\rangle \;=\; \langle{{\Psi ^2} _i (\bm{\xi})}\rangle\;\delta_{ij}.
\end{equation}
The random field for the acoustic pressure $u(\mathbf{x},t,\theta)$ can also be expanded in terms of PCE as:
\begin{equation} \label{Eq. uPCE}
u(\mathbf{x},t,\theta) \approx \sum_{j=0}^{N} u_{j}(\mathbf{x},t) \Psi_j(\bm{\xi}).
\end{equation}
where $u_{j}$ are the deterministic polynomial chaos coefficients to be computed 
using the stochastic Galerkin method. 
Applying the PCE for input and output terms in Eq.~(\ref{Eq.randODEchapt2}) provides:
\begin{equation}
\sum_{l=0} ^{L} \mathbf{{M}}_l \Psi_{l}(\bm{\xi}) \sum_{j=0} ^{N}   \ddot{\mathbf{u}}_j   \Psi_{j}(\bm{\xi}) + \sum_{q=0} ^{Q} \mathbf{{C}}_q  \Psi_{q}(\bm{\xi})    \sum_{j=0} ^{N} \dot{\mathbf{u}}_j \Psi_{j}(\bm{\xi})+ \sum_{i=0} ^{M} \mathbf{{K}}_{i}  \Psi_{i}(\bm{\xi})\sum_{j=0} ^{N} \mathbf{u}_{j}  \Psi_{j}(\bm{\xi}) \approx \mathbf{f}
\end{equation}
where $\mathrm{\mathbf{{M}}}_{i}, \mathbf{{C}}_{i},\mathbf{{K}}_{i}$ are the deterministic coefficient matrices. 
The Galerkin projection of the above stochastic system of equations onto a PCE basis leads to:
\begin{align}\label{Eq.galerproject_chap2}
  \sum_{j=0} ^{N} \sum_{l=0} ^{L} \langle  \Psi_{l} \Psi_{j}\Psi_{k} \rangle  \mathbf{M}_l  \ddot{\mathbf{u}}_j   +  \sum_{j=0} ^{N} \sum_{q=0} ^{Q}  \langle  \Psi_{q} \Psi_{j}\Psi_{k}\rangle  \mathbf{C}_q \dot{\mathbf{u}}_j  + \sum_{j=0} ^{N} \sum_{i=0} ^{M}   \langle \Psi_{i} \Psi_{j} \Psi_{k} \rangle \mathbf{K}_i  \mathbf{u}_{j}   \nonumber \\
= \langle \mathbf{f}  \Psi_{k} \rangle   \quad k = 0,1,2,....N
\end{align}
It leads to a large deterministic system of ODEs as:
\begin{equation}
\mathbfcal{M} \ddot{\mathcal{U}}  + \mathbfcal{C}  \dot{\mathcal{U}} + \mathbfcal{K} \mathcal{U} = \mathcal{F}
\end{equation}
where $\mathbfcal{M}, \mathbfcal{C}$ and $\mathbfcal{K}$ are the block matrices with each block defined as:

\begin{equation}\label{Eq.sto_masschap2}
[\mathbfcal{M}]_{jk} = \sum_{l=0} ^{L} G_{ljk} \mathbf{M}_{l}  
\end{equation}

\begin{equation}
[\mathbfcal{C}]_{jk} = \sum_{q=0} ^{Q} G_{qjk} \mathbf{C}_{q} 
\end{equation}

\begin{equation}\label{Eq.stok}
[\mathbfcal{K}]_{jk}  = \sum_{i=0} ^{M} G_{ijk} \mathbf{K}_{i} 
\end{equation}
where
\begin{equation}
 G_{ijk} =  \langle \Psi_{i} \Psi_{j} \Psi_{k} \rangle
 \end{equation}
 and
\begin{equation}
[\mathcal{F}]_{k}  = \langle \bf{f} \mathrm{\Psi_{k}} \rangle.
\end{equation} 
The solution vector and its higher order derivatives are represented as $\mathcal{U} = [ u_0, u_1, u_2 \cdots u_N ] $. Note, each $u_j$ contains the solution for all degrees of freedom in the grid corresponding to the PCE coefficient. For the current model, a deterministic mass matrix is used which then reduces the Eq.~(\ref{Eq.sto_masschap2}) to a block diagonal matrix as:
 \begin{equation}
[\mathbfcal{M}]_{jk}  =  \langle\Psi_{j}\Psi_{k}\rangle \delta_{jk} \; \mathbf{M},
\end{equation}
where $\delta_{jk}$ is the kronecker delta function. The damping matrix coefficients for Rayleigh damping using the mean mass and random stiffness matrix can be computed as (see Appendix \ref{subsec:rayleigh} for the derivation):
 \begin{equation}
 \mathbf{C}_{i}  = \alpha_0 \mathbf{M} + \alpha_1 \mathbf{K}_{i},
\end{equation}
where $\alpha_0, \alpha_1$ are the coefficients computed as in Appendix \ref{subsec:rayleigh}. Time discretization of this system can be carried out using the Newmark-beta scheme. From the subdomain level mass, damping and stiffness matrices we can build the transient stiffness matrix as:
\begin{equation}\label{Eq:K_T_Sto}
\mathbfcal{\tilde{{K}}}  = \bf{\mathbfcal{M}} \frac{\mathrm{1}}{\zeta \mathrm{\Delta t^2}} + \bf{\mathbfcal{C}}  \frac{\gamma}{\zeta \mathrm{\Delta t} } + {\mathbfcal{K}}.	
\end{equation}
Now, the DD-based assembly of these stochastic matrices and the construction of a two-level Neumann-Neumann preconditioner is presented.

\section{Probabilistic Two-level Neumann-Neumann Preconditioner}

The formulation of the probabilistic two-level Neumann-Neumann solver is explained in this section corresponding to the acoustic wave propagation in random media explained in Section \ref{sec:acwave_random_c2}. The spatial discretization of the stochastic problem in Eq.~(\ref{Eq.acous_sto}) can be assembled into random mass, damping and stiffness matrices as:
\begin{equation}\label{Eq.randODEchapt2_2}
\bf{M} (\theta) \ \ddot{\bf{u}}(\theta) +  {\bf{C}} (\theta) \dot{\bf{u}}(\theta) + {\bf{K}} (\theta) \bf{u}(\theta) \approx \bf{f}
\end{equation}
where each of these random matrices admits a PCE in the form as:
\begin{equation}
{\bf{K}} (\theta)  = \sum_{i=0} ^{M} \mathbf{{K}}_{i}  \Psi_{i}(\bm{\xi}).
\end{equation}
The spatial domain of the problem can be decomposed into several non-overlapping domains with interior and interface nodes as shown in Fig.~\ref{fig.interface_corner}. This decomposition allows the stiffness matrix to be assembled (Note this decomposition is applied to mass and damping matrices as well but not shown explicitly for brevity) at the subdomain level as:
\begin{equation}\label{Eq:ki}
\mathrm{\mathbf{{K}}}_{i} =
{\begin{bmatrix}
     {\mathbf{K}}_{II,i}^{\it{1}} & \dots  &   0    \ \  &  {\mathbf{K}}_{I {\it{\Gamma}},i}^1 \mathbf{R}_1
\\[0.3em]
     \vdots   & \ddots & \vdots \ \  &  \vdots
\\[0.3em]
     0        & \dots  &  {\mathbf{K}}_{II,i}^{n_s}  \ \ &  {\mathbf{K}}_{I {\it{\Gamma}},i}^{n_s}  \mathbf{R}_{n_s}
\\[0.3em]
    \mathbf{R}^{\it{T}}_{\it{1}} {\mathbf{K}}_{\it{\Gamma} I,i}^{\it{1}} & \dots  &  \mathbf{R}^T_{n_s} {\mathbf{K}}_{\it{\Gamma} I,i}^{n_s}  \ \ & \displaystyle\sum_{s=1}^{n_s} \mathbf{R}^T_s {\mathbf{K}}_{{\it{\Gamma}} {\it{\Gamma}},i}^s \mathbf{R}_s
\end{bmatrix}}
\end{equation}
After Galerkin projection of the system as in Eq.~(\ref{Eq.galerproject_chap2}) each of the stochastic subdomain level matrices can be assembled from Eq.~(\ref{Eq:ki}) as:
\begin{equation}\label{Eq.stoksub}
[\mathbfcal{K}_{\alpha \beta}]_{jk} ^s  = \sum_{i=0} ^{M} G_{ijk} \mathbf{K}_{\alpha \beta,i} ^s
\end{equation}
where $\alpha$ and $\beta$ correspond to the interior and interface nodes respectively and $s$ represent the subdomain.
Thus, the transient stiffness matrix can be assembled using stochastic subdomain blocks of mass, damping and stiffness as:
\begin{equation}\label{Eq:K_T_Stosub}
[\mathbfcal{\tilde{{K}}}_{\alpha \beta}] ^s = [\bf{\mathbfcal{M}}_{\alpha \beta}] ^s \frac{\mathrm{1}}{\zeta \mathrm{\Delta t^2}} + [\bf{\mathbfcal{C}}_{\alpha \beta}] ^s  \frac{\gamma}{\zeta \mathrm{\Delta t} } + [{\mathbfcal{K}}_{\alpha \beta}] ^s.	
\end{equation}
Notice the difference in the construction of stochastic matrices in Eq.~(\ref{Eq.stoksub}) with respect to Eq.~(\ref{Eq.stok}) and the difference of Eq.~(\ref{Eq:K_T_Stosub}) with Eq.~(\ref{Eq:K_T_Sto}).

Performing global assembly of these subdomain level transient stiffness matrices one obtains:
\begin{equation}\label{Eq:GAssembly_Sto}
{\begin{bmatrix}
     \tilde{\mathbfcal{K}}_{II}^1 & \dots  &   0    \ \  &  \tilde{\mathbfcal{K}}_{I {\it{\Gamma}}}^1 \mathbfcal{R}_1
\\[0.3em]
     \vdots   & \ddots & \vdots \ \  &  \vdots
\\[0.3em]
     0        & \dots  &  \tilde{\mathbfcal{K}}_{II}^{n_s} \ \ &  \tilde{\mathbfcal{K}}_{I {\it{\Gamma}}}^{n_s} \mathbfcal{R}_{n_s}
\\[0.3em]
    \mathbfcal{R}^T_1 \tilde{\mathbfcal{K}}_{{\it{\Gamma}} I}^1 & \dots  &  \mathbfcal{R}^T_{n_s} \tilde{\mathbfcal{K}}_{{\it{\Gamma}} I}^{n_s} \ \ & \displaystyle\sum_{s=1}^{n_s} \mathbfcal{R}^T_s \tilde{\mathbfcal{K}}_{{\it{\Gamma}} {\it{\Gamma}}}^s \mathbfcal{R}_s
\end{bmatrix}}
\begin{Bmatrix}
     \mathcal{U}_I^1
\\[0.3em]
     \vdots
\\[0.3em]
      \mathcal{U}_{I}^{n_s}
\\[0.3em]
      \mathcal{U}_{\it{\Gamma}}
\\[0.3em]
\end{Bmatrix}  =
\begin{Bmatrix}
     \tilde{ \mathcal{F} }_I^1
\\[0.3em]
     \vdots
\\[0.3em]
     \tilde{\mathcal{F}}_{I}^{n_s}
\\[0.3em]
     \displaystyle\sum_{s=1}^{n_s} \mathbfcal{R}^T_s  \tilde{\mathcal{F}}_{\it{\Gamma}}^s
\end{Bmatrix}
\end{equation}
where,
\begin{equation}
\mathcal{U}_I^s = [{\bf{u}}^s_{I,0}, \dots, {\bf{u}}^s_{I,j} ]^T, \ \ \ \ \ \mathcal{U}_{\it\Gamma} = [{\bf{u}}_{{\it{\Gamma}},0}, \dots, {\bf{u}}_{{\it{\Gamma}},j}]^T,
\end{equation}
\begin{equation}\label{Eq.sto_restrict}
\mathbfcal{R}_s = blockdiagonal ( {\bf{R}}_{s,0},\dots,{\bf{R}}_{s,j}).
\end{equation}
Similar to the deterministic case, applying a Gaussian block elimination allows the construction of Schur complement as:
\begin{equation}\label{eq:Sto_Schur}
\mathbfcal{S} \mathcal{U}_{\it\Gamma} = \mathcal{G}_{\it\Gamma},
\end{equation}
where,
\begin{align}
\mathbfcal{S} &=  \sum_{s=1}^{n_s} \mathbfcal{R}_s^T [ \tilde{\mathbfcal{K}}_{{\it\Gamma} {\it\Gamma}}^s - \tilde{\mathbfcal{K}}_{\it{\Gamma} I}^s [{\tilde{\mathbfcal{K}}_{II}^s}]^{-1} \tilde{\mathbfcal{K}}_{I \it{\Gamma}}^s] \mathbfcal{R}_s =  \sum_{s=1}^{n_s} \mathbfcal{R}_s^T \mathbfcal{S}_{s} \mathbfcal{R}_{s}, \\
\mathcal{G}_{\it\Gamma} &= \sum_{s=1}^{n_s} \mathbfcal{R}_s^T [\tilde{\mathcal{F}}_{\it{\Gamma}}^s - \tilde{\mathbfcal{K}}_{\it{\Gamma} I}^s [\tilde{\mathbfcal{K}}_{II}^s]^{-1} \tilde{\mathcal{F}}_I^s ] =  \sum_{s=1}^{n_s} \mathbfcal{R}_s^T {\mathcal{G}_{\it\Gamma}}_s.
\end{align}
To construct the coarse grid, the solution vector is decomposed again to create the global coarse problem on corner nodes as (similar to deterministic setting in Eq.~(\ref{eq:uruc}))
\begin{equation}\label{eq:uruc_sto}
\begin{Bmatrix}
     {{\mathcal{U}^s_r}}
\\[0.3em]
     {{\mathcal{U}^s_c}}
\\[0.3em]
\end{Bmatrix}  =
\begin{Bmatrix}
     {{\mathbfcal{R}_s^r}}
\\[0.3em]
     {{\mathbfcal{R}_s^c}}
\\[0.3em]
\end{Bmatrix}
\mathcal{U}^s_{\it\Gamma},
\end{equation}
\begin{equation}\label{eq:coarsegrid2}
\mathbfcal{F}_{cc} \; \mathcal{U}_c = \mathcal{Q}_{c},
\end{equation}
where the operator $\mathbfcal{F}_{cc}$ and corresponding force vector $\mathcal{Q}_{c}$ for corner nodes is defined as:
\begin{align}\label{eq:Fcc_2}
\mathbfcal{F}_{cc} &= \sum_{s=1}^{n_s} {\mathbfcal{B}_c^s}^T (\mathbfcal{S}_{cc}^s - \mathbfcal{S}_{cr}^s [\mathbfcal{S}_{rr}^s]^{-1} \mathbfcal{S}_{rc}^s)\mathbfcal{B}_c^s, \\
\mathcal{Q}_{c}&= \sum_{s=1}^{n_s} {\mathbfcal{B}_c^s}^T (\mathbfcal{F}_{c}^s - \mathbfcal{S}_{cr}^s [\mathbfcal{S}_{rr}^s]^{-1} \mathbfcal{F}_{r}^s).
\end{align}
where,
\begin{equation}
\mathbfcal{B}_c ^s = blockdiagonal ( {\bf{B}}_{c,0}^s, \dots ,\; {\bf{B}}_{c,j}^s).
\end{equation}
and $\mathbfcal{S}^s_{\alpha\beta} = \tilde{\mathbfcal{K}}_{\alpha\beta}^s - \tilde{\mathbfcal{K}}_{\alpha I}^s [\tilde{\mathbfcal{K}}^s _{II}]^{-1} \tilde{\mathbfcal{K}}_{I \beta}^s$ is the Schur complement for corner and remaining nodes. Finally, the stochastic two-level Neumann-Neumann preconditioner can be written as:
\begin{equation}\label{eq:NNCP2}
\mathbfcal{M}_{NNC}^{-1} = \sum_{s=1}^{n_s} \mathbfcal{R}_s^T \mathbfcal{D}_s({\mathbfcal{R}_s^r}^T [\mathbfcal{S}^s_{rr}]^{-1} \mathbfcal{R}_s^r) \mathbfcal{D}_s \mathbfcal{R}_s  + \mathbfcal{R}_0^T [\mathbfcal{F}_{cc}]^{-1} \mathbfcal{R}_0,
\end{equation}
where $\mathbfcal{R}_0$ is defined as:
\begin{equation}
\mathbfcal{R}_0 = \sum_{s=1}^{n_s} {\mathbfcal{B}_c^s}^T ({\mathbfcal{R}_s^c} - \mathbfcal{S}_{cr}^s [\mathbfcal{S}^s_{rr}]^{-1} \mathbfcal{R}_s^r) \mathbfcal{D}_s \mathbfcal{R}_s.
\end{equation}
Note, $\mathbfcal{D}_s$ is the subdomain level stochastic counterpart of the scaling matrix which can be computed using the stochastic restriction matrix in Eq.~(\ref{Eq.sto_restrict}) similar to the deterministic case in Eq.~(\ref{Eq.det_scaling}). \textcolor{ss}{Even though, the construction of the two-level Neumann-Neumann preconditioner in the stochastic setting is similar to the deterministic setting, each component of the stochastic preconditioner now encompasses the information pertaining to all PCE coefficients in them.} More details on the implementations and algorithms for the probabilistic two-level Neumann-Neumann preconditioner can be found in \cite{subber_thesis, ajit_thesis, ajit_CMAM}.

\section{Numerical Results}

This section discusses the numerical experiments conducted for acoustic wave propagation on a two-dimensional random media. The spatial domain is a unit square with Dirichlet boundary conditions on all four sides (fixed, $u = 0$). The CFL number for all experiments uses a value of $0.65$ according to Eq.~(\ref{Eq.CFL}). All the scalability studies are conducted for a duration of $0.5$ s and the time step size is adjusted accordingly with changing mesh sizes.
The number of iterations for all experiments reported is the average value for the selected number of time steps and the time to solution is for the complete duration of the simulation. Note, the acoustic pressure has units of $Pa$ ($\frac{N}{m^2}$ or $dB$), but not explicitly mentioned in numerical results for simplicity. The verification of the stochastic Galerkin method is illustrated by comparing the results to MCS. Further, the numerical and parallel scalabilities of the probabilistic two-level Neumann-Neumann DD-based solver are presented. The scalabilities with respect to the stochastic parameters (increasing number of random variables and order of expansion) are also illustrated. For all numerical experiments concerning the stochastic case in this section, an exponential covariance kernel as in Eq.~(\ref{Eq.CovarKer2D_chapt2}) is used for representing the underlying Gaussian process with zero mean and standard deviation of $0.1$. The PCE for the input has $3$ random variables with $2^{nd}$ order while the output PCE uses $3^{rd}$ order unless explicitly specified. The PCE basis functions are normalized with respect to their standard deviation in all numerical experiments.

\subsection{Comparison with One-level Preconditioners }
\begin{figure}[htbp]
        \centering
        \includegraphics[scale=0.5]{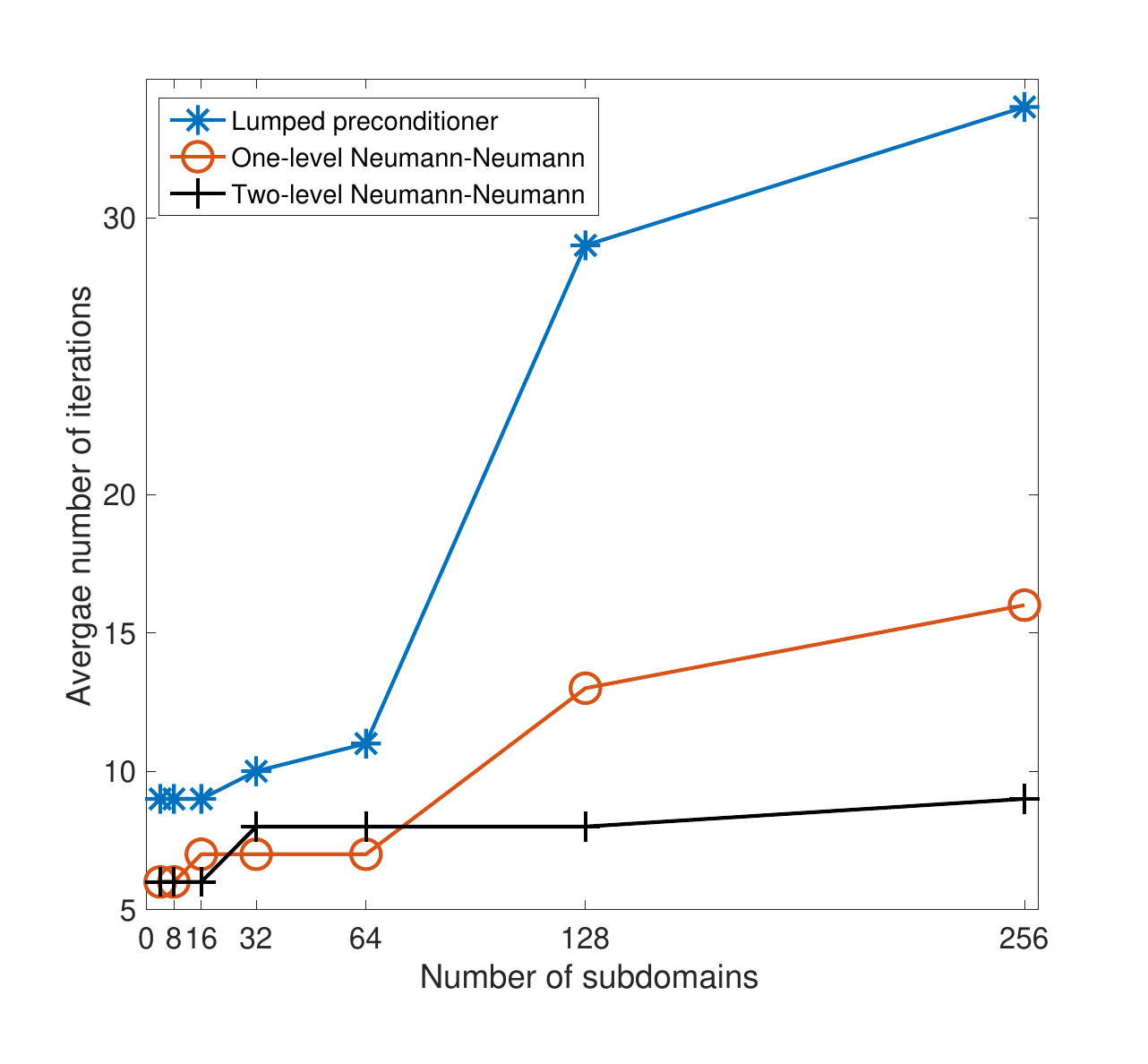} 
    \caption{Comparison of DD-based preconditioners applied to acoustic wave propagation problem}
    \label{Fig:compareDD_wave}
\end{figure}

This subsection compares the different types of DD-based preconditioners for the acoustic wave propagation problem in the deterministic medium. This subsection compares the performance of different preconditioners (both one-level and two-level) to illustrate the need for a two-level method for acoustic wave propagation problems in the deterministic setting. Two one-level preconditioners, Lumped preconditioner and Neumann-Neumann preconditioner (see \cite{subber_thesis} for details on these preconditioners) are chosen to test against the performance of two-level Neumann-Neumann preconditioner. One-level preconditioners are shown to lose scalability 
with an increasing number of subdomains because of the lack of information exchange across several subdomains \cite{book_DD_tarek,book_DD_widlund,book_DD_barryFsmith}. This is rectified for two-level methods (such as two-level Neumann-Neumann) using a coarse grid which couples all the subdomains. 
A square domain with $1563$ vertices and a time step of $0.01$ is chosen for the study. Fig.~\ref{Fig:compareDD_wave} reports the average iteration count among all the time steps for the different preconditioners with an increasing number of subdomains. As the number of subdomains increases, the iteration counts of the Lumped preconditioner increase rapidly. Even though the one-level Neumann-Neumann preconditioner has lower iteration counts than Lumped preconditioner, it also shows an increase in iteration count. 
Two-level Neumann-Neumann preconditioner in contrast has the lowest iteration counts with modest increase for increasing number of subdomains. An interesting point considering one-level Neumann-Neumann preconditioner is that the floating subdomains \cite{subber_thesis} generally found for static problems are implicitly removed for the current problem with the addition of mass and damping terms to the stiffness operator.

\subsection{Verification for Acoustic Wave in a Stochastic Medium}
\begin{figure}[htbp]
    \centering
    \begin{subfigure}[t]{0.48\textwidth}
        \centering
        \includegraphics[height=2.2in]{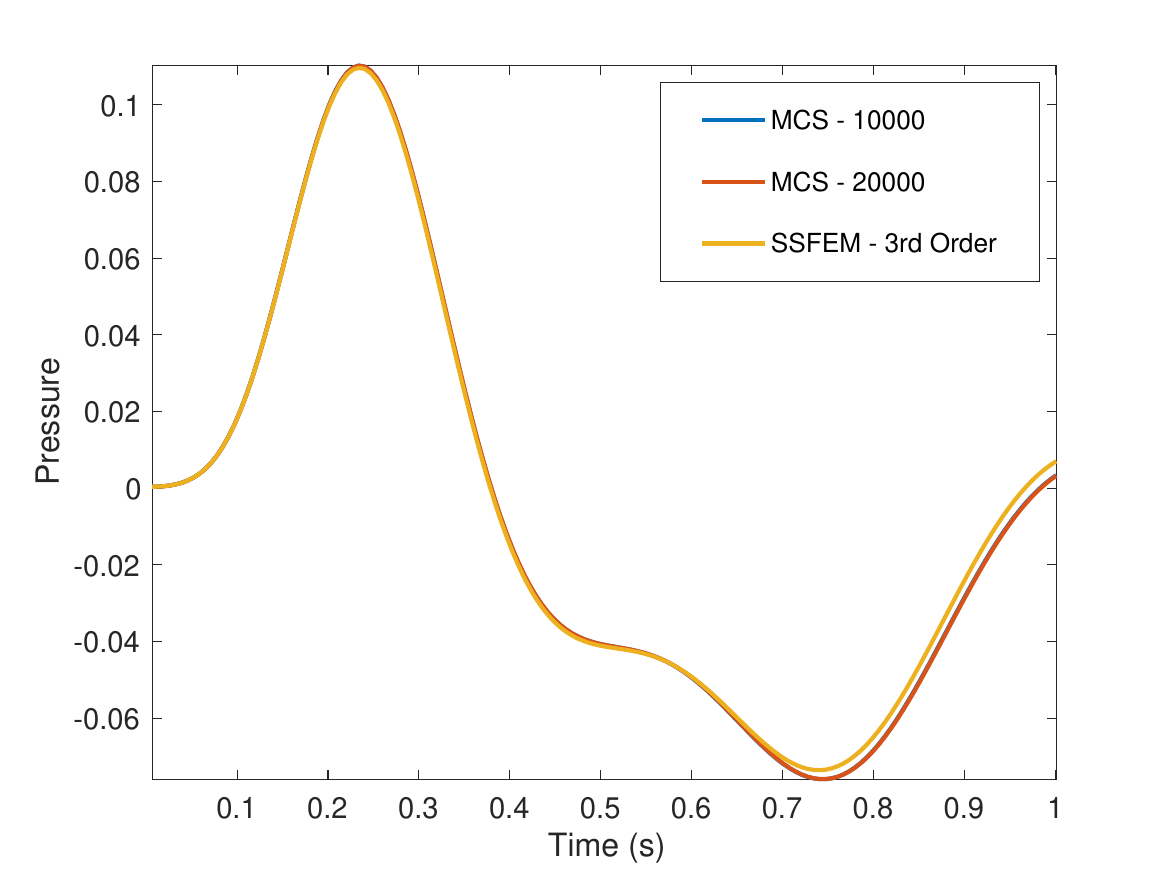}{a}
    \end{subfigure}
 ~   
    \begin{subfigure}[t]{0.48\textwidth}
        \centering
        \includegraphics[height=2.2in]{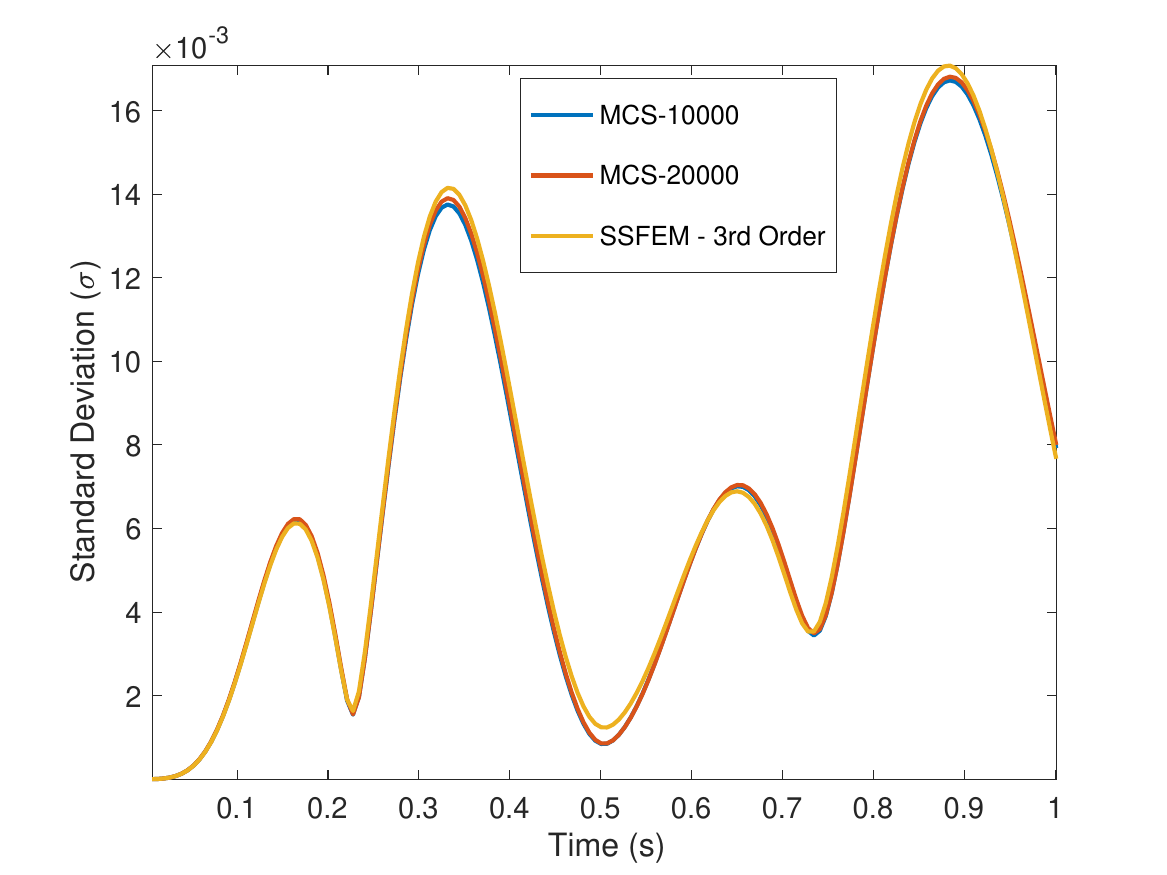} 
    \end{subfigure}
    ~
    \begin{subfigure}[t]{0.48\textwidth}
        \centering
        \includegraphics[height=2.2in]{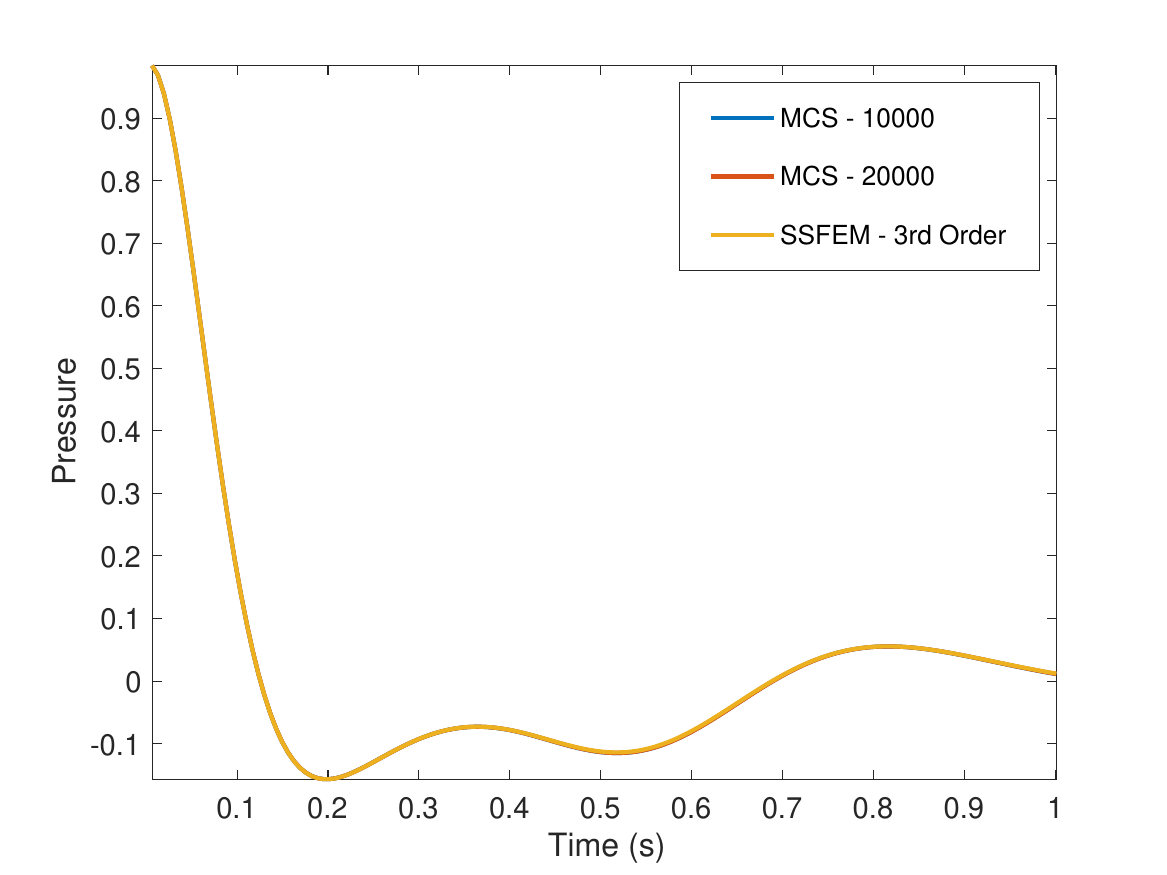}{b}
    \end{subfigure}
    ~
     \begin{subfigure}[t]{0.48\textwidth}
        \centering
        \includegraphics[height=2.2in]{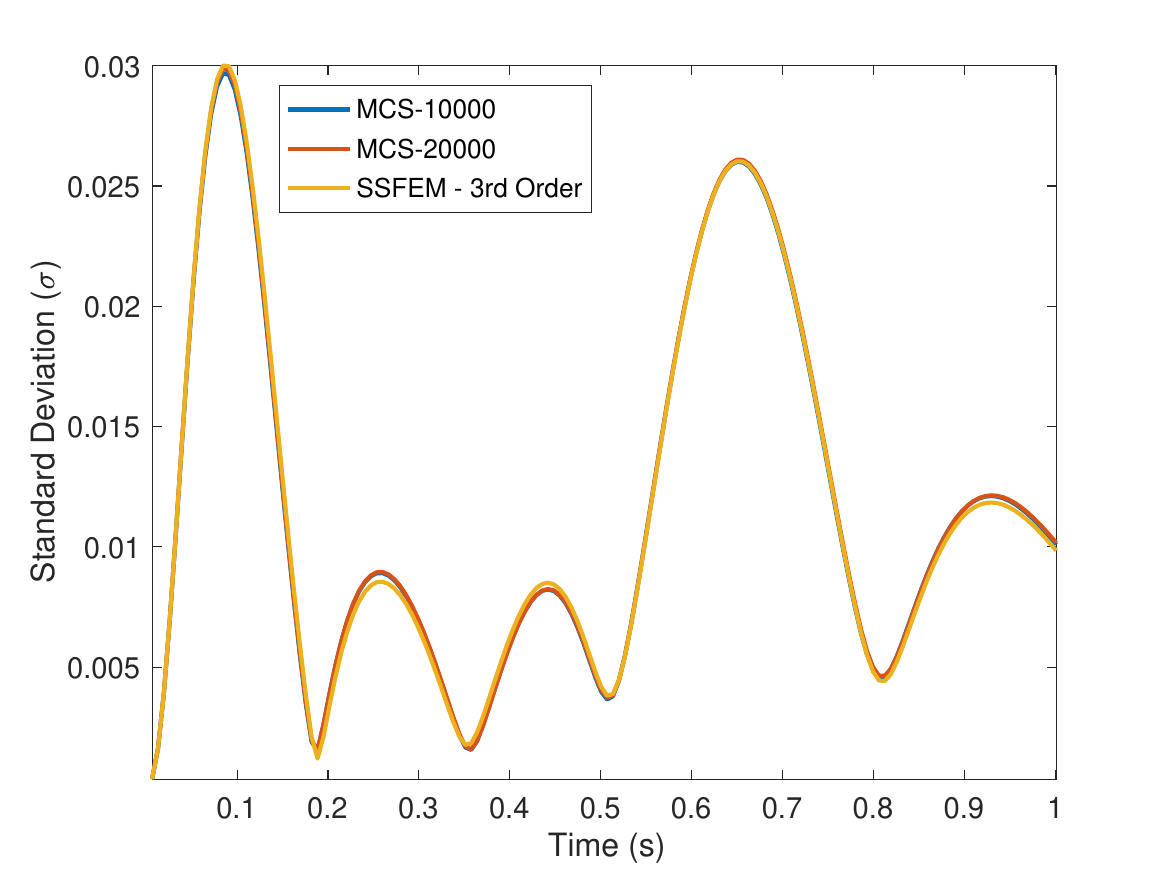} 
    \end{subfigure}
    ~
    \begin{subfigure}[t]{0.48\textwidth}
        \centering
        \includegraphics[height=2.2in]{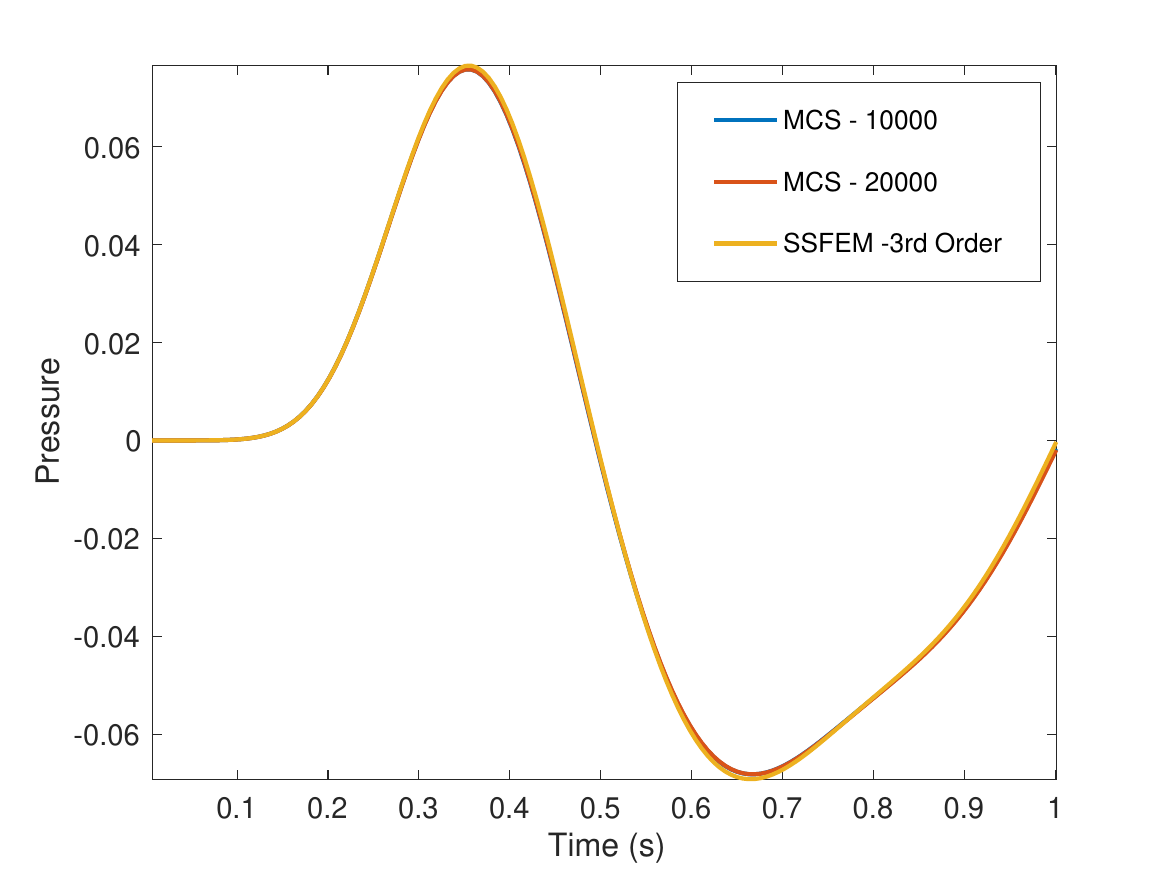}{c}
    \end{subfigure}
    ~
     \begin{subfigure}[t]{0.48\textwidth}
        \centering
        \includegraphics[height=2.2in]{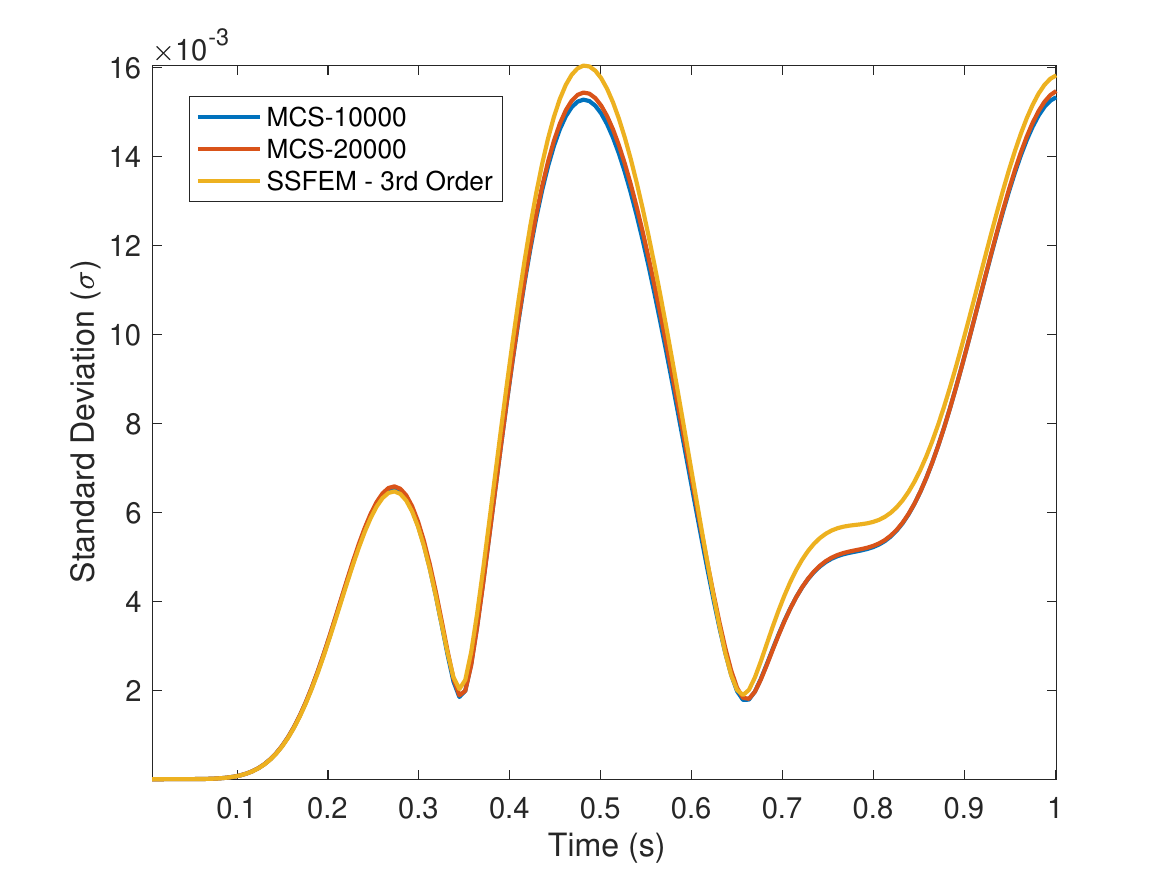} 
    \end{subfigure}
    \caption{Mean (left) and standard deviation (right) of the acoustic pressure field for three different points in the domain. Top row shows $(x = 0.5, y =0.5)$, middle row $(x = 0.7,y = 0.7)$ and bottom row $(x = 0.2,y = 0.7)$.}\label{Fig.wave2D_mcs}
\end{figure}

\begin{figure}[htbp]
    \centering
    \begin{subfigure}[t]{0.48\textwidth}
        \centering
        \includegraphics[height=2.2in]{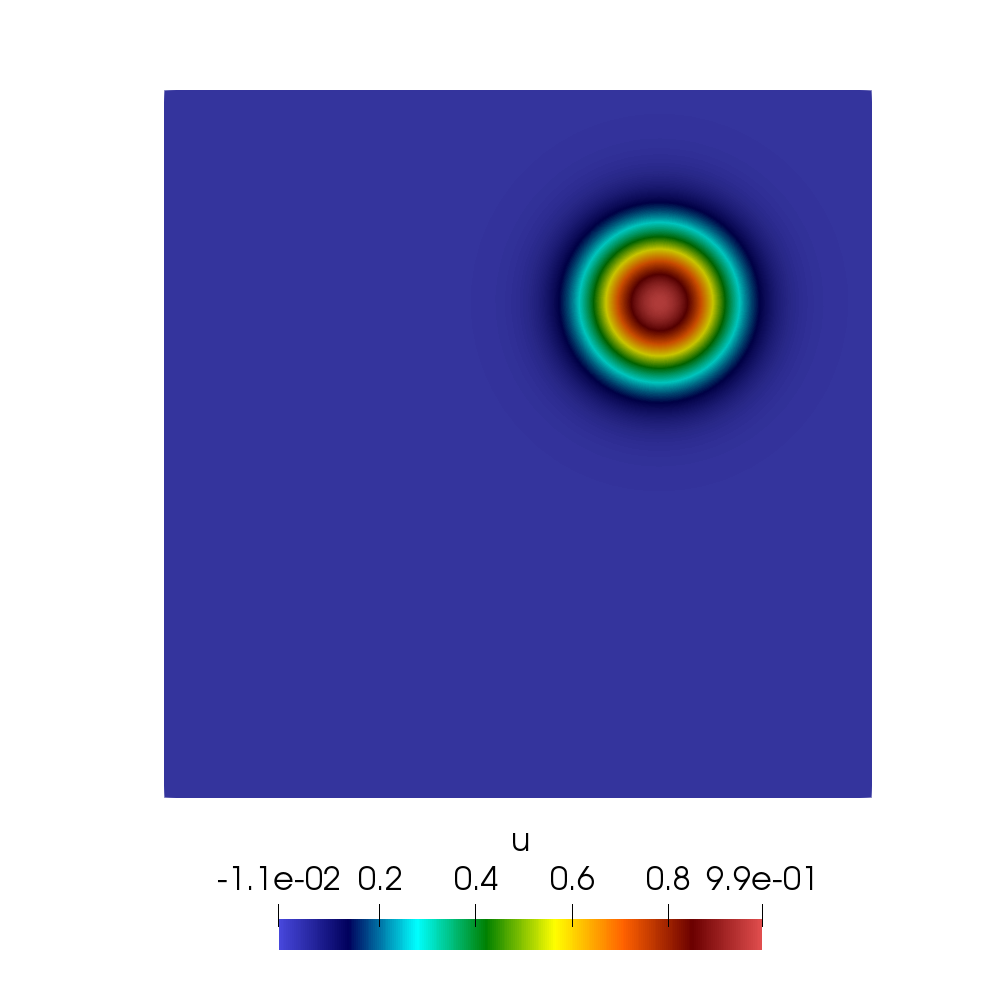}
    \end{subfigure}
 ~   
    \begin{subfigure}[t]{0.48\textwidth}
        \centering
        \includegraphics[height=2.2in]{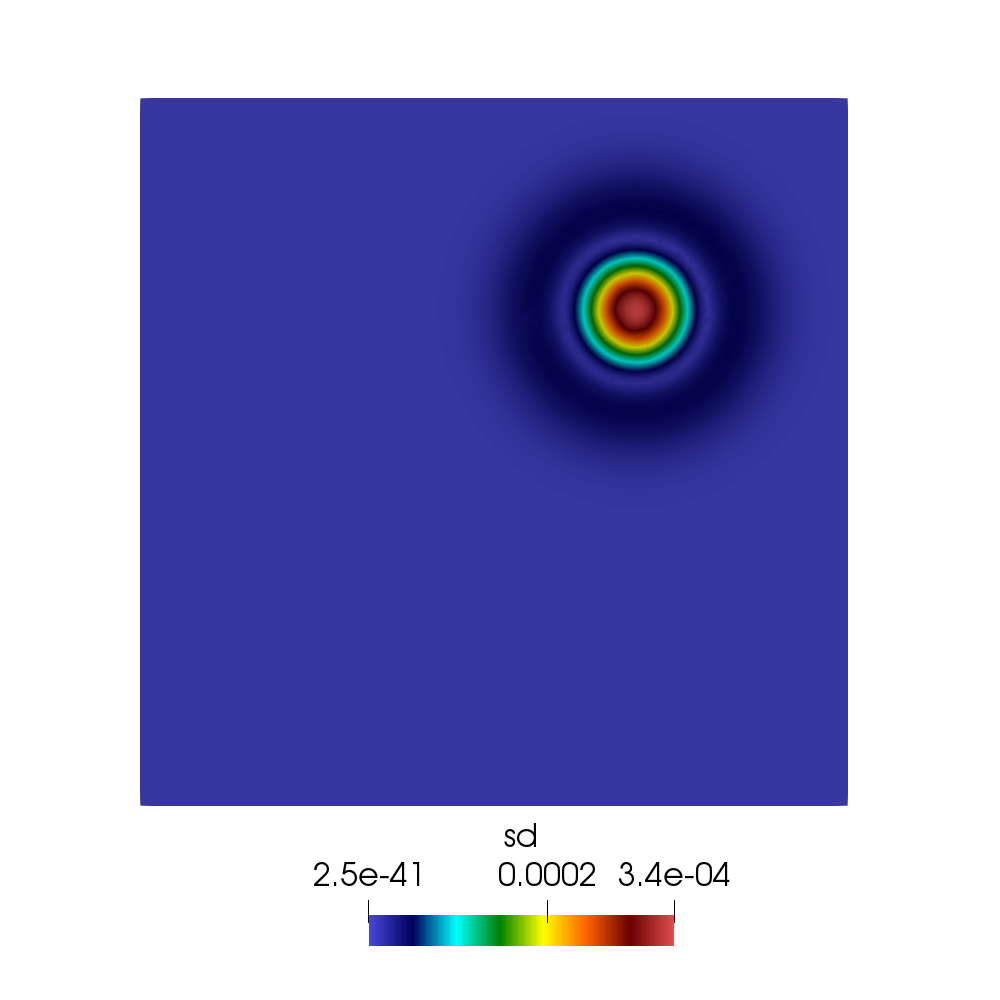} 
    \end{subfigure}
    ~
    \begin{subfigure}[t]{0.48\textwidth}
        \centering
        \includegraphics[height=2.2in]{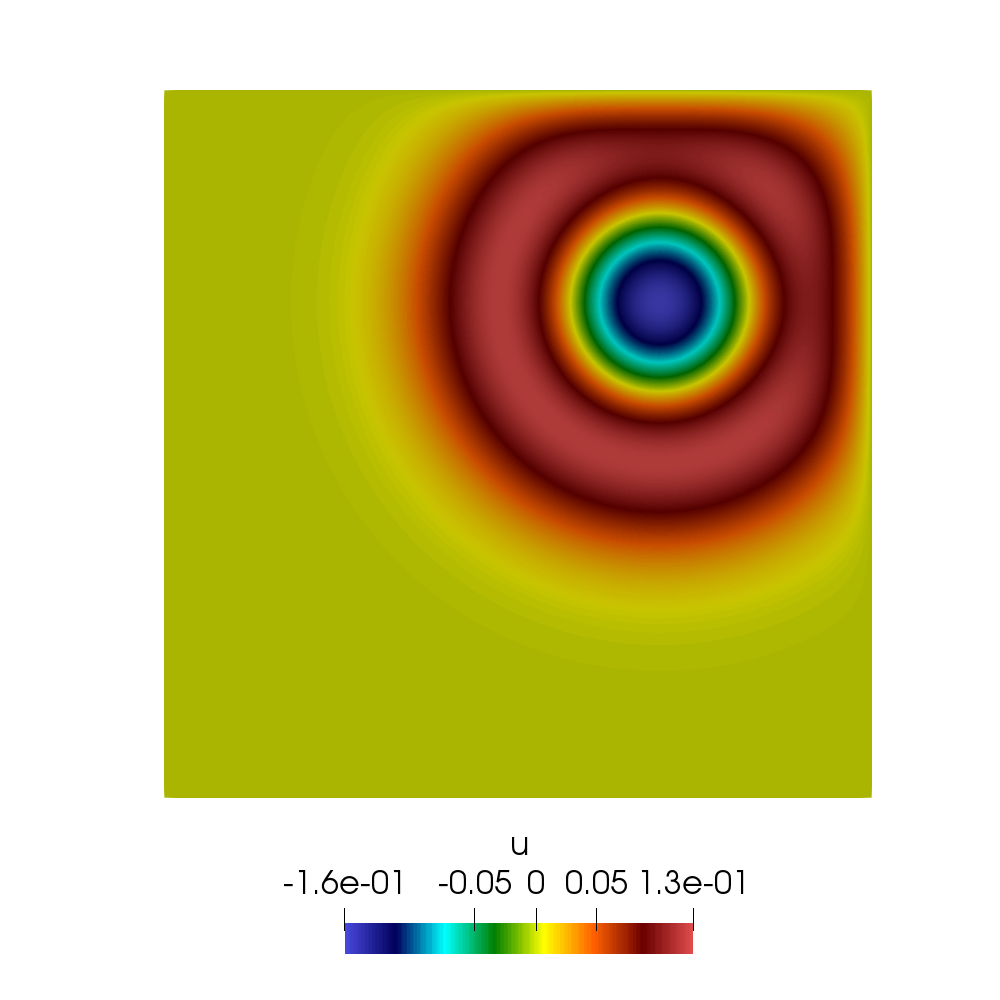}
    \end{subfigure}
    ~
     \begin{subfigure}[t]{0.48\textwidth}
        \centering
        \includegraphics[height=2.2in]{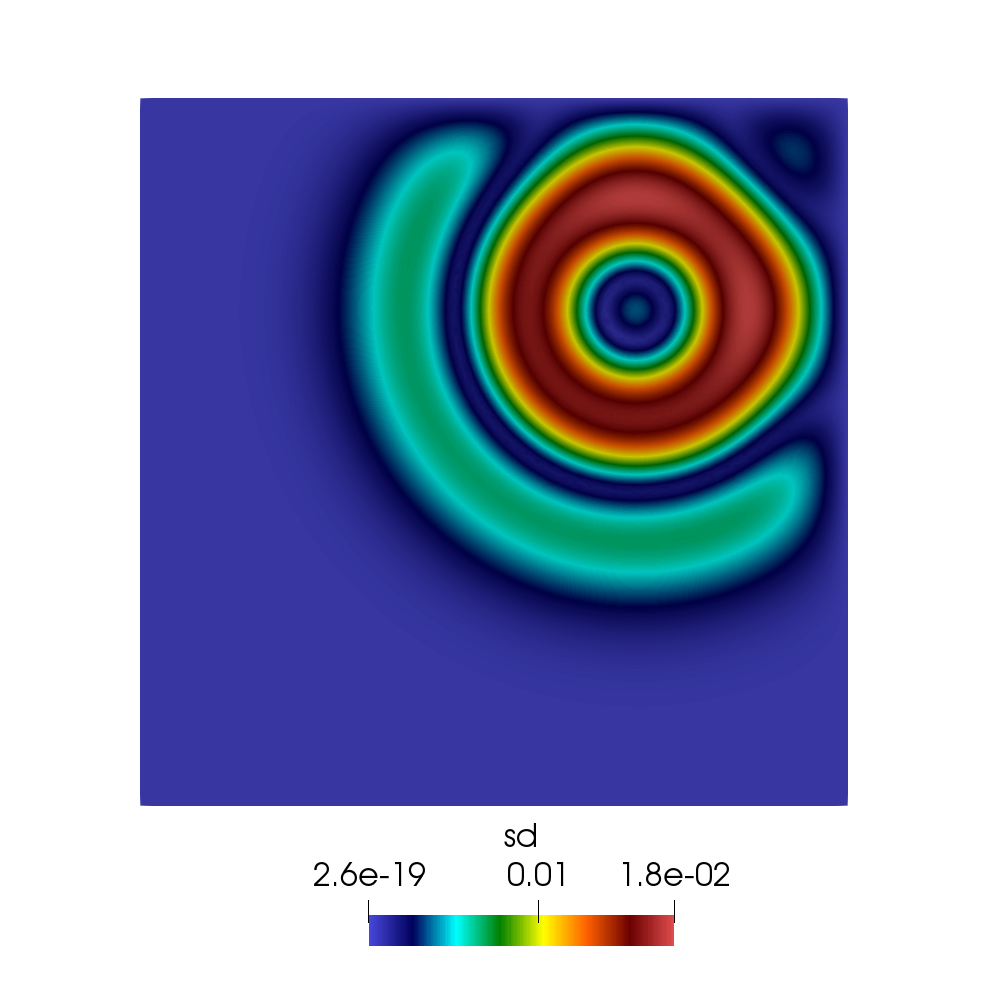} 
    \end{subfigure}
    ~
    \begin{subfigure}[t]{0.48\textwidth}
        \centering
        \includegraphics[height=2.2in]{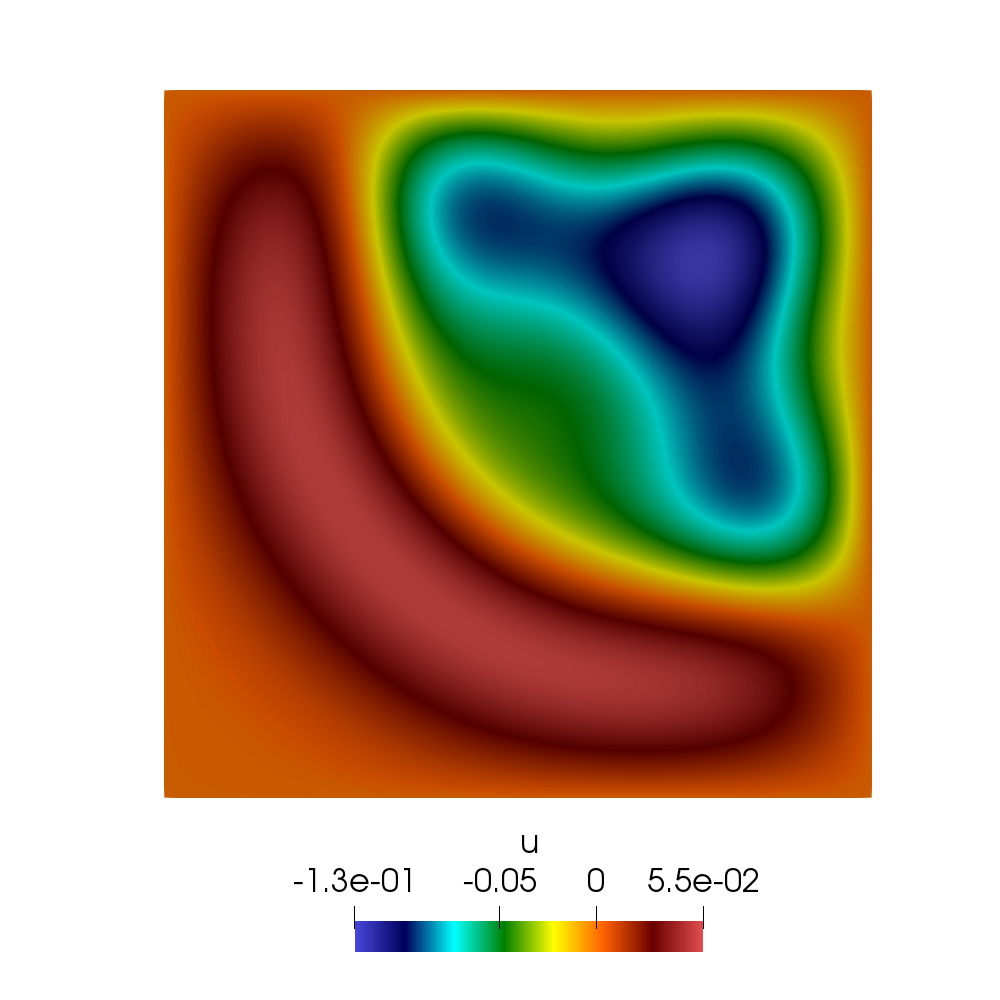}
    \end{subfigure}
    ~
     \begin{subfigure}[t]{0.48\textwidth}
        \centering
        \includegraphics[height=2.2in]{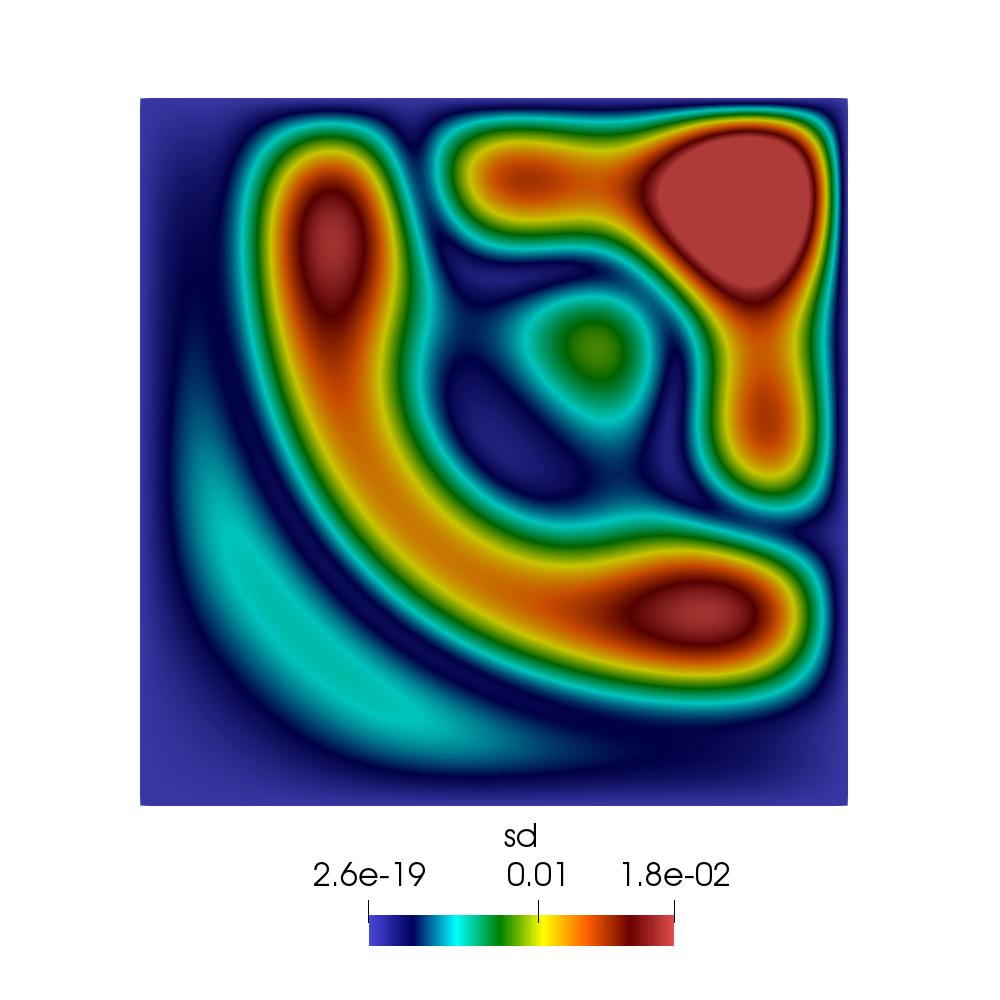} 
    \end{subfigure}
    ~
        \begin{subfigure}[t]{0.48\textwidth}
        \centering
        \includegraphics[height=2.2in]{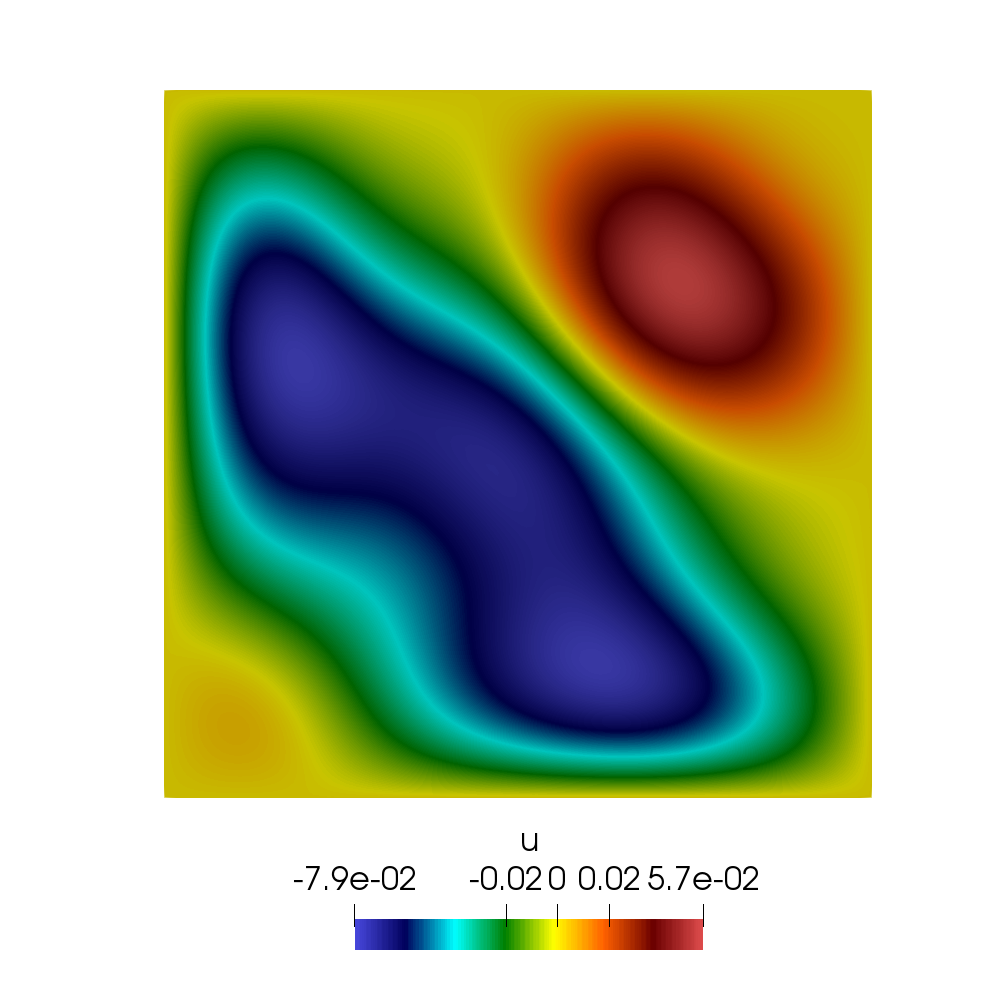} 
    \end{subfigure}
    ~
        \begin{subfigure}[t]{0.48\textwidth}
        \centering
        \includegraphics[height=2.2in]{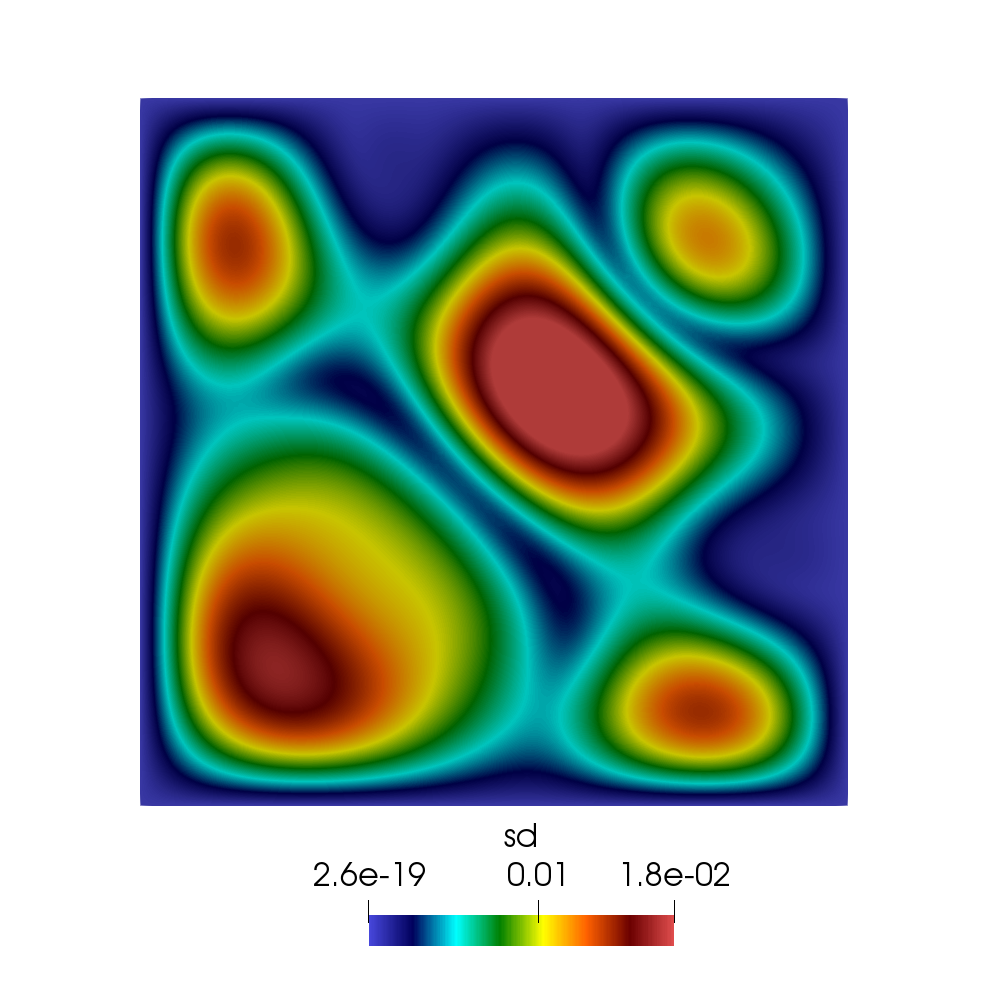} 
    \end{subfigure}
    \caption{Mean (left) and standard deviation (right) of pressure field at time $0.0065 s, 0.2 s, 0.5s$ and $0.8 s$.}\label{Fig.wave2D_vtu}
\end{figure}

The intrusive stochastic Galerkin method is compared against MCS for acoustic wave propagation with a random wave speed in a two-dimensional domain as shown in Eq.~(\ref{Eq.acous_sto}). The initial condition is a Gaussian pulse at $(x_0=0.7, y_0=0.7)$ described as:

\begin{equation}\label{Eq.initial_2Dwave}
u_0(x_0=0.7, y_0=0.7) = \beta e^{-(\frac{(x - x_0)^{2} + (y - y_0)^{2}}{\alpha})}
\end{equation}
where $\beta = 1$ and $\alpha = 0.01$. A time step size of $6.5 \times 10^{-3}$ is used with a finite element mesh size of $13472$ vertices. The Rayleigh damping coefficients are computed as $\alpha_0 = 0.5445$ and $\alpha_1 = 0.0174$ (see Appendix \ref{subsec:rayleigh}). The mean and standard deviation of the pressure for the stochastic Galerkin method and MCS at three different points in the domain are shown in Fig.~\ref{Fig.wave2D_mcs}. The results show good agreement between both approaches for all three points verifying the stochastic Galerkin method. The standard deviation of the pressure in time shows large variations in contrast to the mean pressure which is captured well by the stochastic Galerkin method. The contour plots of the pressure field at four different time steps are shown in Fig.~\ref{Fig.wave2D_vtu}. The complex spatial features in standard deviation with the reflection of waves from boundaries can be observed in the right column. This underlines the complexity and importance of uncertainty quantification for wave propagation problems.

\subsection{Scalability with Mesh Size}

\begin{figure}[htbp]
    \centering
    \begin{subfigure}[ht]{0.475\textwidth}
        \centering
        \includegraphics[height=2.5in,width=3.2in]{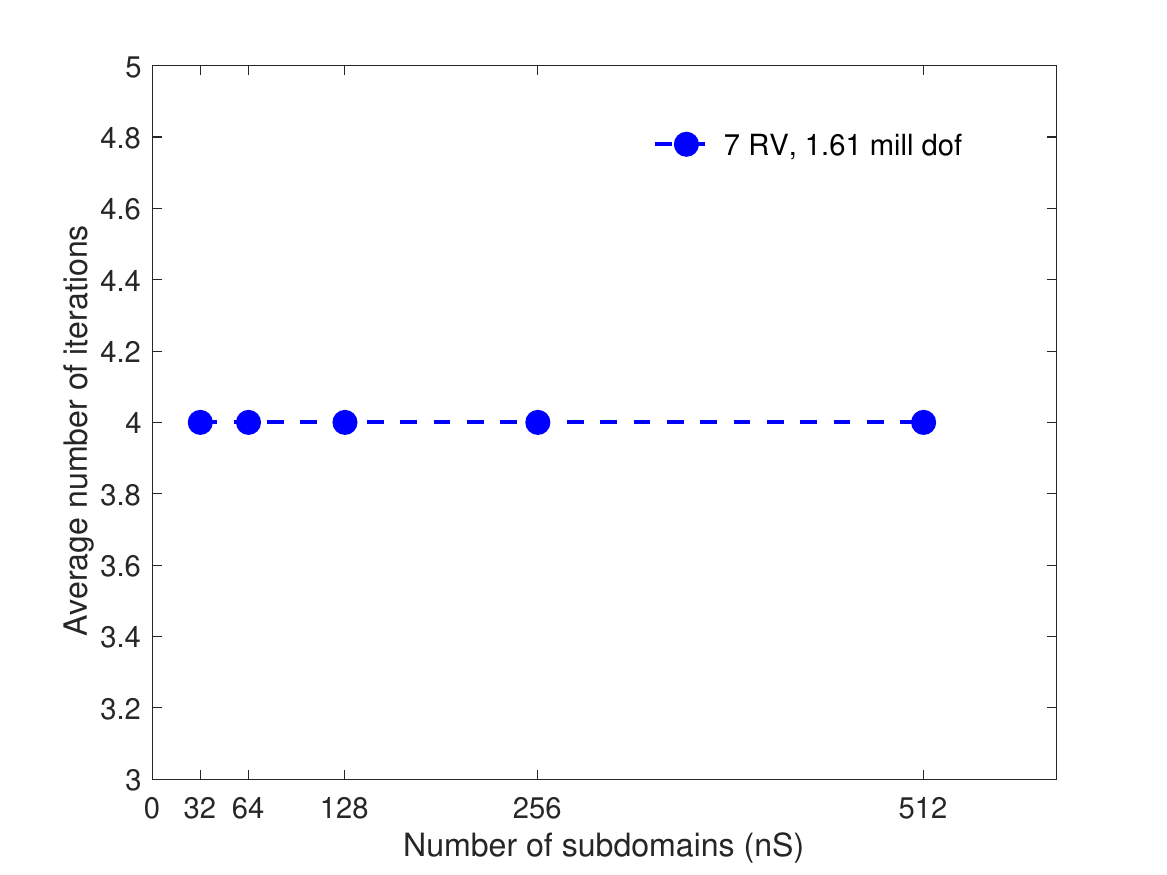} 
        \caption{Strong numerical scalability}
    \end{subfigure}
    \begin{subfigure}[ht]{0.475\textwidth}
        \centering
        \includegraphics[height=2.5in,width=3.2in]{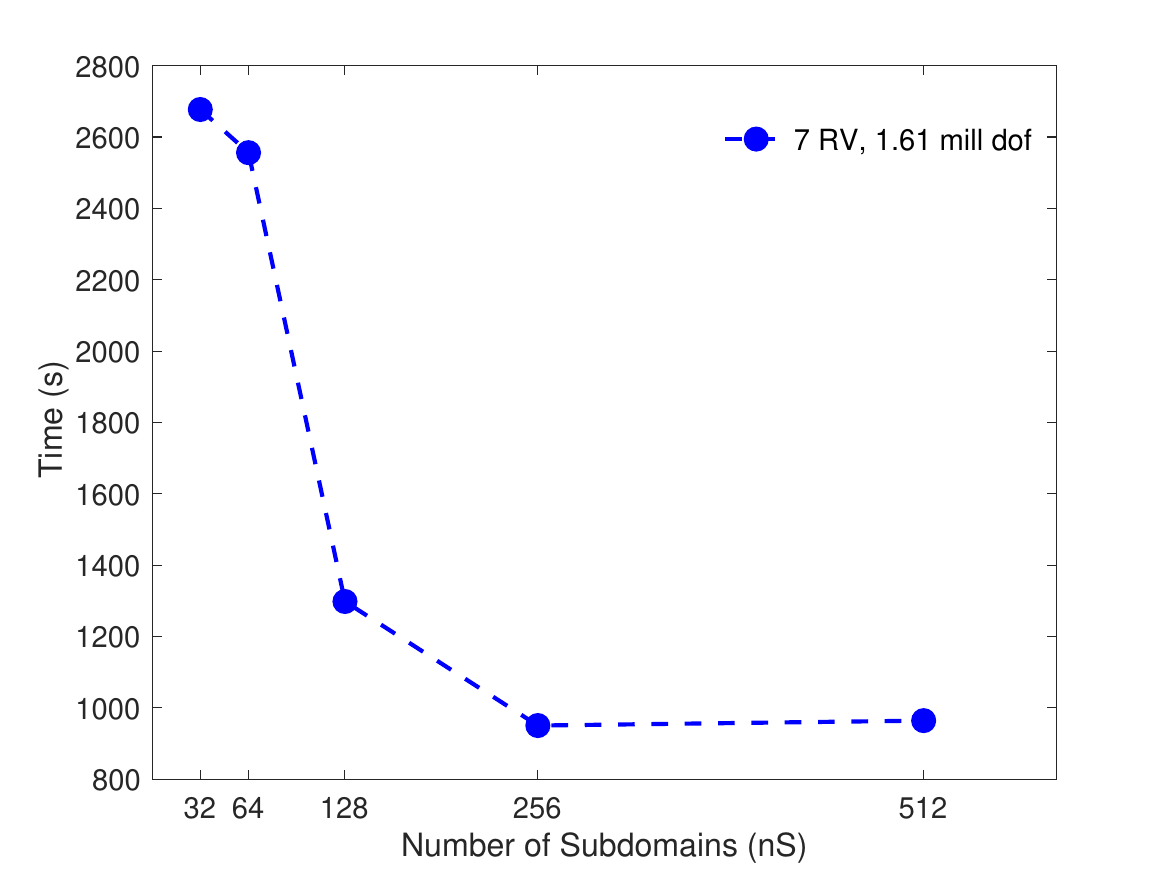} 
         \caption{Strong parallel scalability}
    \end{subfigure}
     \begin{subfigure}[ht]{0.5\textwidth}
        \centering
        \includegraphics[height=2.5in,width=3.2in]{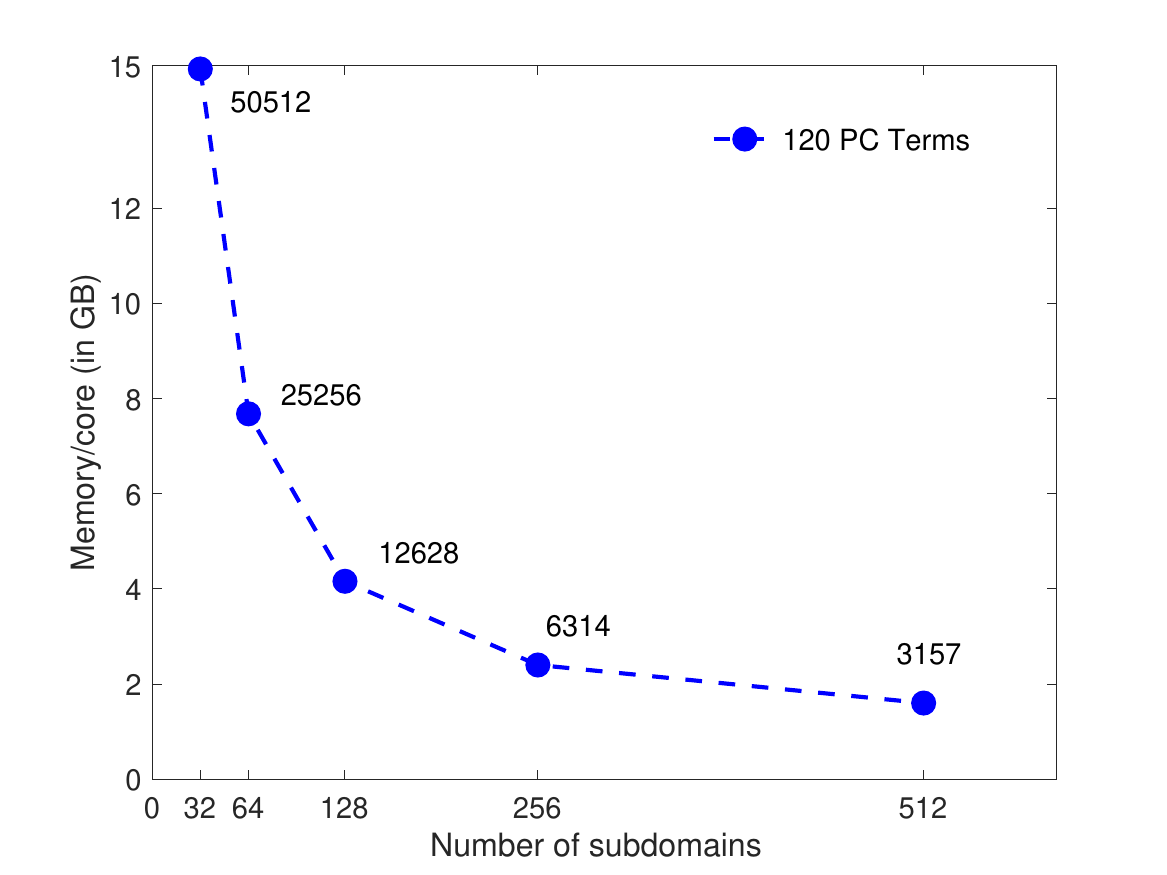} 
         \caption{Memory/core with an increasing number of subdomains}
    \end{subfigure}
    
    \caption{Strong scalability and memory consumption for acoustic wave propagation problem with 7 input random variables.} \label{Fig. strong_rv7}
\end{figure}

The strong and weak scalabilities of the two-level Neumann-Neumann solver for wave propagation on a random media are discussed in this section. The strong numerical and parallel scalabilities of the solver for a $7$ random variable case ($120$ PCE terms) with 
a fixed mesh size of $13472$ vertices (leading to a total problem size of $1.6$ million) is shown in Fig.~\ref{Fig. strong_rv7}.
The average number of iterations is constant with an increasing number of subdomains in Fig.~\ref{Fig. strong_rv7}a and the time to solution decreases in Fig.~\ref{Fig. strong_rv7}b. A significant reduction in memory consumption per core with an increasing number of subdomains for a $7$ random variable PCE is shown in Fig.~\ref{Fig. strong_rv7}c. The memory/core for the problem with $32$ cores reaches $15$GB. The need for using DD-based solvers for stochastic problems is evident.

\begin{figure}[htbp]
    \centering
    \begin{subfigure}[ht]{0.49\textwidth}
        \centering
        \includegraphics[height=2.5in,width=3.2in]{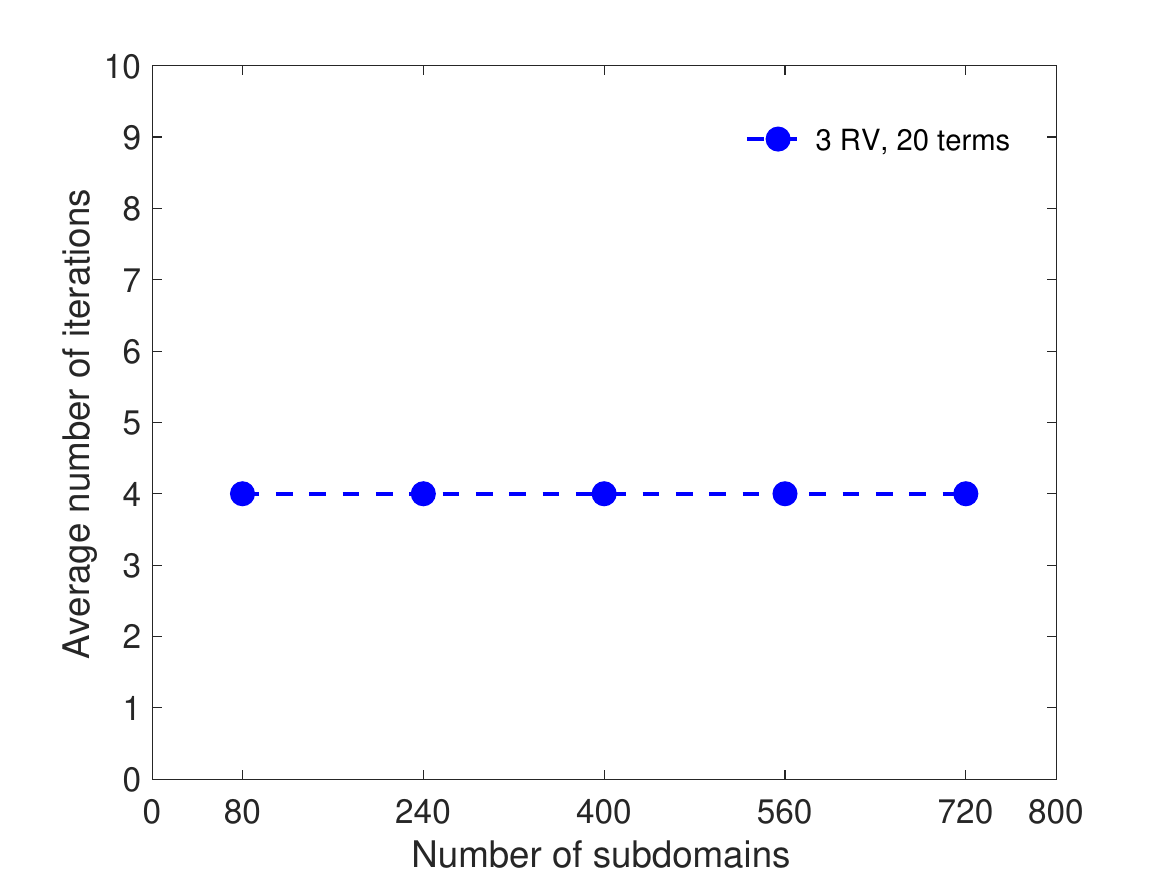} 
    \end{subfigure}%
    ~ 
    \begin{subfigure}[ht]{0.49\textwidth}
        \centering
        \includegraphics[height=2.5in,width=3.2in]{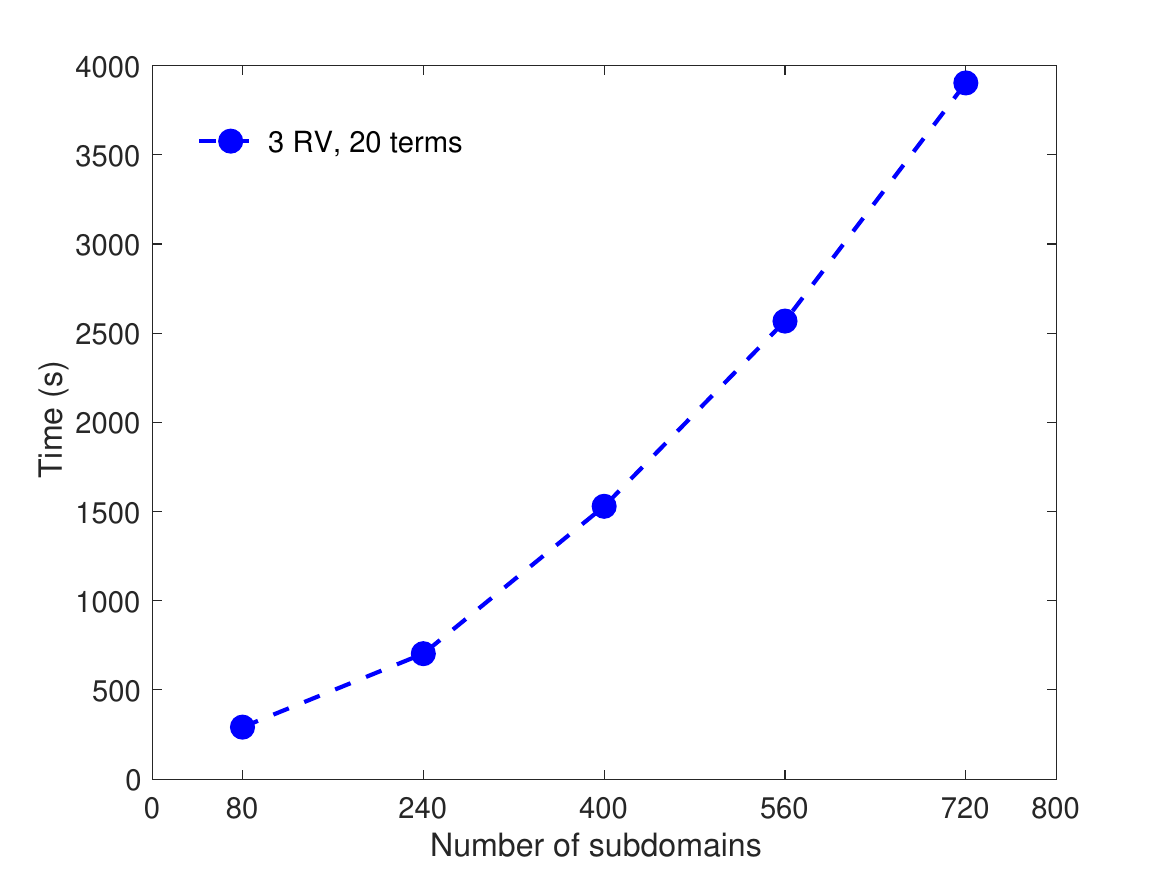} 
    \end{subfigure}
    
    \caption{Weak numerical (left) and parallel (right) scalability with fixed problem size per subdomain and increasing global mesh size.} \label{Fig.weak_rv7}
\end{figure}

The weak numerical and parallel scalabilities of the solver are illustrated in Fig.~\ref{Fig.weak_rv7}. The global problem size and number of subdomains are increased simultaneously with the problem size per core fixed ($\approx 3400$ degrees of freedom). A $3$ random variable case is used here with a number of processes increasing from $80$ to $720$ (leading to a maximum problem size of $2.4$ million). The average number of iteration counts remains constant with an increasing number of processes showing excellent weak scaling. However, the time to solution shows a rapid increase. This may be due to the increased time consumed in parallel communications for all time steps and the increased time consumed for coarse solve.

\subsection{Scalability with Stochastic Parameters}

\begin{figure}[htbp]
    \centering
        \includegraphics[height=2.5in,width=3.2in]{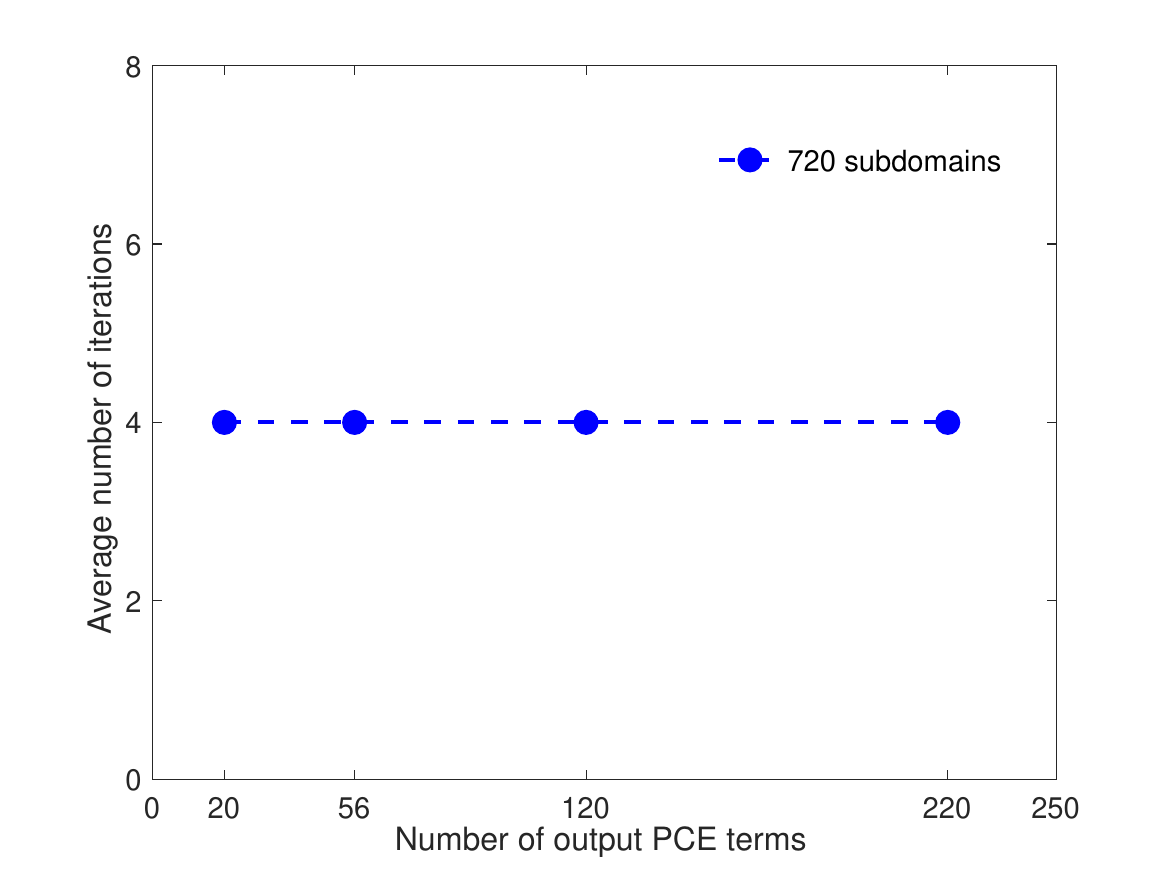}
    \caption{Numerical scalability with increasing number of random variables for a fixed number of subdomains} \label{Fig.stoscal_720}
\end{figure}

This section discusses the scalability of the probabilistic two-level Neumann-Neumann solver with an increasing number of random variables and order of expansions. This study uses a discretized mesh size of $37267$ vertices with a time step size of $3.9\times 10^{-3}$. The time to solution is reported for a total of $128$ time steps. The number of PCE terms is increased from $20$ ($3$ random variable) to $220$ ($9$ random variables) case with a fixed number of subdomains ($720$). The total problem size in this case is $8.19$ million.
The average iteration counts for the solver are reported in Fig.~\ref{Fig.stoscal_720} which shows a constant value of $4$ iterations. 

\begin{figure}[htbp]
    \centering
    \begin{subfigure}[ht]{0.49\textwidth}
        \centering
        \includegraphics[height=2.5in,width=3.2in]{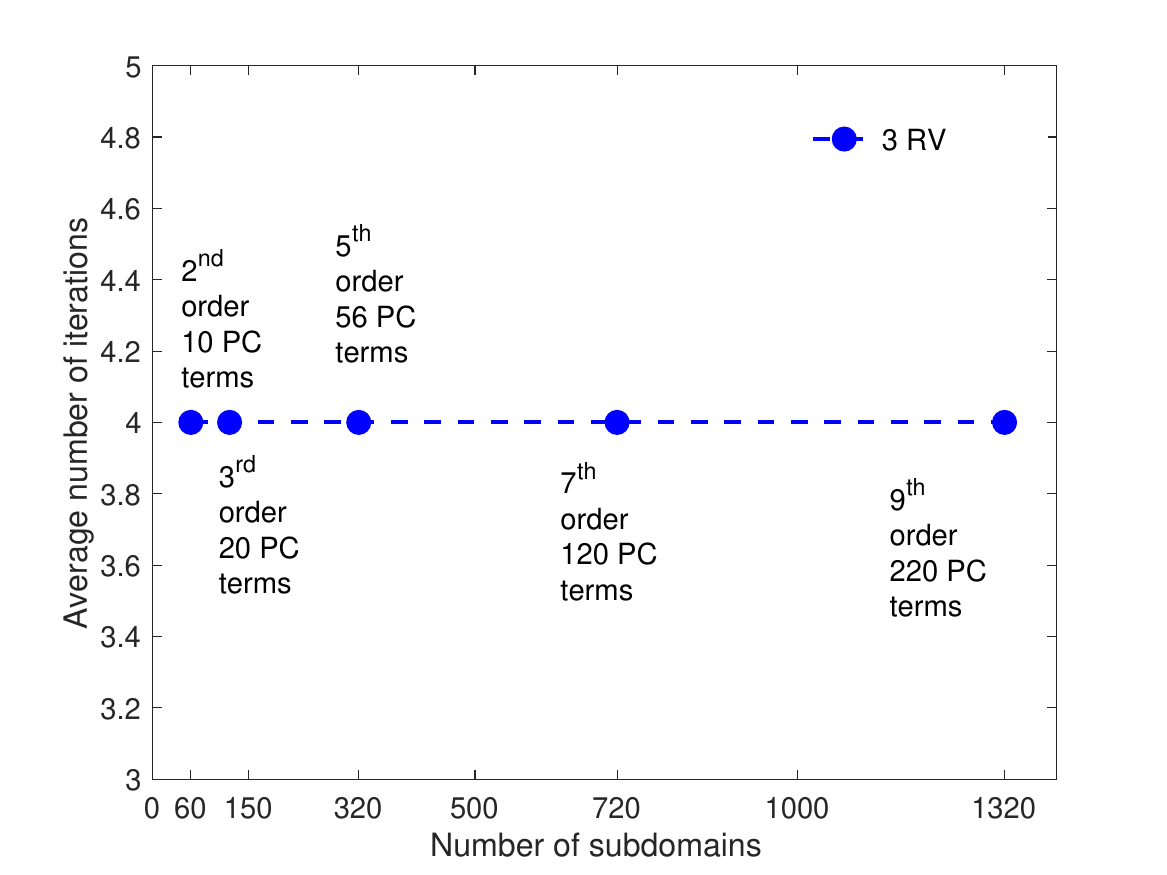} 
    \end{subfigure}%
    ~ 
    \begin{subfigure}[ht]{0.49\textwidth}
        \centering
        \includegraphics[height=2.5in,width=3.2in]{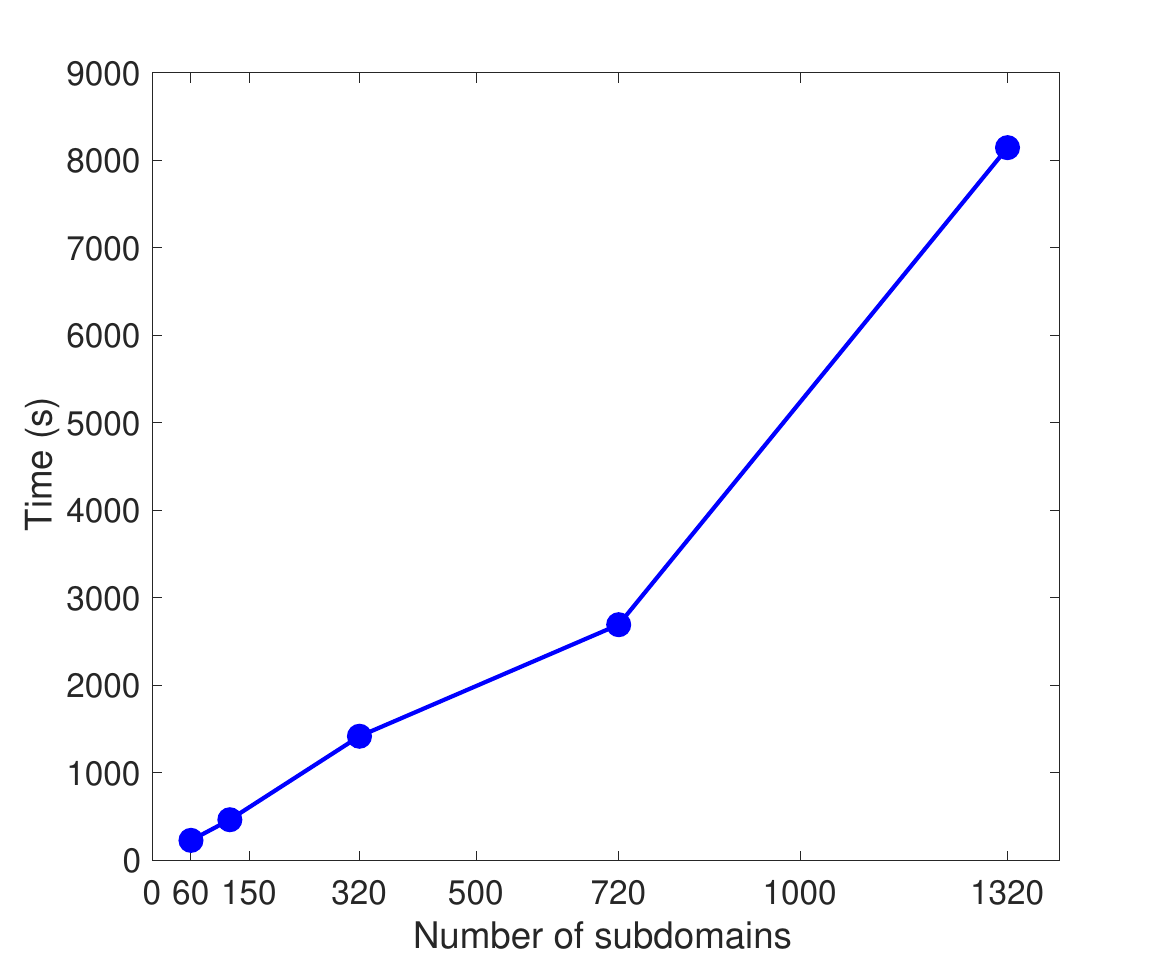} 
    \end{subfigure}
    
    \caption{Numerical and parallel scalability with increasing order and fixed problem size per core} \label{Fig.sto_order}
\end{figure}

The scalability with increasing order of expansions is shown in Fig.~\ref{Fig.sto_order}. For a fixed number of random variables ($3$), the order of output PCE is increased from $2$ to $9$ leading to 
$220$ terms and a total problem size of $8.19$ million. The number of processes is increased to keep the problem size per core fixed at $6200$ dof. Fig.~\ref{Fig.sto_order} shows excellent weak numerical scalability with increasing order of expansion. However, the time to solution increases even with a fixed problem size per core which is due to the strong stochastic coupling introduced by higher-order PCE. Moreover, the use of collective communication operations in parallel algorithms also contributes significantly to the increase in time. This can be improved by point-to-point communications which only transfer data between individual processes as necessary. Even with an increase in the time to solution, the application of DD-based solvers for stochastic problems is necessary to solve these problems demanding large memory and floating-point operations.

\section{Conclusion}
\textcolor{ss}{Uncertainty quantification for the acoustic wave propagation through a two-dimensional random media is carried out using sampling-free intrusive stochastic Galerkin method. The fine temporal and spatial resolutions required to capture high frequency oscillations increases the computational cost and memory requirements for these systems. Moreover, the changes in the uncertainty of response with time demands higher order expansions. Non-overlapping domain decomposition-based methods offer a way to decompose the spatial domain and distribute the memory and computations to several cores/processes. A conjugate gradient iterative solver is used for the coupled system of equations involving symmetric positive definite stochastic coefficient matrix generated from the stochastic Galerkin projection. To expedite the convergence of the solver, we proposed a probabilistic version of the two-level Neumann-Neumann preconditioner. The solver has excellent numerical scalabilities with respect to mesh size and stochastic parameters. However, improvements to parallel scalabilities can be carried out through effective parallel communication algorithms. The combination of intrusive stochastic Galerkin method and domain decomposition-based solvers are effective in solving the uncertainty quantification problems for high-resolution time dependent systems. }

\section*{Acknowledgements}

The author thanks the supervisors and collaborators for their mentorship and advice during the course of this work.  The author also acknowledges the support of the Ontario Trillium Scholarship for International Doctoral Students. This research was enabled in part by support provided by \href{https://www.calculquebec.ca/}{\textit{Calcul Qu\'{e}bec}}, \href{https://www.sharcnet.ca/}{\textit{SHARCNET}}, \href{https://www.scinethpc.ca/}{\textit{SciNet}}, and the \href{https://alliancecan.ca/}{\textit{Digital Research Alliance of Canada}} (alliancecan.ca).

\newpage

\appendix

\section*{Appendix}

\section{Wave Propagation in a One-Dimensional Axially Loaded Bar}

This section describes the numerical experiments for the wave propagation in a one-dimensional axially loaded bar. The elastic modulus of the bar is modelled as a log-normal random variable. Different UQ approaches such as (sampling-free) stochastic Galerkin method, sampling-based non-intrusive spectral projection method and MCS are used to solve this stochastic model.

\subsection{Deterministic Wave Propagation}

\begin{figure}[htbp]
        \centering
        \includegraphics[scale=0.85]{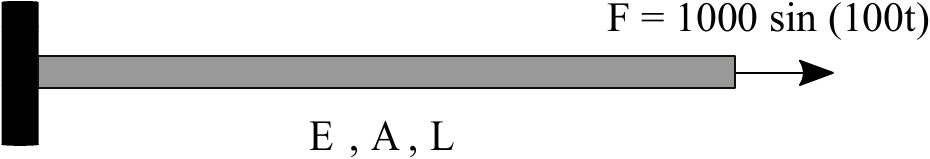} 
        \caption{Cantilever beam with a sinusoidal forcing at the end}\label{Fig.1dbar}
\end{figure}

The axial vibration of a cantilever bar applied with a sinusoidal forcing as shown in Fig.~\ref{Fig.1dbar} is considered for this study. The beam has a constant axial stiffness of $EA$ and density of $\rho$. The equation of motion for the axial displacement at any point on the beam can be written as \cite{humar}:
\begin{equation}\label{Eq.axial1dmain}
\frac{\partial^2 u}{\partial t^2} - c^{2} \frac{\partial^2 u}{\partial x^2} = f
\end{equation}
where $c =  \sqrt{\frac{E}{\rho}} $ is the wave speed. Finite element discretization of the weak form derived from the governing PDE 
in Eq.~(\ref{Eq.axial1dmain}) using two noded linear finite elements provide the elemental mass and stiffness matrices as \cite{bathe_1,bathe_2}:

\begin{align}
M_{e} &= \frac{\rho A_e l_e}{6}
\begin{bmatrix}
2 & 1 \\
 1 & 2 \\
\end{bmatrix}\\
K_{e} &= \frac{E A_e}{l_e}
\begin{bmatrix}
1 &-1  \\
 -1& 1 \\
\end{bmatrix}
\end{align}
where $l_e$ and $A_e$ are the element length and area respectively, and $M_{e}$ and $K_{e}$ are the elemental mass and stiffness matrices respectively. The elemental matrices can be assembled to form the global mass and stiffness matrices $M$ and $K$ respectively.

\begin{figure}[htbp]
    \centering
    \begin{subfigure}[ht]{0.51\textwidth}
        \centering
        \includegraphics[height=2.5in]{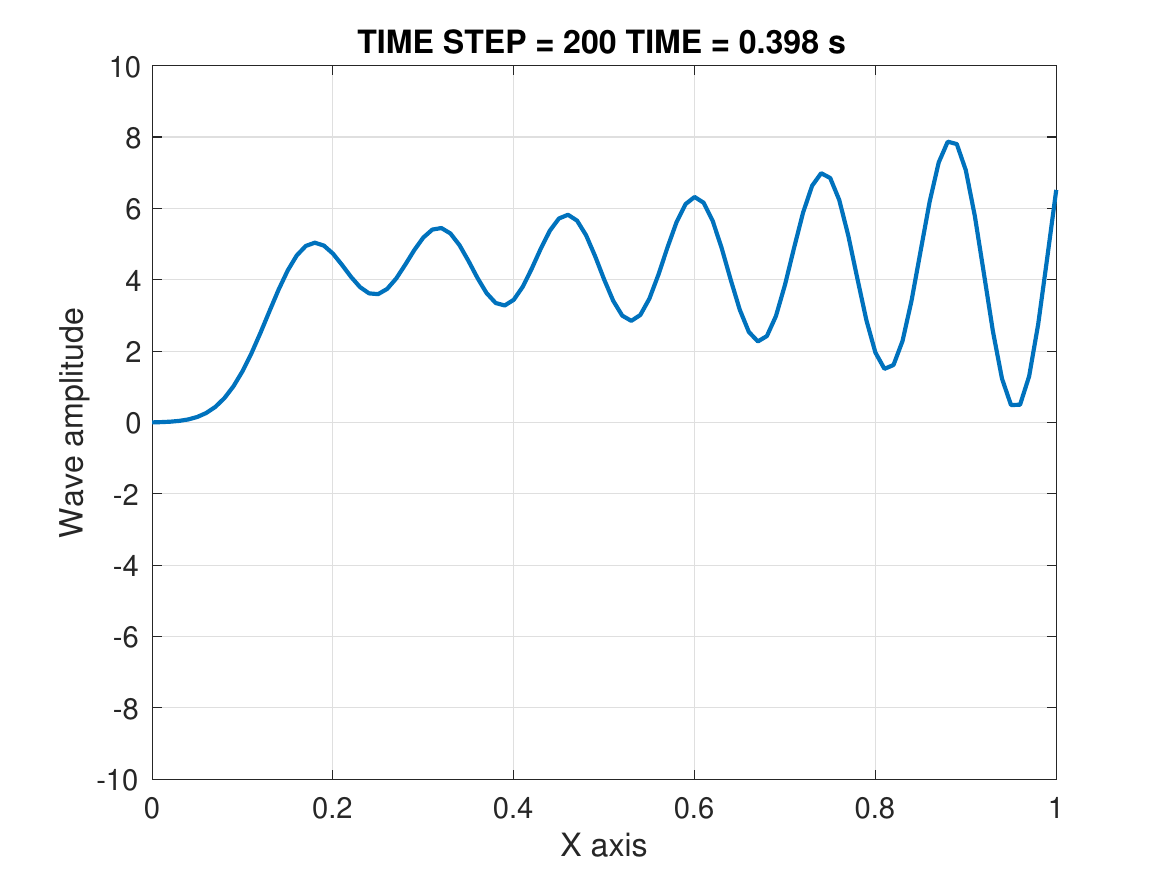} 
    \end{subfigure}%
    ~ 
    \begin{subfigure}[ht]{0.49\textwidth}
        \centering
        \includegraphics[height=2.5in]{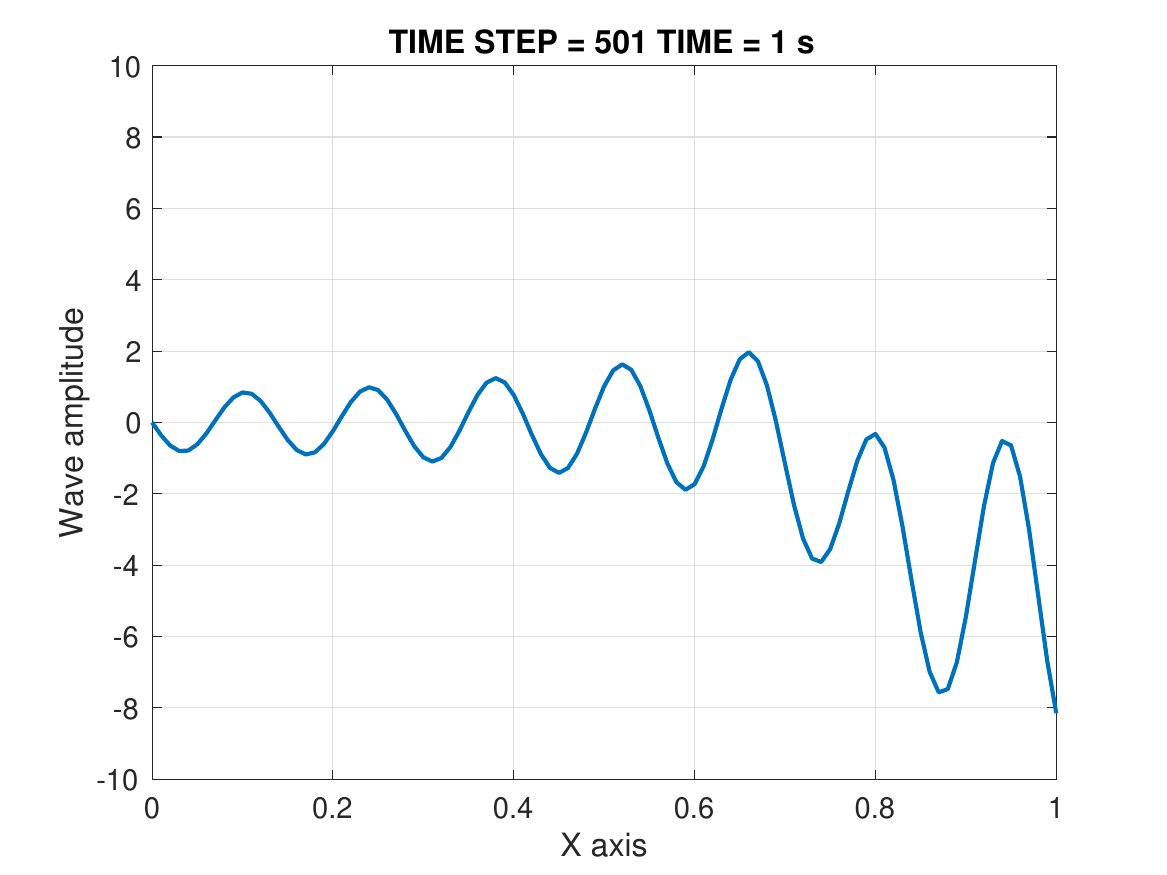} 
    \end{subfigure}
    
    \caption{Deterministic wave propagation of an axially loaded bar at different time steps} \label{Fig.1dwave_det}
\end{figure}

\begin{figure}[htbp]
        \centering
        \includegraphics[scale=0.5]{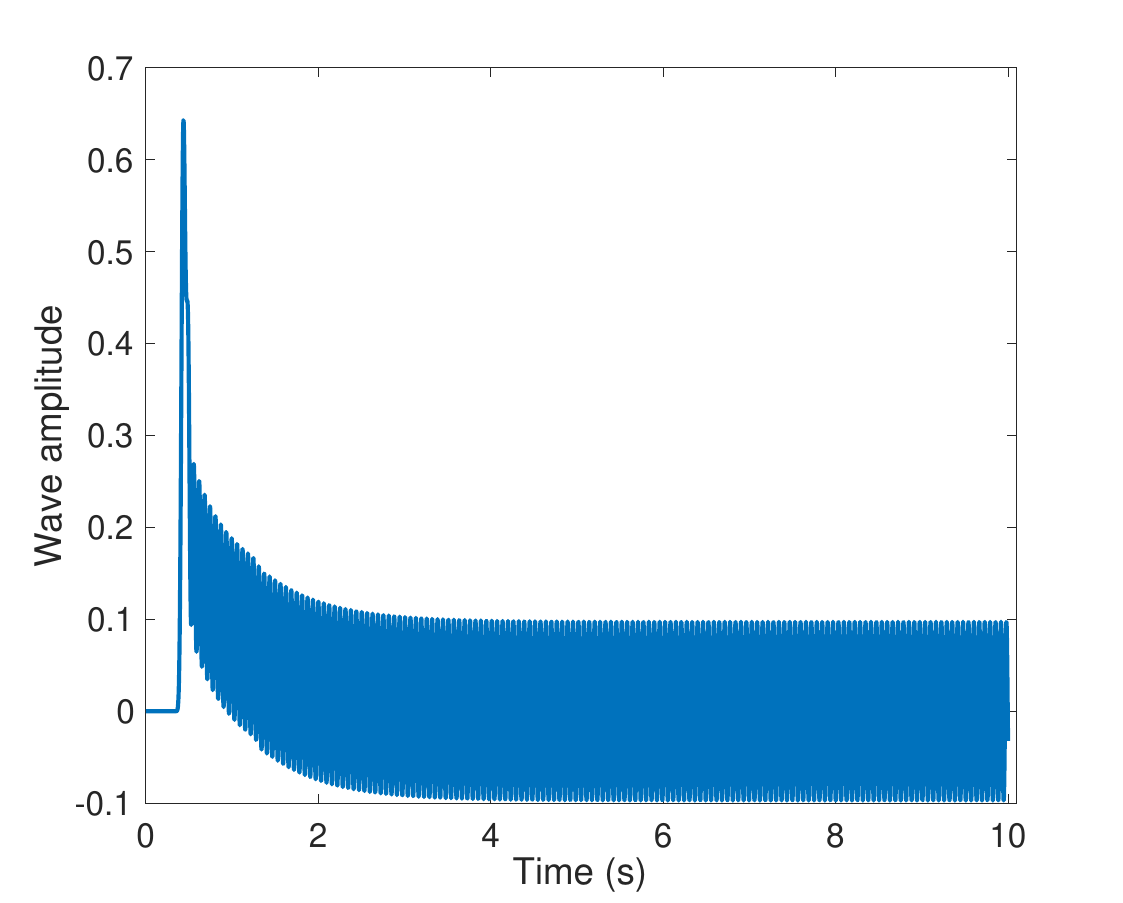} 
        \caption{Steady state of the system with higher damping for $10s$}\label{Fig.1dwave_steady}
\end{figure}

The damping for the system is assumed to follow a Rayleigh damping as \cite{humar}:
\begin{equation}
C = \alpha_0 M + \alpha_1 K
\end{equation} 
where $\alpha_0 = 0.01$ and $\alpha_1 = 0.001$ (see section. \ref{subsec:rayleigh}). The waves at two particular time instants on the bar using values $ E = 5$, $\Delta T = 0.002$, $h = 0.01, \rho = A = L = 1$ are shown in Fig.~\ref{Fig.1dwave_det}. The steady-state response of the axial vibration of the bar (with higher damping) is shown in Fig.~\ref{Fig.1dwave_steady}.

\subsection{Stochastic Wave Propagation}

The equation of motion of an axially loaded bar having random elastic modulus can be written as \cite{elasticw_sto}:
\begin{equation}
\frac{\partial^2 u(x,t,\theta)}{\partial t^2} - \frac{\partial} {\partial x}  \{ c_s(\theta) \frac{\partial u(x,t,\theta)}{\partial x} \} = f(x,t)
\end{equation}
where $c_s(\theta) = \frac{E(\theta)}{\rho}$ is the square of the wave speed. Note, the density is assumed to be deterministic for simplicity. 
The stochastic formulations for the axial vibration on a one-dimensional bar can be derived similar to the two-dimensional acoustic wave propagation (see section. \ref{sec:acwave_random_c2}). For brevity we discuss only the numerical results pertaining to the stochastic formulations.

\begin{figure}[htbp]
    \centering
    \begin{subfigure}[ht]{0.51\textwidth}
        \centering
        \includegraphics[height=2.5in]{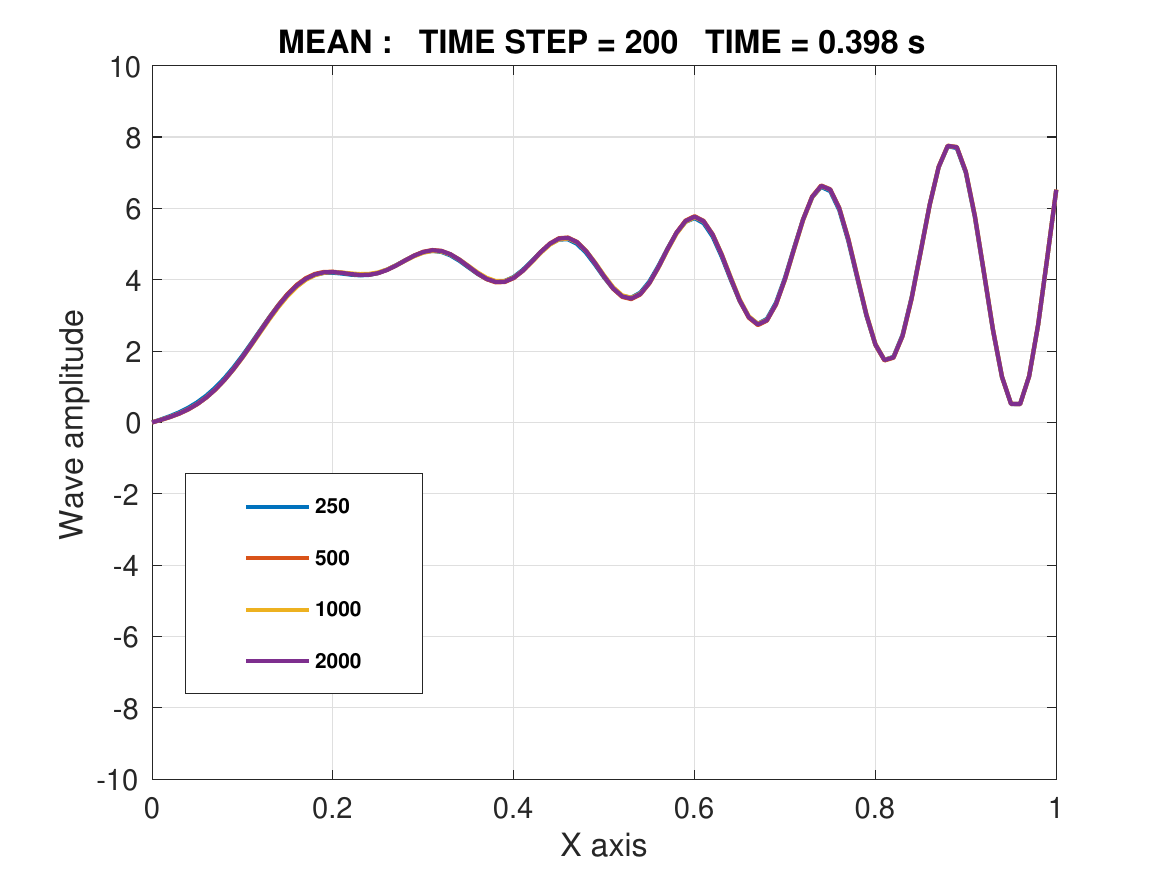} 
    \end{subfigure}%
    ~ 
    \begin{subfigure}[ht]{0.49\textwidth}
        \centering
        \includegraphics[height=2.5in]{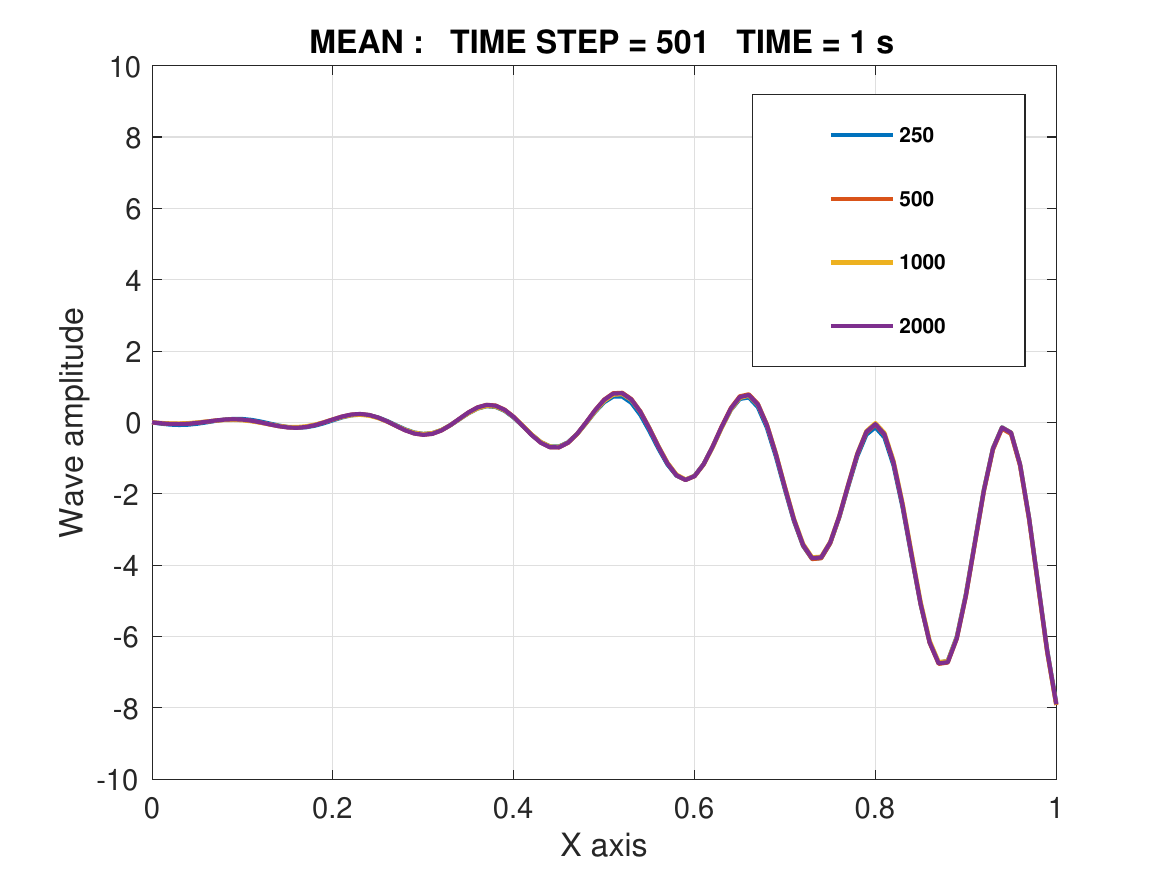} 
    \end{subfigure}
    
    \caption{MCS convergence using different sample sizes}\label{Fig:MC_1D_Convergence}
\end{figure}

\begin{figure}[htbp]
    \centering
    \begin{subfigure}[ht]{0.51\textwidth}
        \centering
        \includegraphics[height=2.5in]{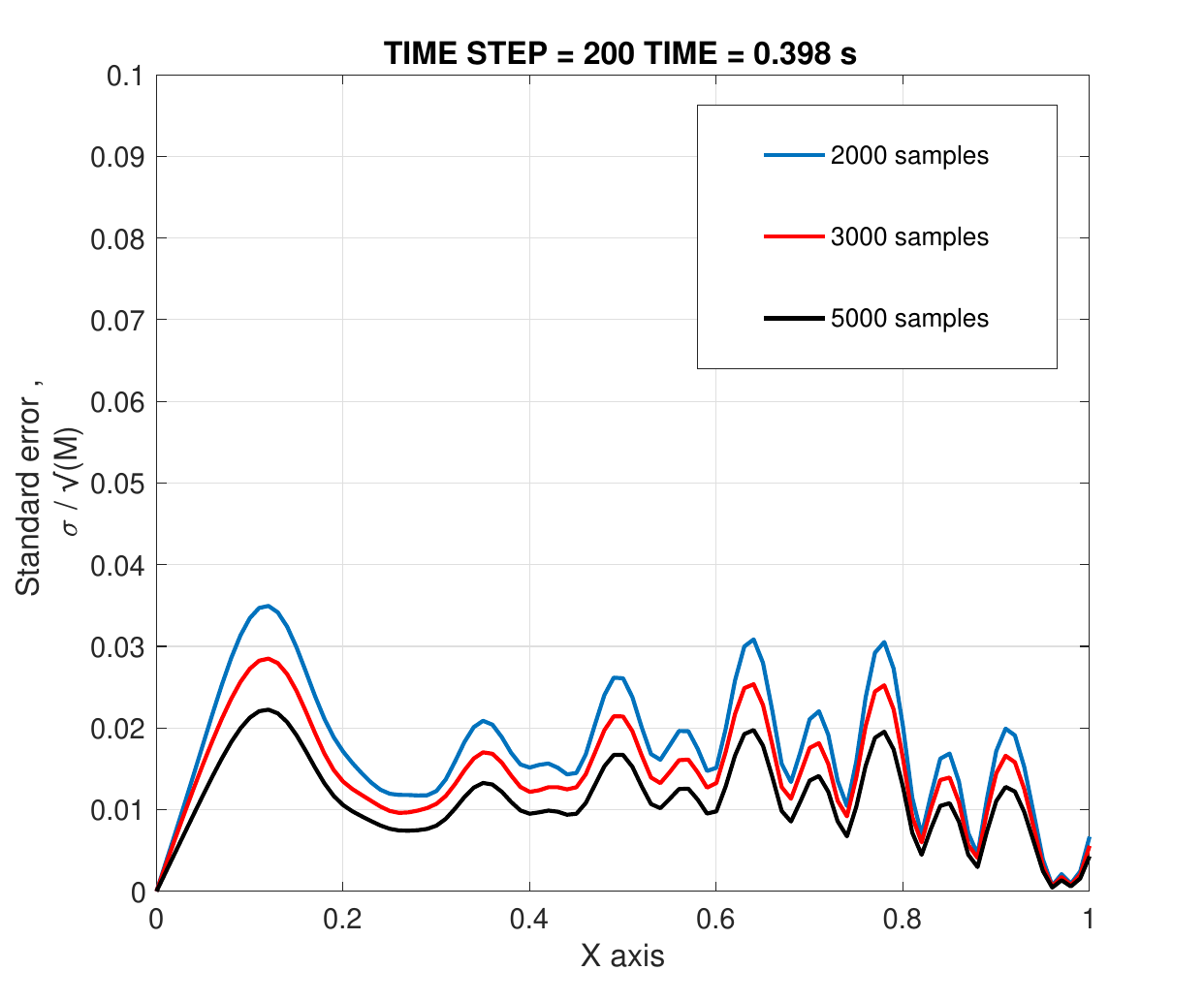} 
    \end{subfigure}%
    ~ 
    \begin{subfigure}[ht]{0.49\textwidth}
        \centering
        \includegraphics[height=2.5in]{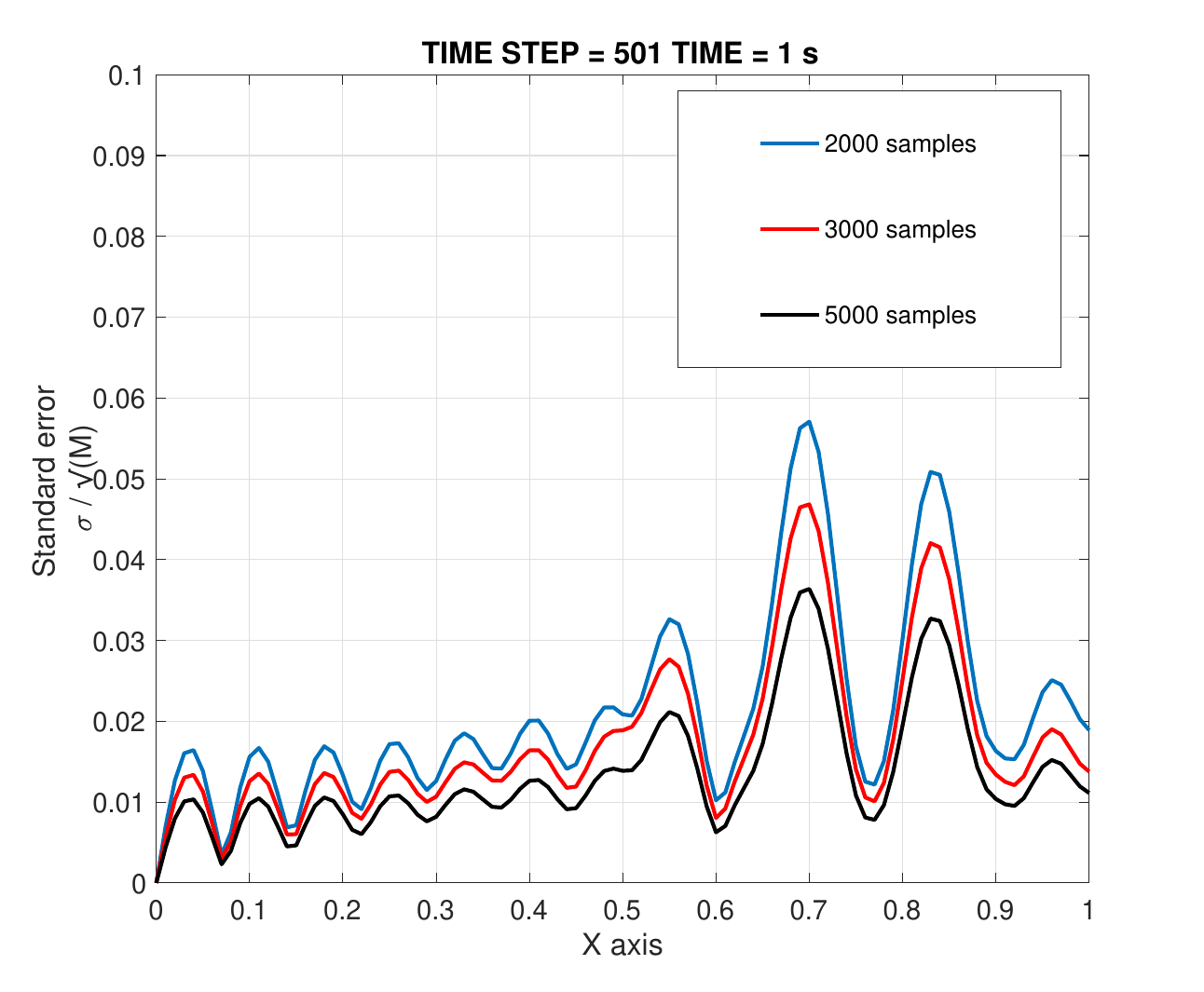} 
    \end{subfigure}
    
    \caption{Standard error along length at different times}\label{Fig:MC_1D_StdError}
\end{figure}

Numerical experiments using Monte Carlo simulations (MCS), stochastic Galerkin (intrusive) method and the non-intrusive sampling method are presented below for the one-dimensional axial bar with random wave speed. The parameter values are tsame as described in the deterministic case. MCS for the stochastic wave propagation of the bar is carried out by propagating several samples of the log-normal modulus of elasticity. A single sample evaluation of MCS involves finding the response of the bar for the whole interval of time. The average solution of a large number of such realizations gives us the mean estimate using MCS. Fig \ref{Fig:MC_1D_Convergence} shows the mean solution at different time steps for various sample sizes. 
Note that the difference between the true mean of the population and the MCS estimate is a random variable which has an expected value and associated variance. If the mean estimate of Monte Carlo samples are $\tilde{\mu}$ and the true mean is $\mu$, the Central Limit Theorem (CLT) states that the error in the estimate of mean by MCS converges to a Gaussian distribution as \cite{book_murphy,book_raulphsmith}:
\begin{equation}
    (\tilde{\mu} - \mu) \rightarrow \mathcal{N}(0, \frac{\sigma^2}{M})
\end{equation}
where $M$ is the number of samples and $\sigma$ is the original standard deviation of the population which is unknown. Generally, this value could also be estimated from the same sample set as \cite{book_murphy,book_raulphsmith}:
\begin{equation}
\tilde{\sigma}^2 =  \frac{1}{M-1} \sum _{i=1} ^ M (y_i - \tilde{\mu})^2
\end{equation}
where $y_i$ are the Monte Carlo outputs and $\tilde{\sigma}^2$ is the estimate of variance. The standard deviation of the error in mean estimate by MCS, $\frac{\sigma}{\sqrt(M)}$ (which can be approximated as $\frac{\tilde{\sigma}}{\sqrt(M)}$) is known as the standard error of the estimator. To reduce the variability in the mean estimate by ten times, the number of samples has to be increased by hundred times. Fig.~\ref{Fig:MC_1D_StdError} shows the standard error values found using the MCS for different sample sizes. 

\begin{figure}[htbp]
    \centering
    \begin{subfigure}[b]{\textwidth}
       \centering
        \includegraphics[width=0.475\linewidth]{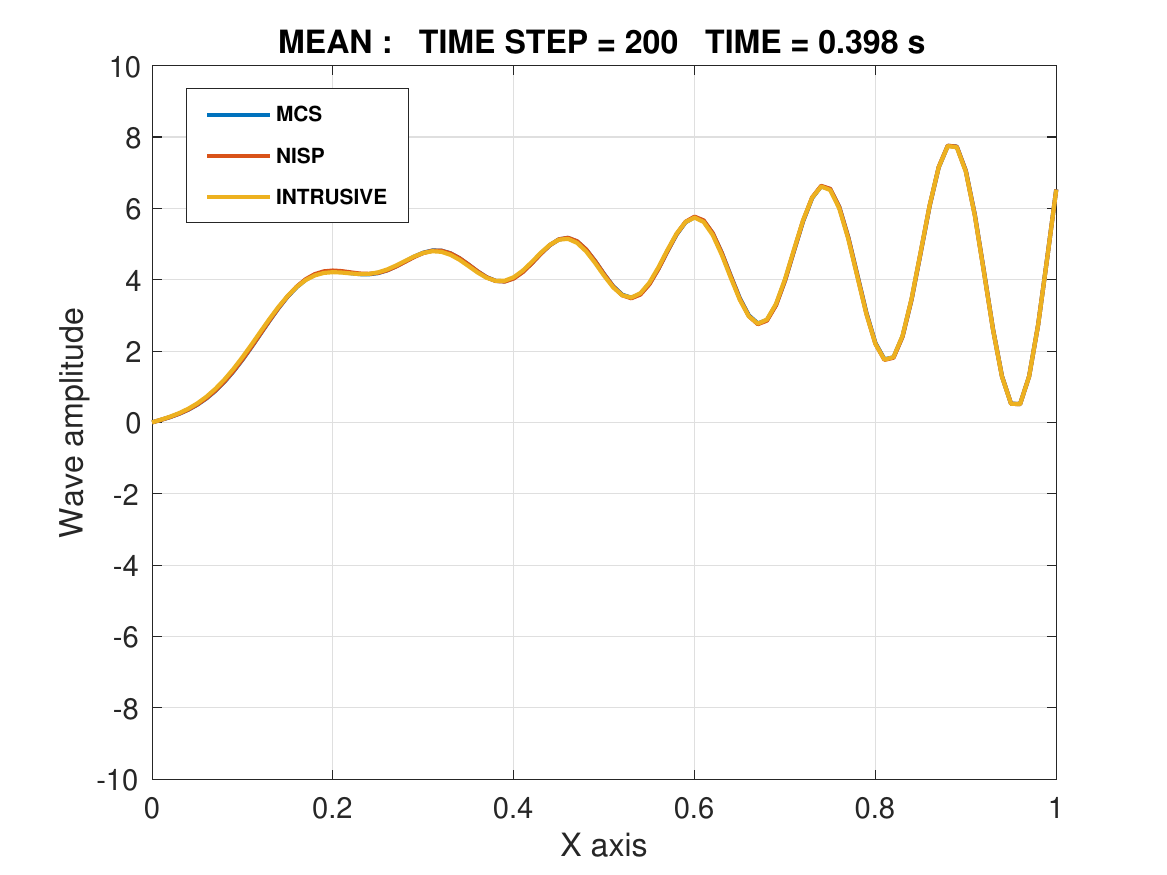}              
        \hfill
        \includegraphics[width=0.475\linewidth]{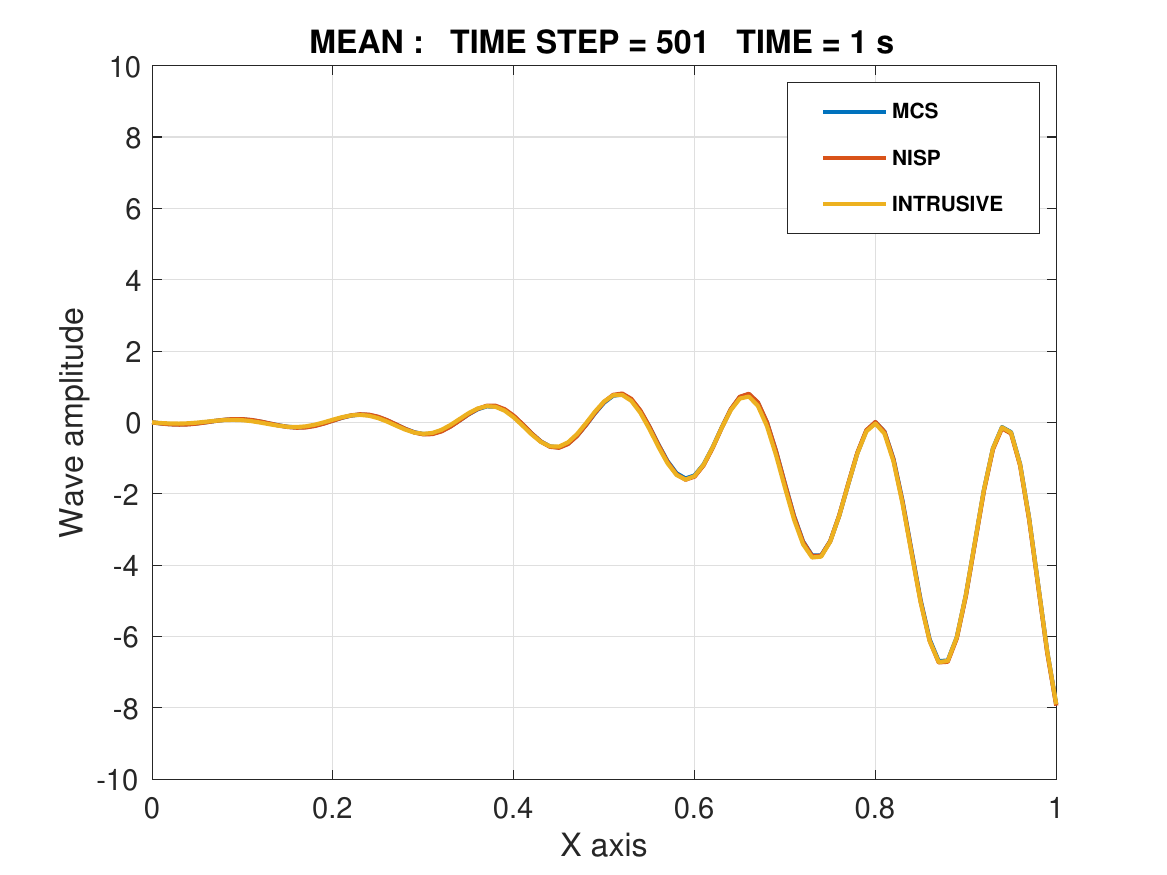} 
        \caption{Mean solution at different time steps for all methods} \label{1.a}   
    \end{subfigure}

    \centering
    \begin{subfigure}[b]{\textwidth}
        \centering
        \includegraphics[width=0.475\linewidth]{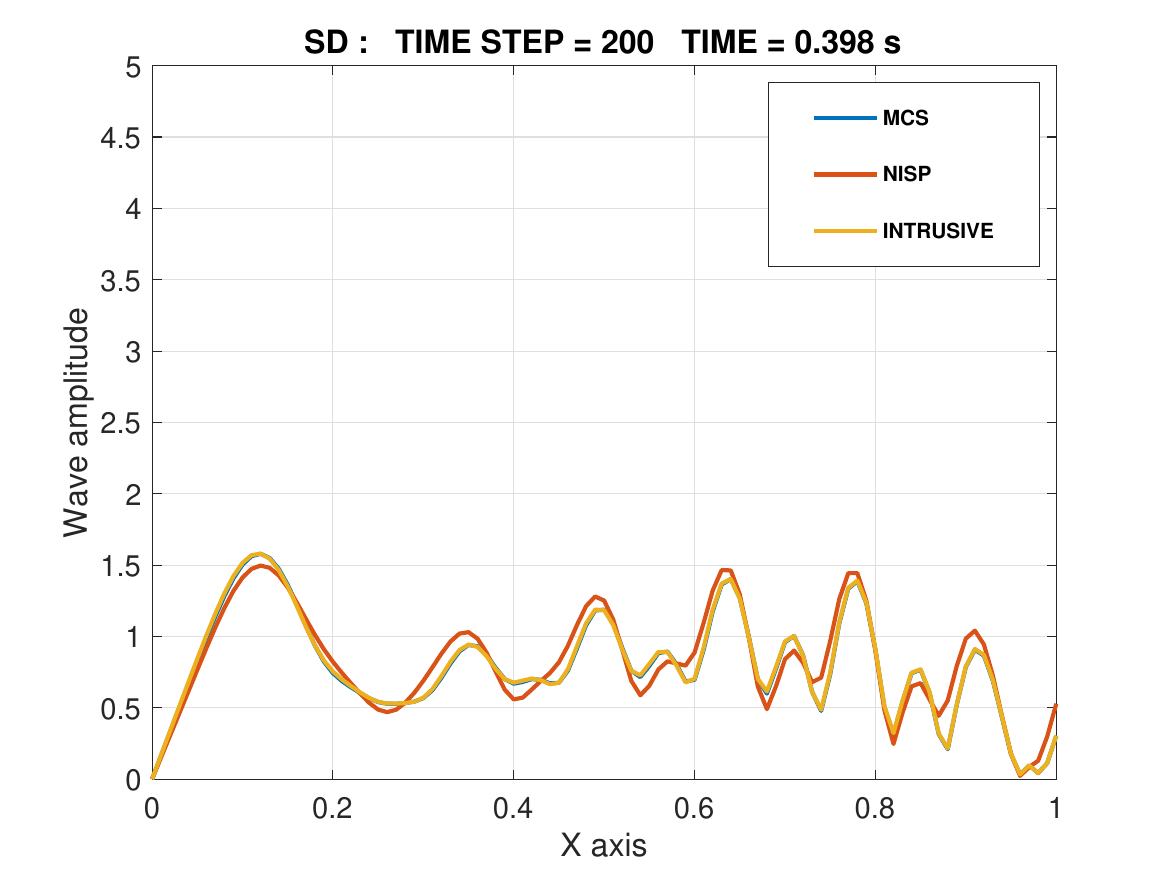}      
\hfill
        \includegraphics[width=0.475\linewidth]{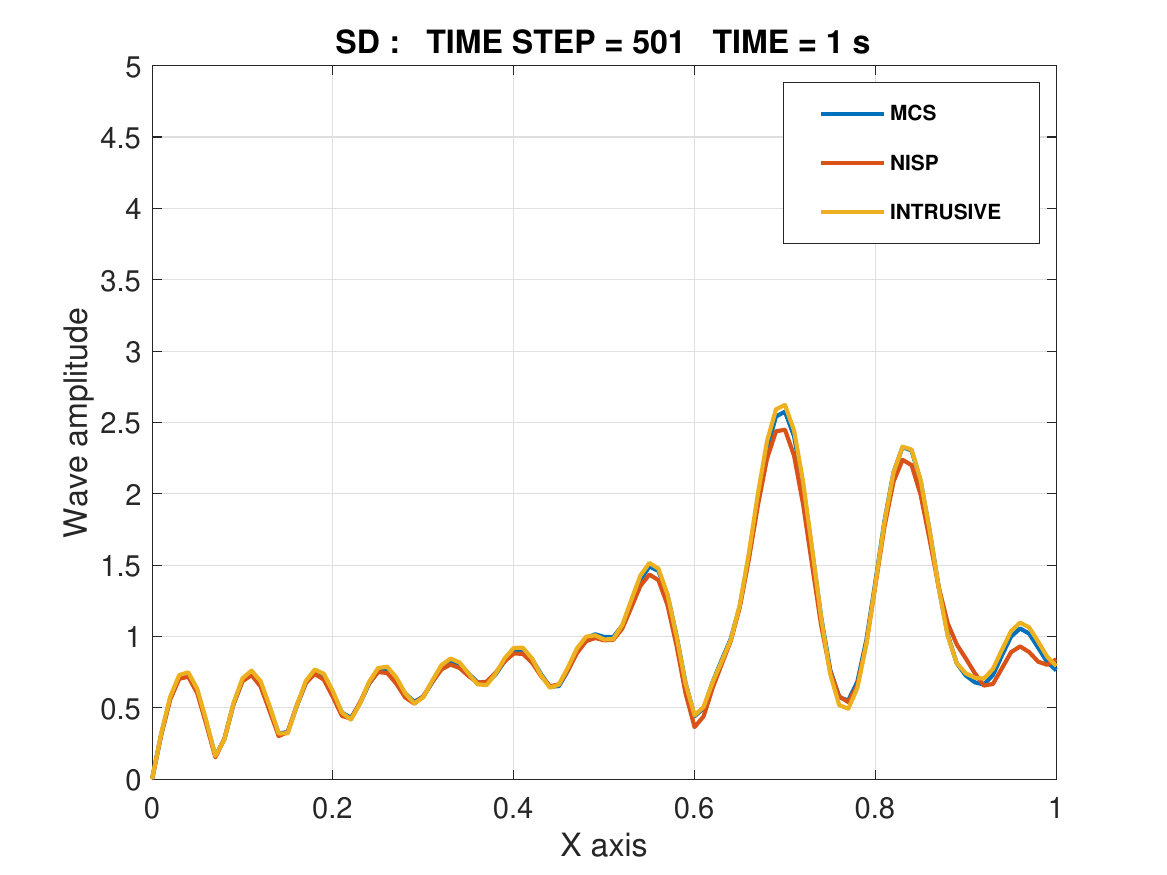} 
        \caption{Standard deviation at different time steps for all methods} \label{1.b} 
    \end{subfigure}
    
  \caption{Mean and standard deviation for axial stress wave in a 1D bar due to a sinusoidal forcing}\label{Fig. 1dwave_compare}
\end{figure}

A comparison of mean and standard deviation using stochastic Galerkin and non-intrusive method using sampling is shown in Fig.~\ref{Fig. 1dwave_compare}. Both intrusive and non-intrusive methods use seventh-order input and tenth-order output polynomial chaos expansions. The MCS and the calculation of PCE coefficients of the non-intrusive method use a sample size of 2000. It can be noted that the mean and standard deviation of stochastic Galerkin method converges to the MCS solution while the non-intrusive method has higher differences with MCS.

\begin{figure}[htbp]
    \centering
    \begin{subfigure}[b]{\textwidth}
       \centering
        \includegraphics[width=0.475\linewidth]{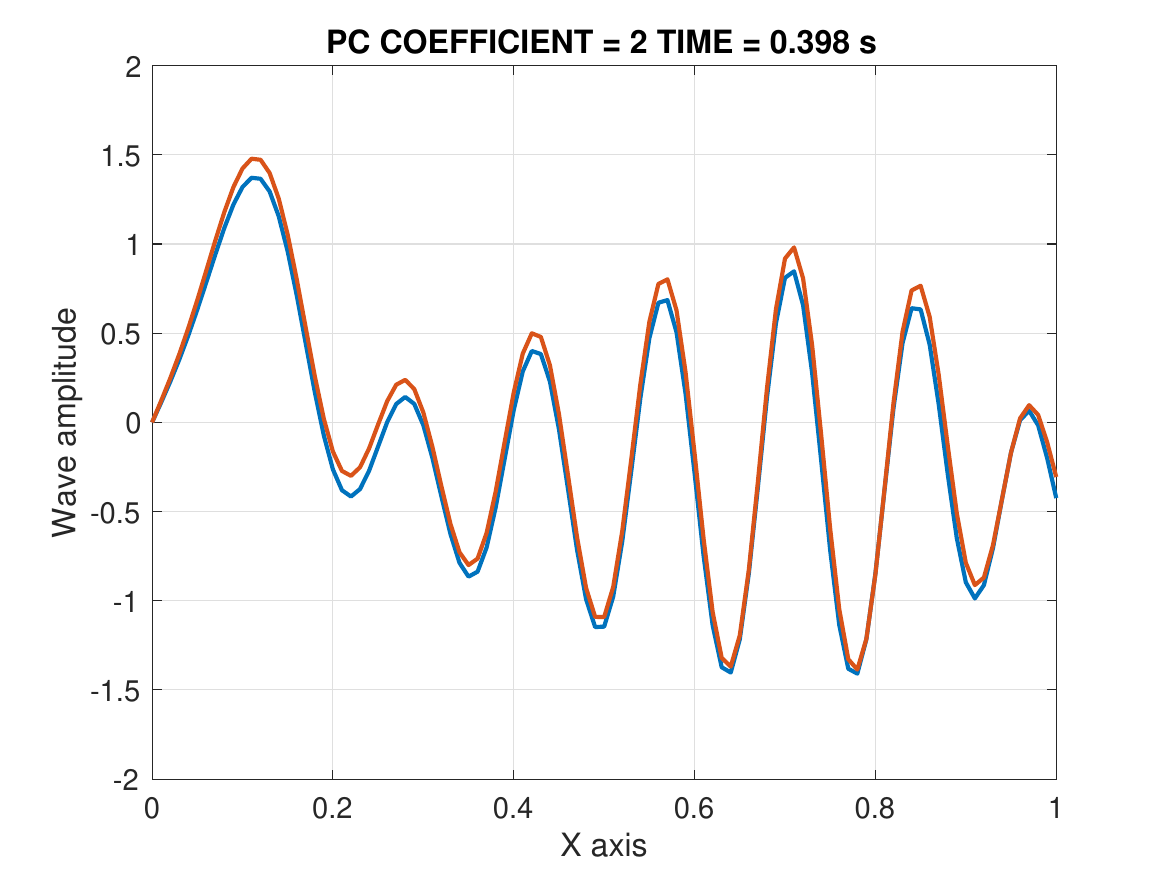}              
        \hfill
        \includegraphics[width=0.475\linewidth]{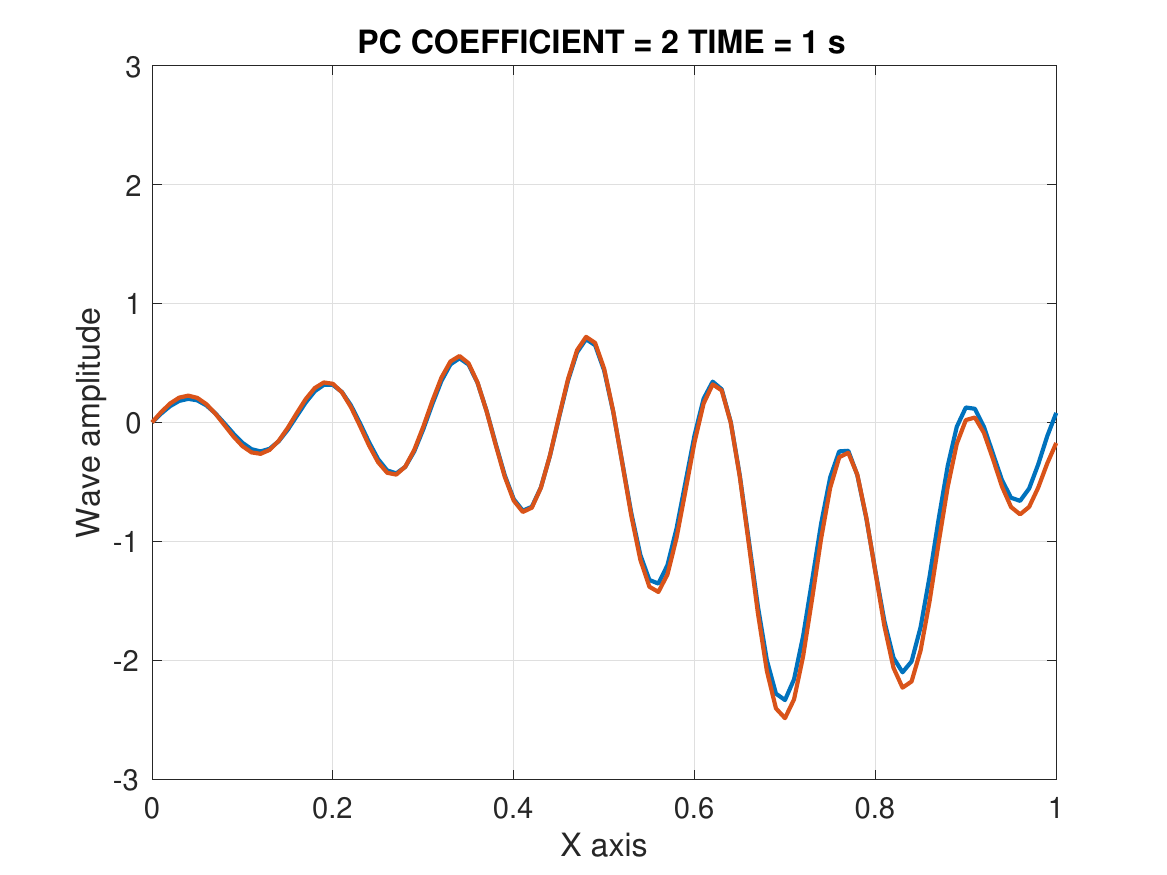} 
     \end{subfigure}
    \centering
    \begin{subfigure}[b]{\textwidth}
        \centering
        \includegraphics[width=0.475\linewidth]{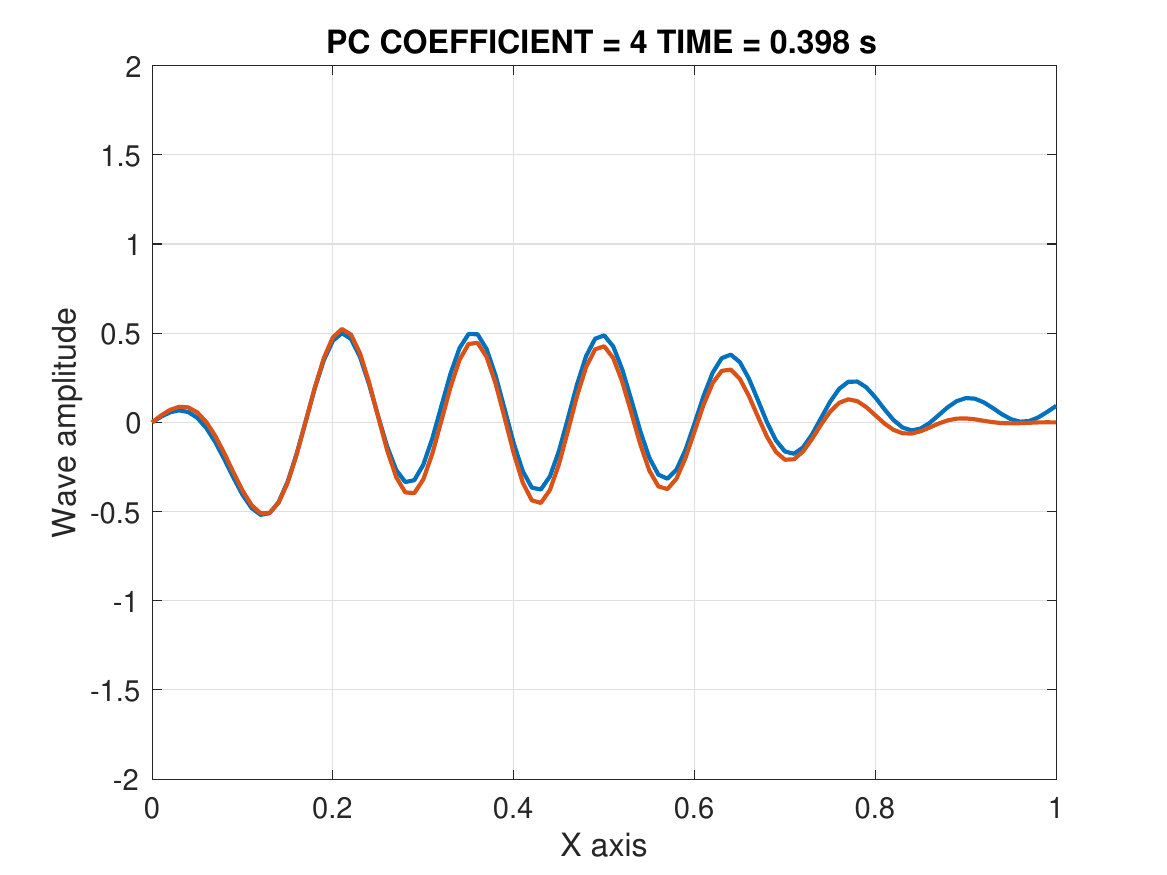}      
\hfill
        \includegraphics[width=0.475\linewidth]{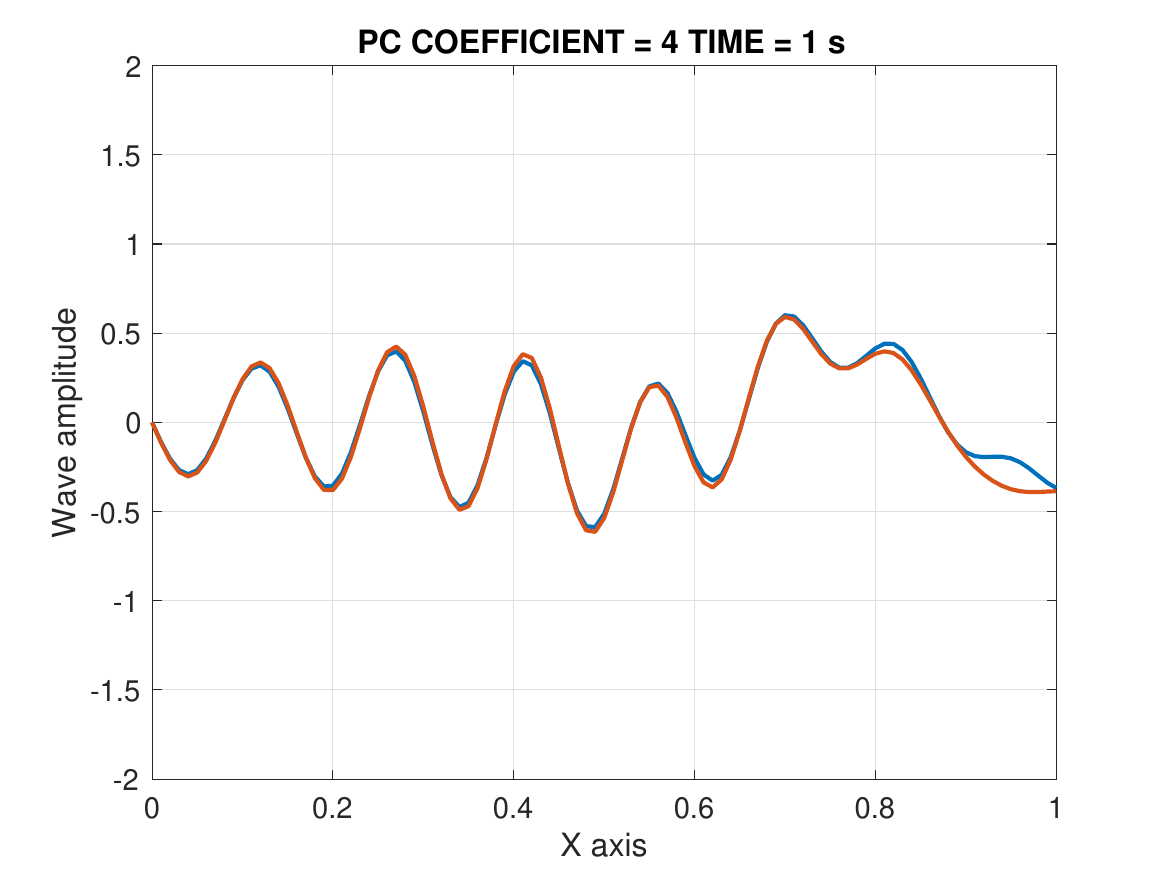} 
    \end{subfigure}
\centering
    \begin{subfigure}[b]{\textwidth}
        \centering
        \includegraphics[width=0.475\linewidth]{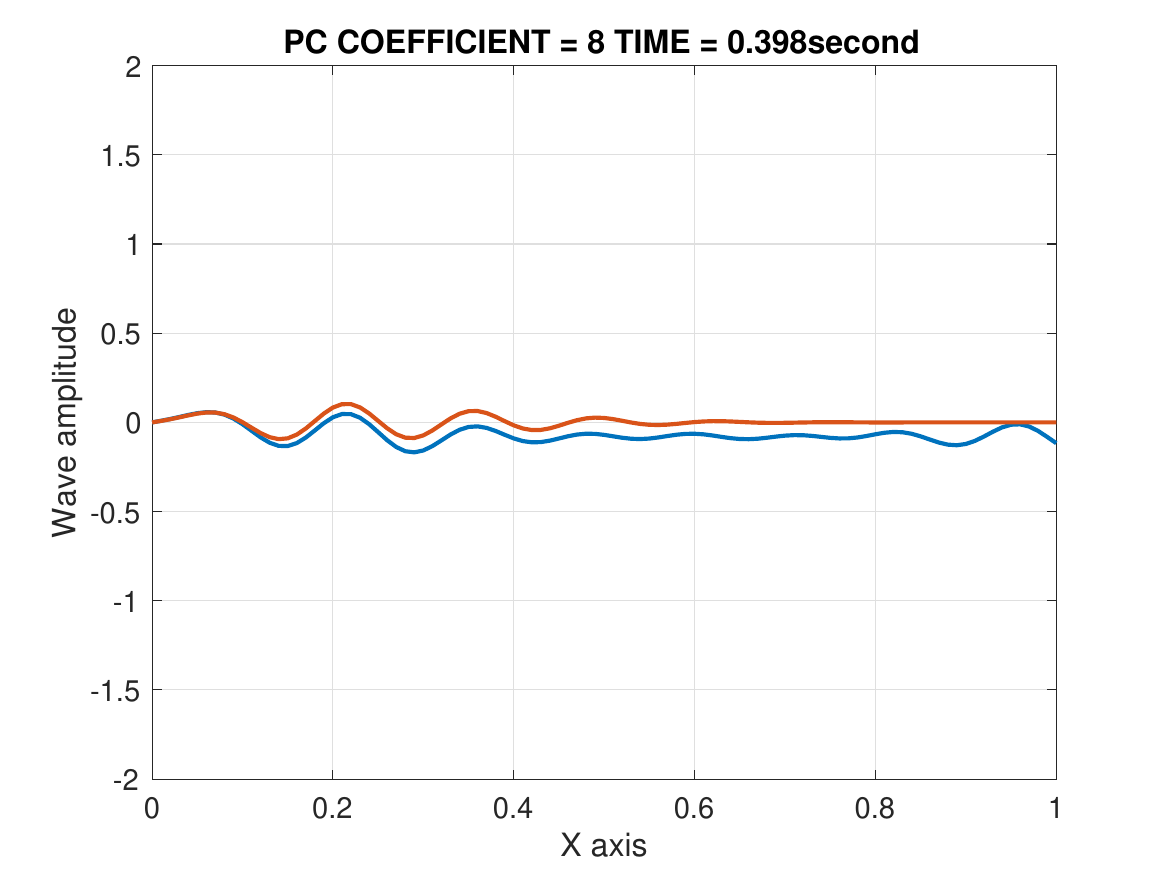}      
\hfill
        \includegraphics[width=0.475\linewidth]{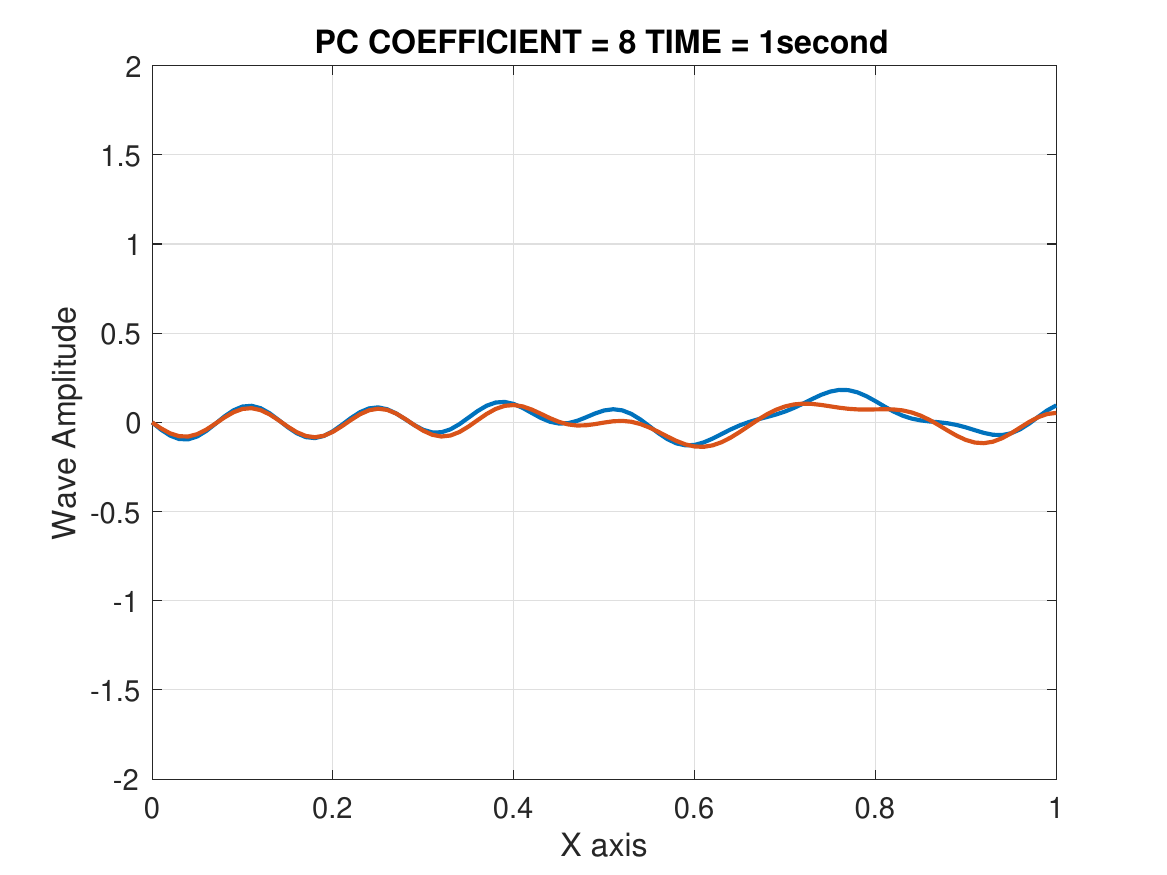} 
    \end{subfigure}
  \caption{PC coefficients for axial stress wave in a 1D bar due to a sinusoidal forcing}\label{Fig.1dwave_pccompare}
\end{figure}

PCE coefficients calculated using both intrusive and non-intrusive method show a close match between both methods in Fig.~\ref{Fig.1dwave_pccompare}. The magnitude of PCE coefficients decrease with increasing order.

\section{Acoustic Wave Propagation in Two-Dimensional Media}

This section discusses some relevant numerical experiments for the acoustic wave propagation on two-dimensional media. A verification for the wave propagation in the deterministic setting is presented along with an analysis on the effect of CFL conditions. The pdf of the wave pressure at different time steps computed using a log-normal random wave speed is also shown.

\subsection{Verification for Acoustic Wave in Deterministic Media}\label{Sec.2dwave_det}
 
\begin{figure}[htbp]
    \centering
    \begin{subfigure}[ht]{0.5\textwidth}
        \centering
        \includegraphics[height=2.5in]{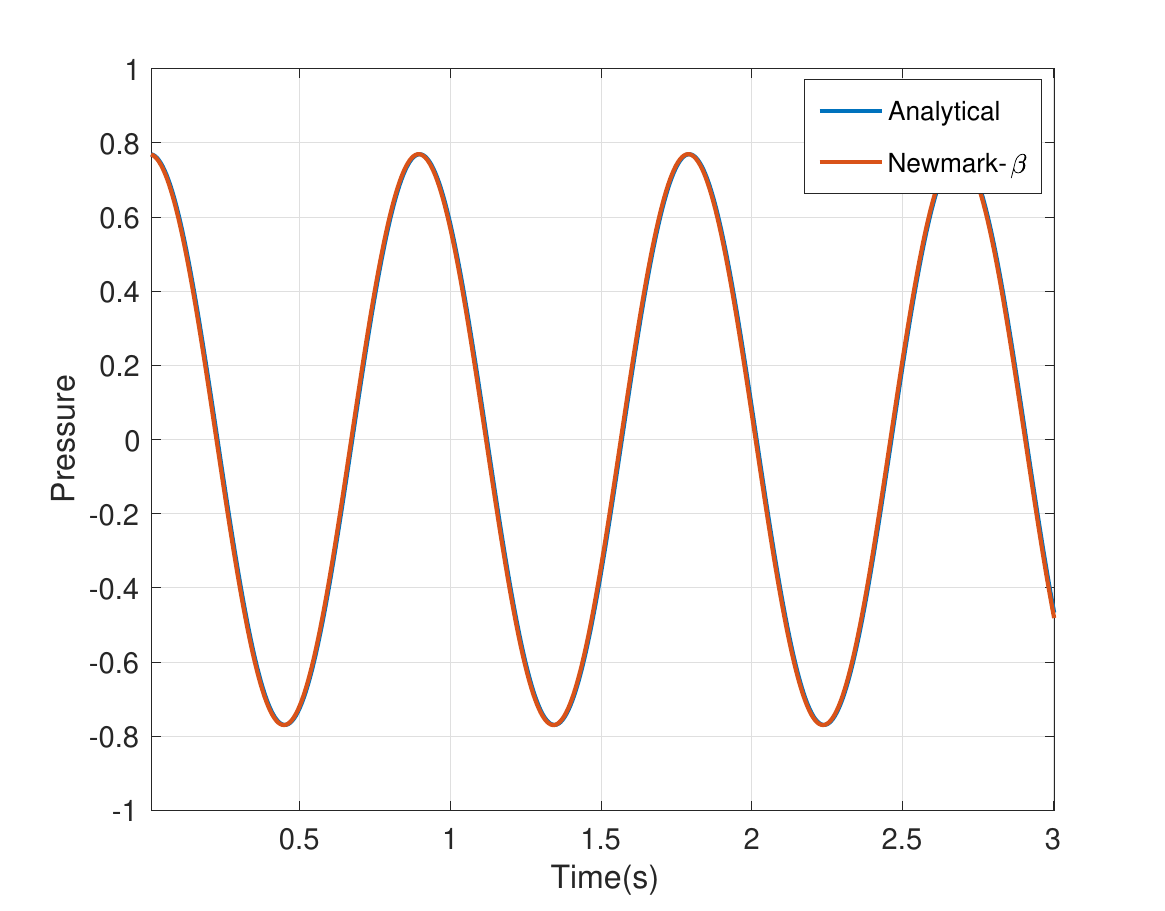} 
    \end{subfigure}
        
    \begin{subfigure}[ht]{0.5\textwidth}
        \centering
        \includegraphics[height=2.5in]{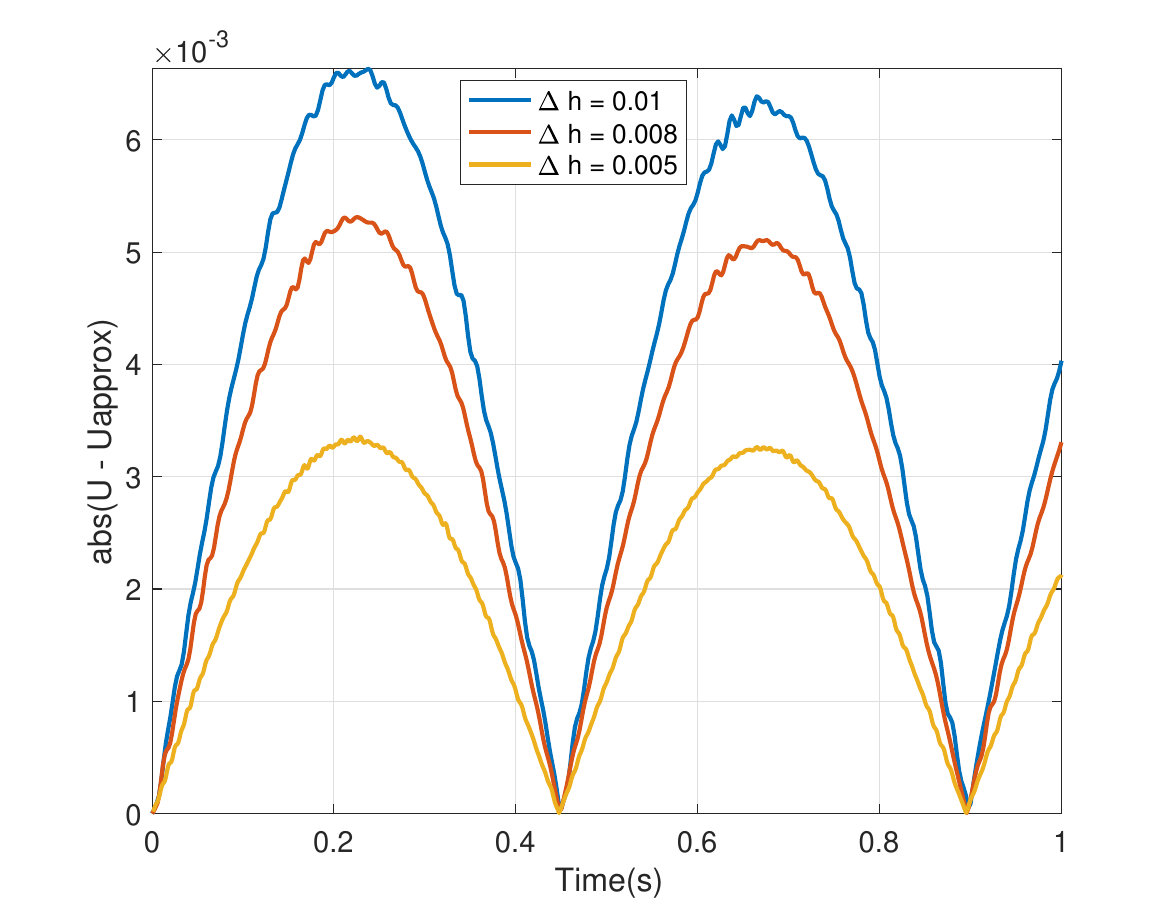} 
    \end{subfigure}
    \begin{subfigure}[ht]{0.5\textwidth}
        \centering
        \includegraphics[height=2.5in]{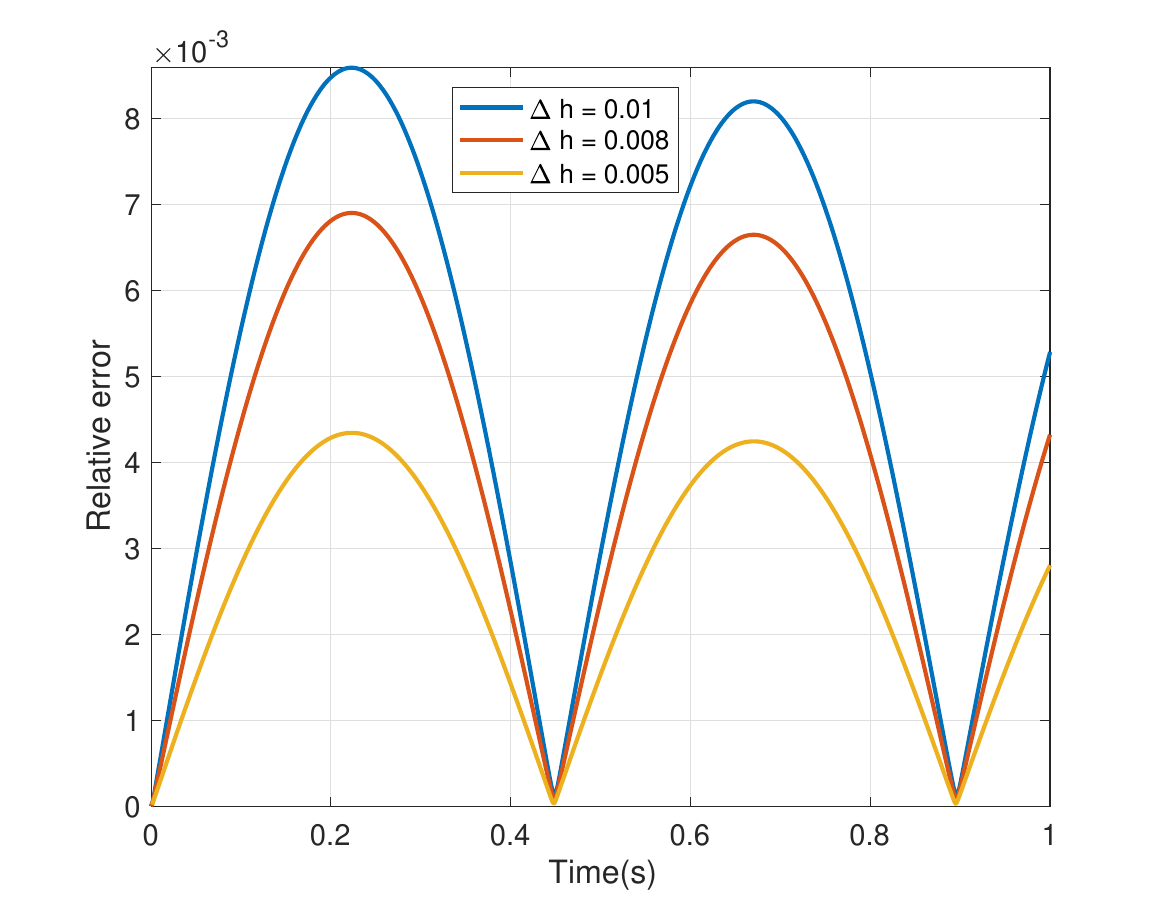} 
    \end{subfigure}
    
    \caption{Verification for acoustic wave propagation in a two-dimensional domain (a) solution, (b) absolute error and (c) relative error at $(x =0.292, y = 0.703)$ with CFL number of $0.25$}\label{Fig.2dwave_verify}
\end{figure}

The verification for the acoustic wave equation in Eq.~(\ref{Eq.acoustic_deter}) in a deterministic medium is presented. The boundary and initial conditions for the wave equation in the two-dimensional unit square domain with vertices at $(0,0), (0,1), (1,0)$ and $(1,1)$ are considered as:
\begin{align}
u(0,y,t) = u(1,y,t) = u(x,0,t) = u(x,1,t) = 0\\
u(x,y,0) = \mathrm{sin}(m \pi x)\,\mathrm{sin}(n \pi y)
\end{align}
where $m=2$ and $n=1$. The wave speed is considered as a deterministic value of $c = 1$ and the source term $f = 0$. For simplicity, damping is neglected. The analytical solution for the wave propagation problem can be found to be \cite{DD_Acoustic_Pavarino}:
\begin{equation}\label{Eq.2danalytical}
u(x,y,t) = \mathrm{sin}(m \pi x) \,\mathrm{sin}(n \pi y) \,\mathrm{cos}(c \pi \times \sqrt{m^2 + n^2} \times t).
\end{equation}
For the numerical solution of the problem, the spatial domain is discretized with $13472$ vertices and a time step of $0.001$s is used for the temporal discretization. The numerical and analytical solution for the wave propagation are compared in Fig.~\ref{Fig.2dwave_verify}a. The absolute error is calculated as the absolute value of the difference between numerical and analytical solutions whereas the relative error at a time step is computed as:
\begin{equation}
e(t) = \frac{\Vert (U(\mathbf{x},t) - \hat{U}(\mathbf{x},t))\Vert}{\Vert U(\mathbf{x},t)\Vert}.
\end{equation}\label{Eq.Norm} 
where $U(\mathbf{x},t)$ is the analytical solution and $\hat{U}(\mathbf{x},t)$ is the numerical solution.

\begin{figure}[htbp]
\centering
\includegraphics[width=0.5\linewidth]{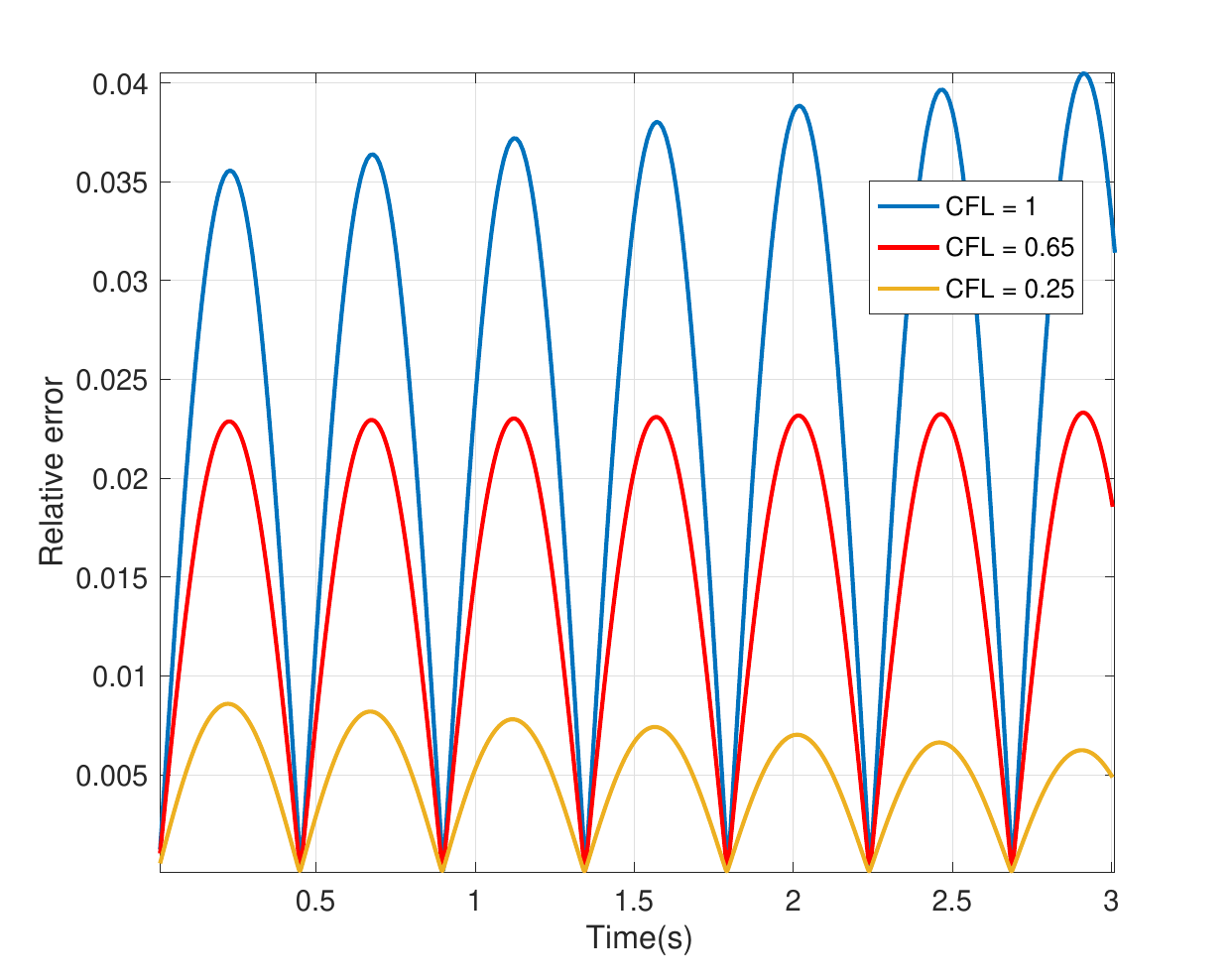} 
\hfill
\includegraphics[width=0.475\linewidth]{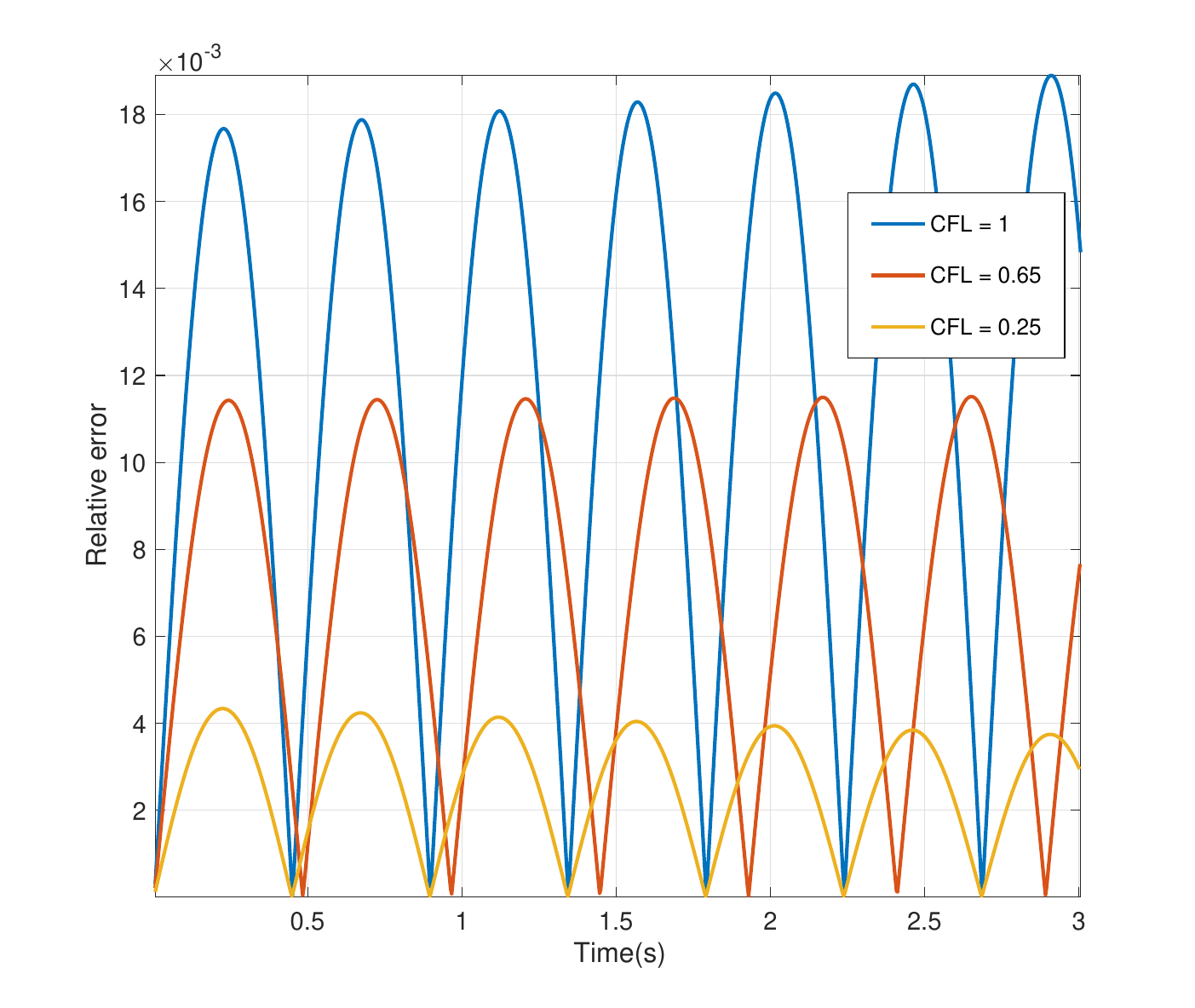}
\caption{Relative error for various CFL numbers with a coarse mesh of $13472$ vertices (left) and a fine mesh of $53263$ vertices (right).}\label{Fig:CFL}
\end{figure}

A numerical experiment showing the effect of using different CFL numbers for mesh sizes of $13472$ and $53263$ vertices is carried out as shown in Fig.~\ref{Fig:CFL}. These plots demonstrate that finer mesh and smaller CFL number give smaller errors. The error for higher CFL numbers in both mesh sizes can be observed to be increasing with time. A CFL number of $0.65$ is chosen for all the numerical experiments in wave propagation models (also in accordance with \cite{bathe_1, bathe_2} to reduce the numerical dispersion error for the Newmark-beta method).

%

\subsection{Computation of Rayleigh Damping Coefficients}\label{subsec:rayleigh}
\begin{figure}[htbp]
        \centering
        \includegraphics[height=2.5in]{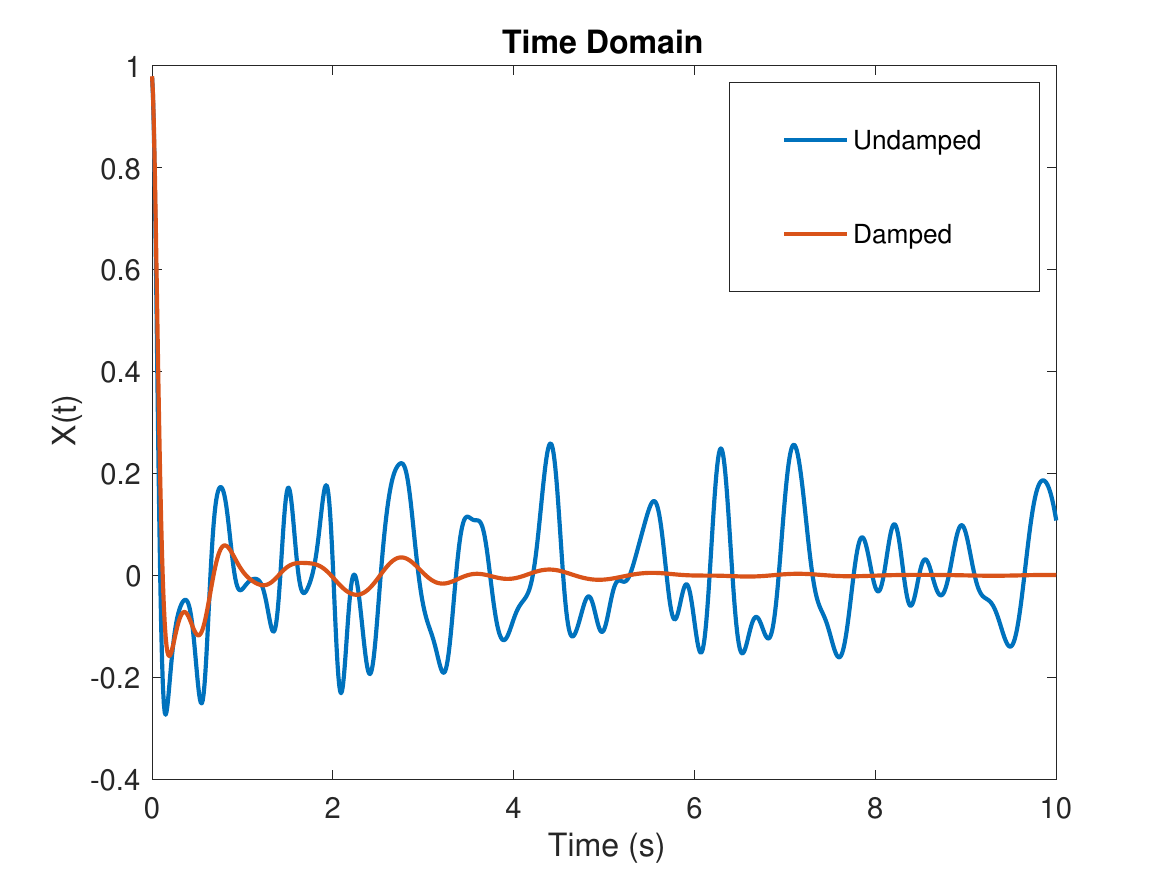} 
         \caption{Damped and undamped response at point $(0.7,0.7)$ in time domain.}\label{Fig.RD_TD}
\end{figure}

\begin{figure}[htbp]
       \centering
        \includegraphics[height=2.5in]{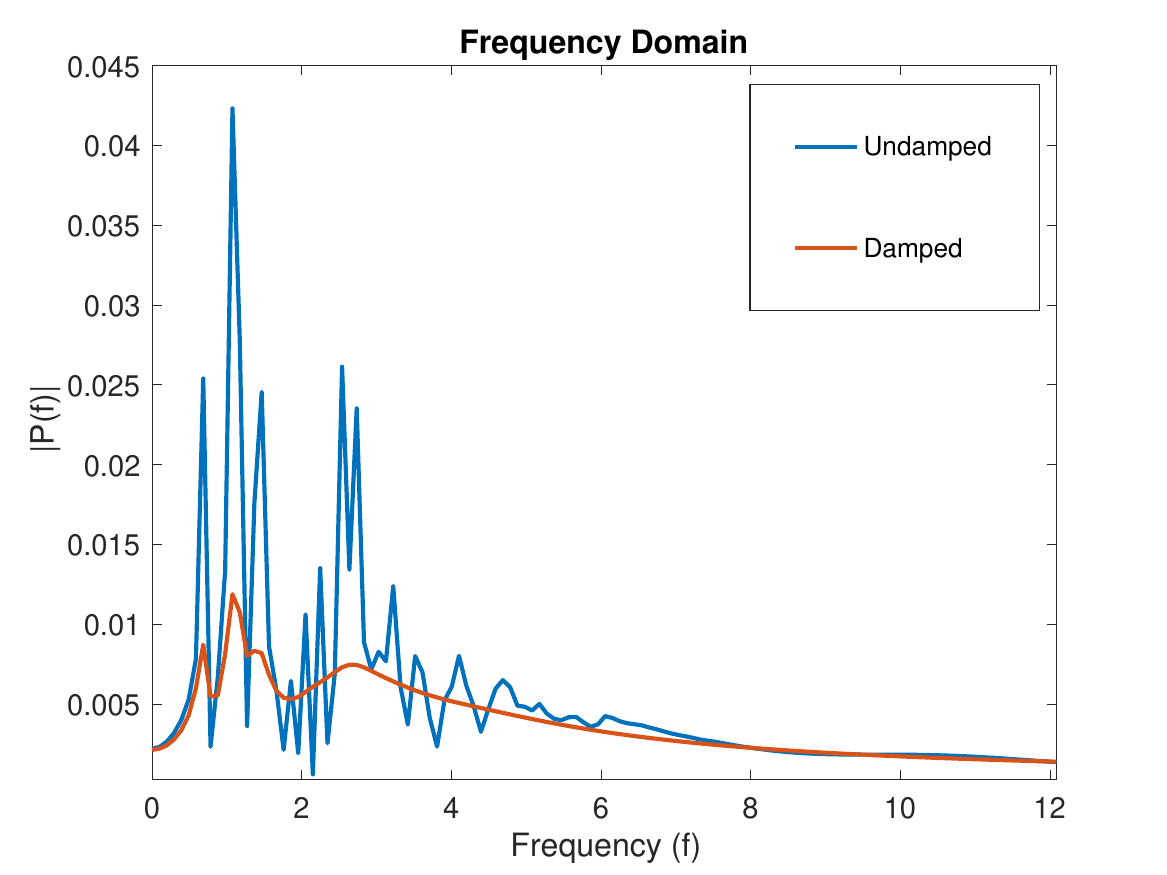} 
         \caption{Damped and undamped response at point $(0.7,0.7)$ in frequency domain.}\label{Fig.RD_FD}
\end{figure}

For all numerical experiments in this article, the damping matrix is constructed using a Rayleigh damping model.
This section describes the computation of the Rayleigh damping coefficients. The Rayleigh damping can be computed as a combination of mass and stiffness matrices as \cite{humar}:
\begin{equation}
\mathbf{C} = \alpha_0 \mathbf{M} + \alpha_1 \mathbf{K} 
\end{equation}
where $\alpha_0, \alpha_1$ can be computed as:
\begin{equation}
\frac{1}{2}
\begin{bmatrix}
\frac{1}{\omega_i} &  \omega_i \\

\frac{1}{\omega_j} &  \omega_j \\

\end{bmatrix}
\begin{bmatrix}
\alpha_0 \\

\alpha_1 \\
\end{bmatrix} =
\begin{bmatrix}
\xi_i \\

\xi_j \\
\end{bmatrix}
\end{equation}
where $\omega_i$ and $\omega_j$ are the two dominant frequencies in the response and $\xi_i$ and $\xi_j$ are the damping ratios (chosen as $0.1$ in all cases). The time trace of the wave propagation model as explained in section \ref{Sec.2dwave_det} for $10$s is shown in Fig.~\ref{Fig.RD_TD}. The dominant natural frequencies are found to be $0.6836$ Hz and $1.074$ Hz in the Fourier transform of the response as shown in Fig.~\ref{Fig.RD_FD}. The values of $\alpha_0$ and $\alpha_1$ are computed as $0.5445$ and $0.0174$ respectively.
 
\textcolor{ss}{The damping matrix for the stochastic case in this thesis involves a deterministic mass matrix and random stiffness matrix written as},
\textcolor{ss}{
\begin{equation}
\mathbf{C}(\theta) = \alpha_0 \mathbf{M} + \alpha_1 \mathbf{K}(\theta)
\end{equation}
where,
\begin{equation}
 \mathbf{K}(\theta) = \sum_{i = 0} ^{N} \mathbf{K}_i \Psi_i(\bm{\xi})
\end{equation}
and 
\begin{equation}
 \mathbf{C}(\theta) = \sum_{j = 0} ^{N} \mathbf{C}_j \Psi_j(\bm{\xi})
\end{equation}
In order to find the PCE coefficients for the damping matrix $\mathbf{C}_j$ we use Galerkin projection as:
\begin{equation}
 \sum_{j = 0} ^{N} \mathbf{C}_j \langle \Psi_j(\bm{\xi}) \Psi_k(\bm{\xi}) \rangle  = \alpha_0 \mathbf{M} \langle \Psi_k(\bm{\xi}) \rangle + \alpha_1 \sum_{i = 0} ^{N} \mathbf{K}_i \langle \Psi_i(\bm{\xi}) \Psi_k(\bm{\xi}) \rangle \quad k = 0,1, \cdots N
\end{equation}
Using the orthogonality of polynomial chaos each coefficient can be written as:
\begin{equation}
 \mathbf{C}_k = \alpha_0 \mathbf{M} \frac{\langle \Psi_k(\bm{\xi}) \rangle}{\langle \Psi^2 _k(\bm{\xi}) \rangle} + \alpha_1 \sum_{k = 0} ^{N} \frac{\mathbf{K}_k \langle \Psi^2 _k(\bm{\xi}) \rangle }{\langle \Psi^2 _k(\bm{\xi}) \rangle} 
\end{equation}
Simplifying, we can compute the mean damping matrix as the sum of mean mass and stiffness matrices and other PCE coefficients of the damping matrix as:
\begin{align}
 \mathbf{C}_0 &= \alpha_0 \mathbf{M} + \alpha_1 \mathbf{K}_0  \\
 \mathbf{C}_k &= \alpha_1 \mathbf{K}_k
\end{align}
}

\subsection{Probability Density Functions}\label{sec:pdfs}

\begin{figure}[h]
    \centering
    \begin{subfigure}[b]{\textwidth}
       \centering
        \includegraphics[width=0.475\linewidth]{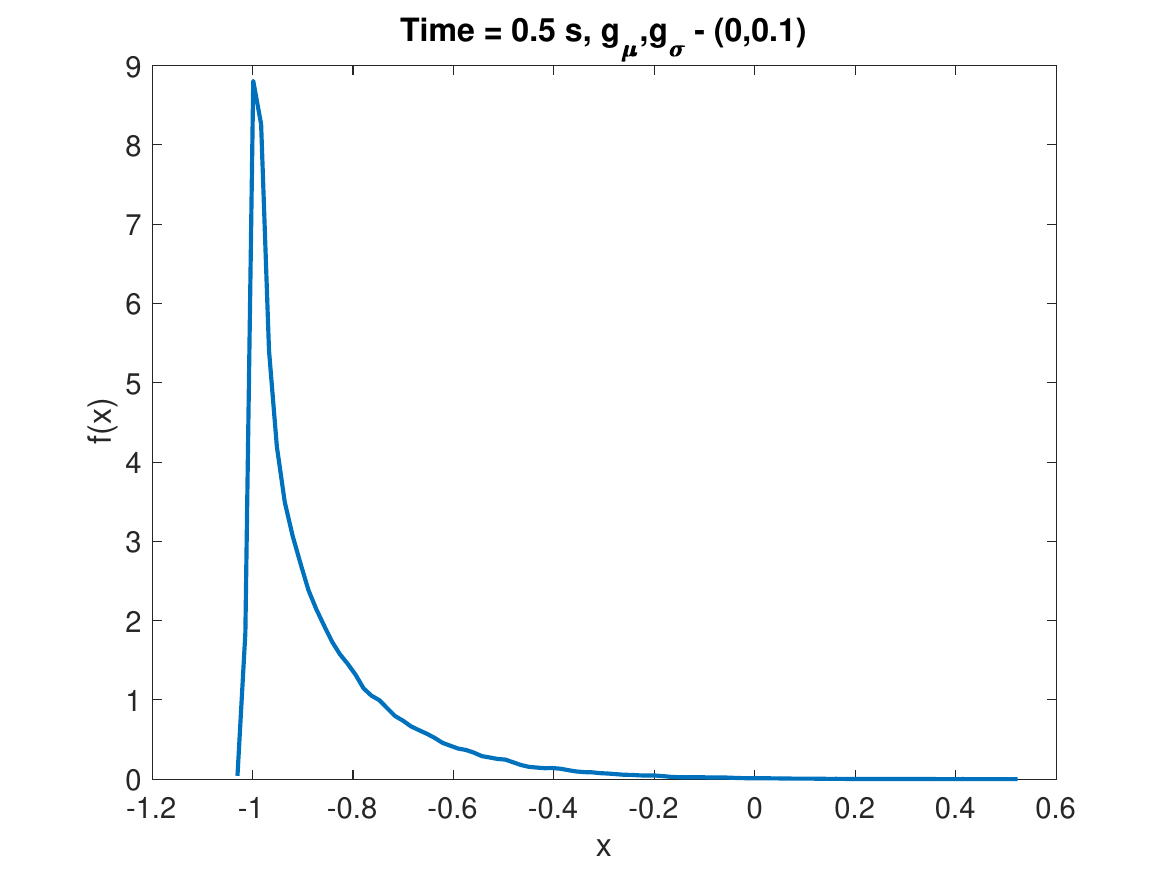}              
        \hfill
        \includegraphics[width=0.475\linewidth]{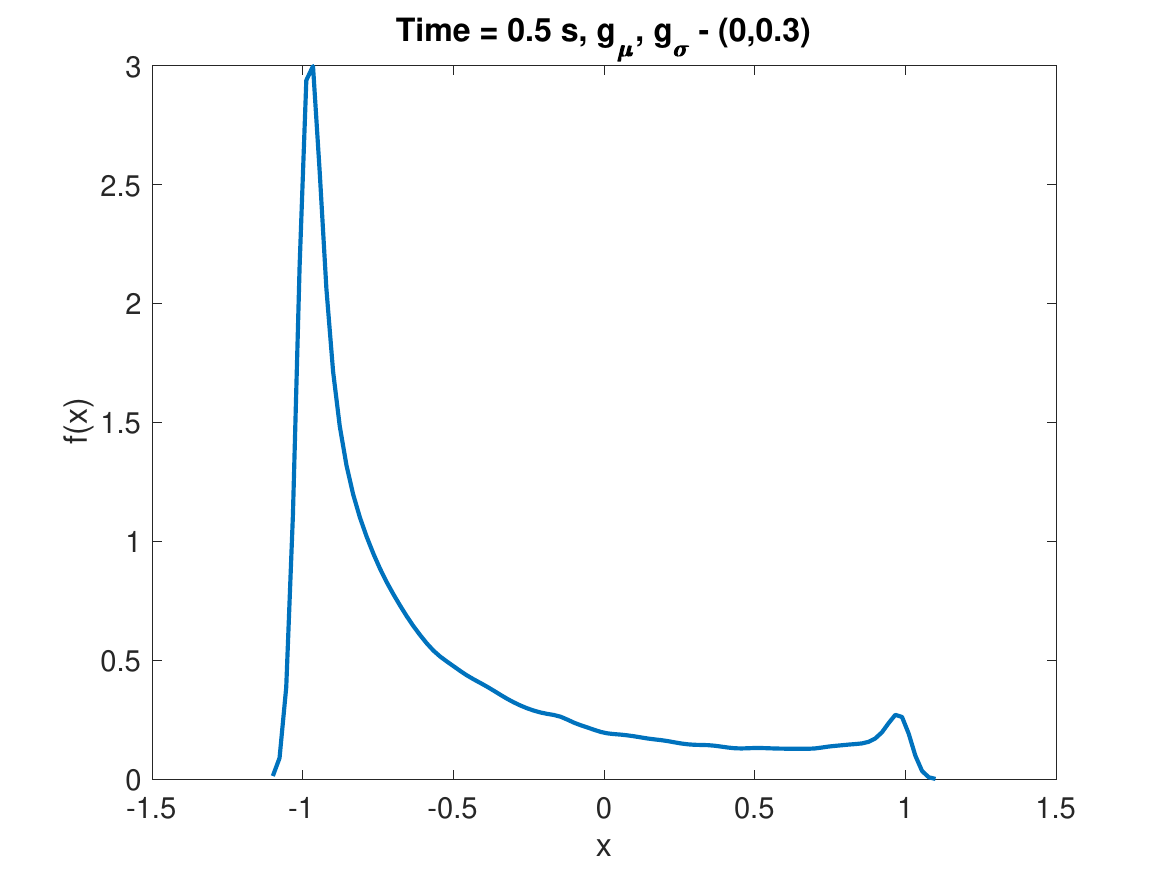}

        \includegraphics[width=0.475\linewidth]{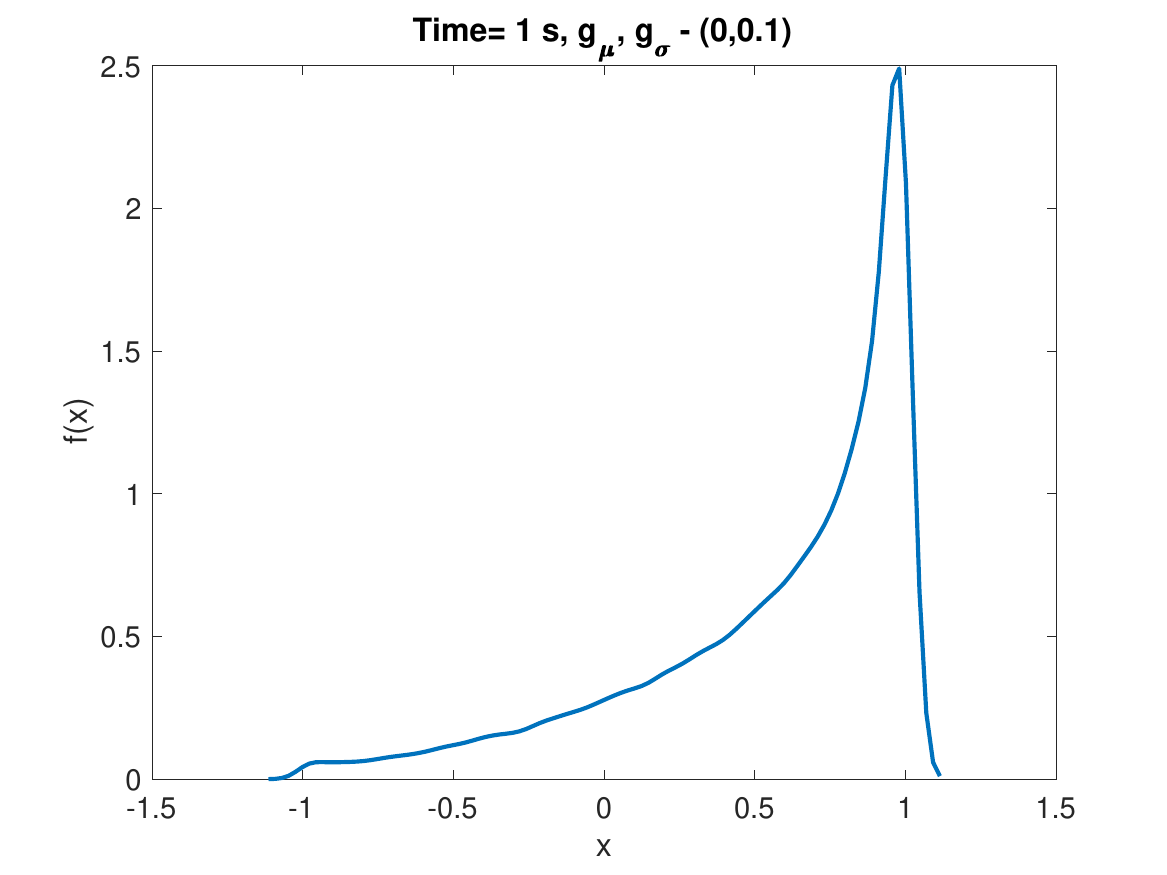}      
		\hfill
        \includegraphics[width=0.475\linewidth]{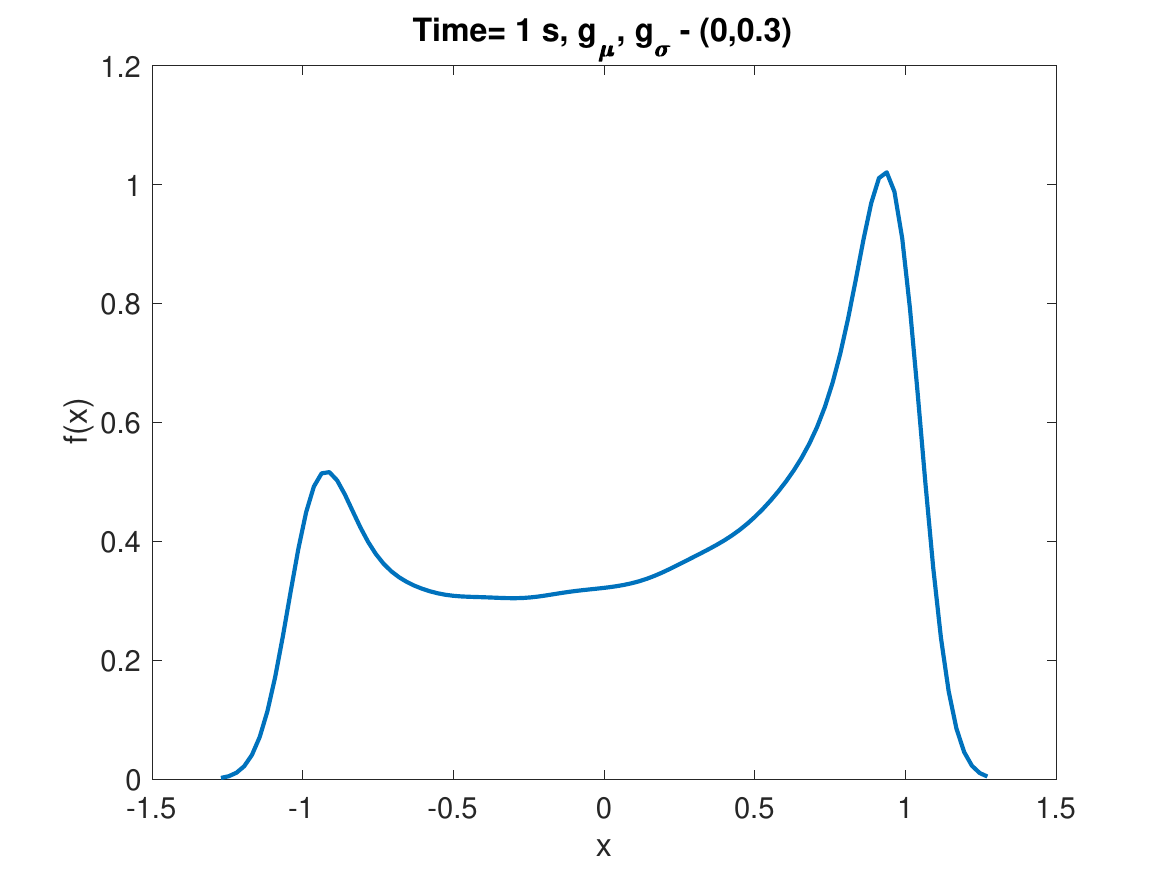} 
        
          \includegraphics[width=0.475\linewidth]{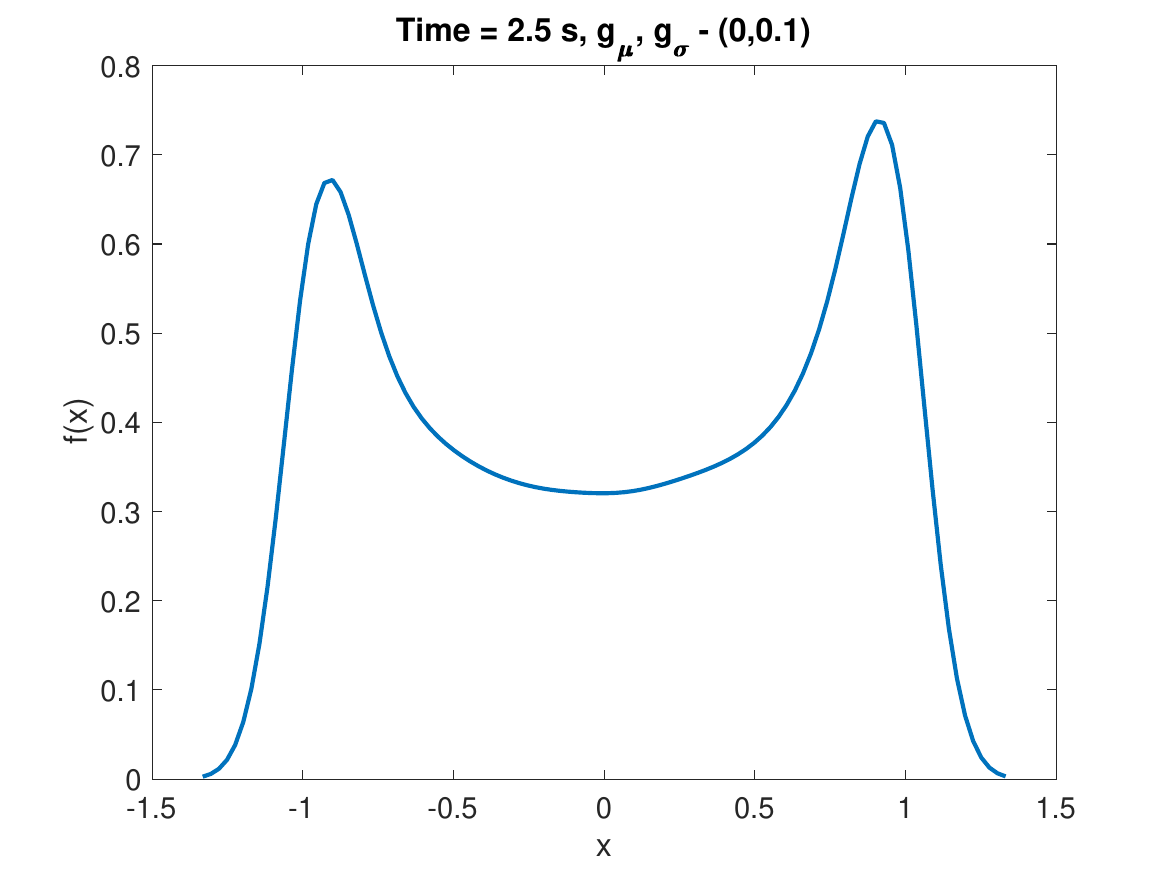}      
		\hfill
        \includegraphics[width=0.475\linewidth]{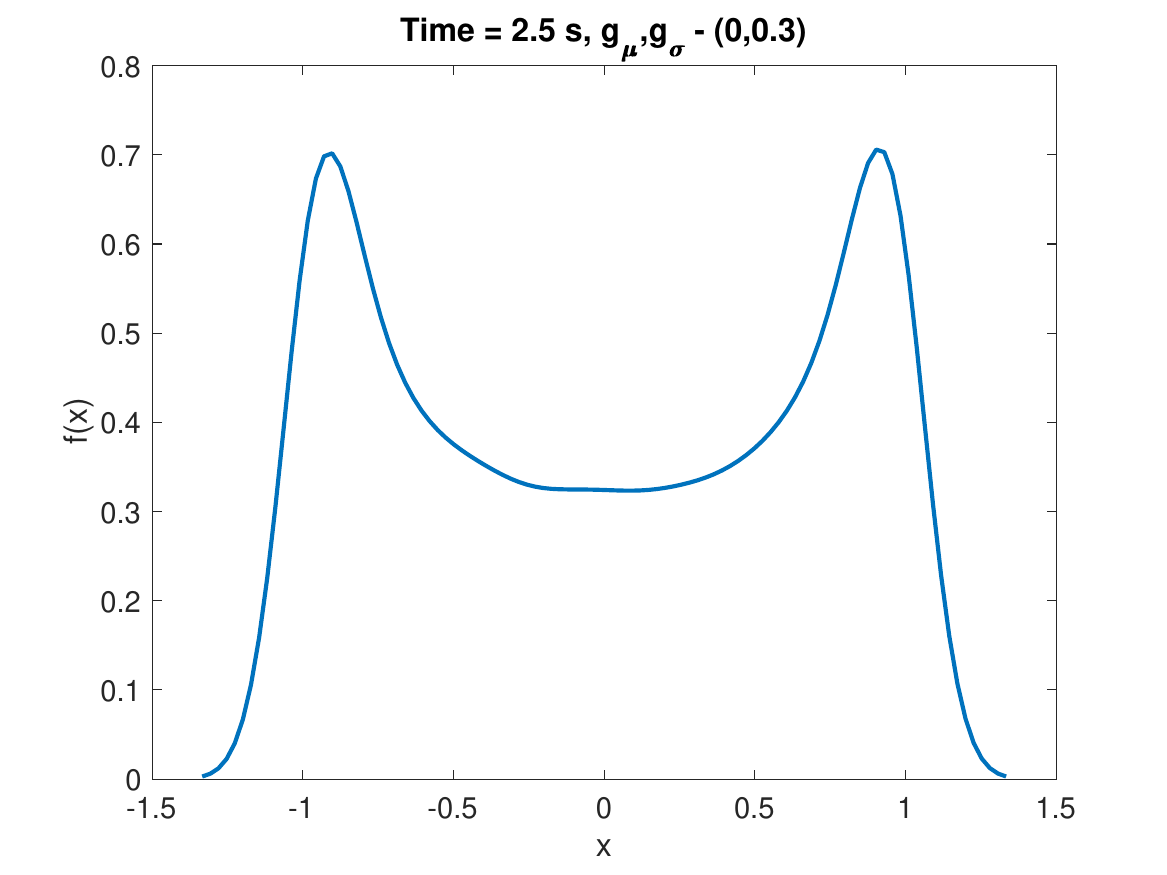} 
        
    \end{subfigure}
    
  \caption{pdfs for the acoustic pressure at $(x = 0.25,y = 0.5)$ for different time steps. Left panels with $(\mu_g = 0, \sigma_g = 0.1)$ and right panels with $(\mu_g = 0, \sigma_g = 0.3)$}\label{Fig:pdf_analytical}
\end{figure}

This section discusses the pdfs of the acoustic pressure on a two-dimensional square domain with log-normal approximation of wave speed. The wave propagation model presented in section \ref{Sec.2dwave_det} is used for the study. From the analytical solution of the wave in Eq.~(\ref{Eq.2danalytical}), $100000$ samples of the log-normal random variable model of the wave speed is used to generate the pdfs at three different time steps as shown in Fig.~\ref{Fig:pdf_analytical}. The pdfs at different time steps are highly non-Gaussian with multiple peaks. With increased input uncertainty ($\mu_g = 0, \sigma_g = 0.3$), the pdfs tend to have multiple peaks at earlier time steps (compared to $\mu_g = 0, \sigma_g = 0.1$). This shows the changing uncertainties for wave propagation problems and the necessity to adapt the PCE for long time integration of wave propagation problems \cite{hazra_TDgPC}. 


\newpage

\bibliographystyle{unsrt}  
\bibliography{references}

\end{document}